\documentclass[12pt,preprint]{aastex}
\begin{document}

\title{Wheels of Fire IV.  Star Formation and the Neutral Interstellar Medium in the Ring Galaxy AM0644-741}

\author{James~L.~Higdon\altaffilmark{1,3,4}, 
        Sarah~J.~U. Higdon\altaffilmark{1}, and
        Richard~J.~Rand\altaffilmark{2,3}
         }

\altaffiltext{1}{Department of Physics, Georgia Southern University, Statesboro, GA 30460-8031, USA}
\altaffiltext{2}{Department of Physics and Astronomy, University of New Mexico, Albuquerque, NM 87131, USA}
\altaffiltext{3}{Visiting astronomer, European Southern Observatory.}
\altaffiltext{4}{Visiting astronomer, Cerro Tololo Inter-American Observatory.}

\begin{abstract} 
We combine data from the Australia Telescope National Facility and the Swedish ESO
Submillimeter Telescope to investigate the neutral interstellar medium (ISM) in
AM0644-741, a large and robustly star-forming ring galaxy. The galaxy's ISM is concentrated 
in the $42$ kpc diameter starburst ring, but appears dominated by atomic gas, with a 
global molecular fraction ($f_{\rm mol}$) of only $0.079 \pm 0.005$. Apart from the 
starburst peak, the gas ring appears stable against the growth of gravitational 
instabilities ($Q_{\rm gas} = 2-7$). Including the stellar component lowers $Q$ 
overall, but not enough to make $Q<1$ everywhere. The ring's global star formation 
efficiency (SFE) appears somewhat elevated relative to early spirals, but 
varies around the ring by more than an order of magnitude, peaking where star 
formation is most intense. AM0644-741's star formation law is peculiar: HI 
follows a Schmidt law while H$_2$ is uncorrelated with star formation rate density.
Photodissociation models yield low volume densities in the ring, particularly in the starburst 
quadrant  ($n \approx 2$ cm$^{-3}$), implying a warm neutral medium dominated ISM.  At 
the same time, the ring's pressure and ambient far-ultraviolet radiation field lead to the 
expectation of a predominantly molecular ISM. We argue that the ring's peculiar
star formation law, $n$, SFE, and $f_{\rm mol}$ result from the ISM's $\ga100$ Myr 
confinement time in the starburst ring, which amplifies the destructive effects of 
embedded massive stars and supernovae. As a result, the ring's molecular ISM becomes 
dominated by small clouds where star formation is most intense, causing $M_{\rm H_2}$ to be 
underestimated by $^{12}$CO line fluxes: in effect $X_{\rm CO} \gg X_{\rm Gal}$ 
despite the ring's solar metallicity. The observed large HI component is primarily 
a low density photodissociation product, i.e., a tracer rather than a precursor of 
massive star formation. Such an ``over-cooked'' ISM may be a general characteristic of 
evolved starburst ring  galaxies.
\end{abstract}

\keywords{galaxies: individual (AM0644-741) $-$ galaxies: individual 
(Lindsay-Shapley Ring) $-$ galaxies: interactions
$-$ galaxies: ISM $-$ galaxies: starburst}

%\slugcomment{Accepted by the Astrophysical Journal}

\vfill \eject

\section{Introduction}

A large and growing body of evidence supports the key role played by gravitational
interactions in the formation and evolution of galaxies. In the hierarchical
structure formation framework, galaxies are assembled over cosmic time 
through multiple mergers \citep[e.g.,][]{white,bell06,jogee}.
Evidence for this can be found in the observed asymmetric, sometimes 
chaotic morphologies of a significant fraction of galaxies detected in
deep surveys \citep[e.g.,][]{williams,ferguson,conselice}. Similarly, there 
are indications that submillimeter galaxies, thought to be the precursors of 
current epoch massive bulge dominated galaxies, experience star formation rates 
(SFR) in excess of $\approx1000$ $M_{\odot}$ yr$^{-1}$ as a direct result of mergers 
\citep[e.g.,][]{smail,richards,tacconi}. Morphological evolution is also clearly 
attributable to close passages and mergers \citep[e.g.,][]{toomre74,dressler,moore,hibbard}. 
Interactions likewise appear involved with the funneling of material 
to the immediate vicinity of supermassive black holes and thus the switching on of 
active galactic nuclei \citep[AGNs; e.g.,][]{heckman,ramos}.
Detailed numerical and observational studies of strongly interacting systems 
in the local universe are thus highly relevant, especially when the 
dynamical history of the encounter can be reasonably well reconstructed, allowing 
detailed numerical simulations capable of following the evolution
of star-forming interstellar medium (ISM) that can be constrained by observations. 

Ring galaxies are striking examples of the ability of
collisions to transform both the morphology and
star-forming activity of a spiral galaxy. Numerical studies since
the mid-1970s argue persuasively that the optically prominent rings are in fact 
outwardly propagating zones of strong orbit crowding within the disk of a spiral 
induced by the near central passage of a companion galaxy
\citep[e.g.,][but see Korchagin et al. 1999, 2001 for a very different formation
mechanism]{lynds,struckhigdon,mapelli}. The expanding rings are transient but
maintain their coherence for roughly one outer-disk rotation 
period ($\sim$300 Myr). H$\alpha$ emission line imaging show the rings to be
nearly always sites of intense massive star formation, 
while star formation within the enclosed disks is greatly
quenched \citep[e.g.,][]{wof1,marston}.\footnotemark[1]
Ring galaxies are thus of interest from the standpoint of both
the triggering and suppression of star formation on galactic scales. HI interferometry
of the Cartwheel ring galaxy provides insight into how star formation is regulated in
these objects \citep{wof2}. Its outer ring possesses $\approx90\%$ of 
the galaxy's total neutral atomic ISM, resulting in very high HI surface densities
($\Sigma_{\rm HI}$), peaking at $\Sigma_{\rm HI} = 75$ $M_{\odot}$ pc$^{-2}$ for 3 kpc 
linear resolution.\footnotemark[2] At the same time, the Cartwheel's 
interior disk, inner ring, and nucleus
are HI poor, with $\Sigma_{\rm HI} \lesssim 2$ $M_{\odot}$ pc$^{-2}$ ($3 \sigma$).
Star formation triggering/suppression can be understood in terms of a 
local gravitational stability criterion, as parameterized by the 
surface density of the neutral atomic ISM.
The Cartwheel's outer HI ring exceeds the gas surface density threshold 
for robust star formation \citep{kennicutt89}, while
the enclosed gas poor disk is stable against the growth of 
large scale gravitational perturbations and subsequent star formation.
Whether simple gravitational stability ideas can explain the distribution
of star formation in other large and evolved ring galaxies remains to be seen.
\footnotetext[1]{One notable exception is the ``Sacred Mushroom'' galaxy (AM1724-622),
whose red and low surface brightness optical ring shows no evidence of recent star 
formation, and indeed suggests that the pre-collision galaxy was an S0 \citep{wallin}.}
\footnotetext[2]{HI surface density and column density 
are related: 1 $M_{\odot}$ pc$^{-2}$ = 1.25 $\times$ 10$^{20}$ atoms cm$^{-2}$.}

Stars form in the cold molecular ISM, so HI 
only tells part of the story. A more complete understanding 
of star formation triggering and suppression in ring galaxies
requires the determination of the {\em molecular} ISM's properties 
as well. This has proven difficult since ring galaxies as a rule 
are not luminous in the $^{12}$CO rotational transitions \citep[e.g.,][]{hor95},
which are the standard tracers of H$_2$ in galaxies. 
This has been attributed to low metal abundances in the ring
from the ``snow-plowing'' of largely unenriched gas from the outer disk
as the ring propagates outward. 
In this paper we examine the atomic and molecular ISM of the southern ring galaxy
AM0644-741 \citep{arp}, also known as the  ``Lindsay-Shapley Ring'', using
data from the Australia Telescope Compact Array (ATCA)\footnotemark[3]
and the 15 m Swedish ESO Submillimeter Telescope (SEST)\footnotemark[4]
at La Silla, Chile. $B$-band and H$\alpha$ images are shown in Figure 1
and a summary of AM0644-741's global properties is given in Table 1.
The ring is very large ($D_{\rm ring} = 95\arcsec$, or
$42$ kpc at the assumed distance), with only those of the Cartwheel and AM1724-622 (the
``Sacred Mushroom'') being larger \citep[][]{wof1,wof2,wallin}.
Optical spectroscopy by \citet[][hereafter FMA]{fma} showed the
ring to be rich in giant star-forming complexes, and expanding in
accord with collisional models. H$\alpha$ imaging by \citet[][hereafter HW97]{wof3}
showed massive star formation to be confined to a pair of (apparently) interlocking rings,
with a total SFR$=3.6$ $M_{\odot}$ yr$^{-1}$ with no extinction correction. (An
unresolved nuclear source contributes $\approx1\%$ of the total $F_{\rm H\alpha}$.)
The offset nucleus and asymmetric distribution of H$\alpha$ emission
point to an off-centered collision with the elliptical $1.8\arcmin$ ($49$ kpc in projection)
to the southeast in Figure 1. $N$-body/smoothed particle hydrodynamics
(SPH) models of this system have been presented in \citet{antunes}. \citet{hor95}
reported the detection of $^{12}$CO($J=1-0$) emission from
the galaxy's nucleus using the SEST (though at a velocity $300$ km s$^{-1}$
smaller than $v_{\rm sys}$). While nuclear $^{12}$CO emission does not
guarantee the detection of molecular gas in the starburst rings \citep[e.g.,][]{hor98},
it motivated the work presented in this paper.
\footnotetext[3]{The Australia Telescope Compact Array is part 
of the Australia Telescope, which is funded by the Commonwealth of Australia for operations 
as a National Facility managed by CSIRO.}
\footnotetext[4]{The SEST is operated jointly 
by the European Southern Observatory (ESO) and the Swedish National Facility
for Radio Astronomy, Chalmers Center for Astrophysics and Space Science.}

We use the ATCA HI data to trace the distribution and kinematics of AM0644-741's 
neutral atomic ISM on $\approx 3$ kpc scales, which is sufficient to allow direct 
comparisons with published VLA HI observations of the Cartwheel. The radio continuum
data additionally provides an obscuration free tracer of current massive star
formation. The SEST is used to measure both 115 GHz $^{12}$CO(J = 1-0) 
and 230 GHz $^{12}$CO(J = 2-1) line emission in the ring galaxy. We take 
advantage of improved receiver sensitivity, particularly at $230$ GHz, to 
reexamine the galaxy's molecular gas content. A major advantage of 230 GHz observations 
is the $22\arcsec$ FWHM beam, which allows the measurement of $^{12}$CO emission 
in the ring, disk, and nucleus separately, permitting the best census of the molecular 
ISM in a robustly star-forming ring galaxy in the pre-ALMA era.

We also utilize far-ultraviolet (FUV) images of AM0644-741 obtained
with the Galaxy Evolution Explorer (GALEX), ground-based optical imaging
obtained at the Cerro-Tololo Inter-American Observatory (CTIO)\footnotemark[5],
and images at 4.5, 24, and 70 $\mu$m from the Spitzer
Space Telescope \citep[$Spitzer$,][]{werner}\footnotemark[6],
as well as optical spectra of AM0644-741's ring using ESO's 3.6 m telescope.
The combined data are used to 
determine basic properties of the neutral atomic and molecular ISM throughout AM0644-741's
ring, nucleus and enclosed disk for comparison with star formation
activity as measured in the FUV, optical, infrared, and radio. 
Among the questions we wish to address is whether the ring's neutral ISM is 
dominated by atomic or molecular gas, and which local property is
primarily responsible for setting the molecular fraction. We also wish to
know if the distribution of star formation in AM0644-741's ring and disk can be
understood in terms of simple gravitational stability arguments as is the case
in the Cartwheel, and whether or not the inclusion of the molecular ISM 
significantly modifies the results of such an analysis. H$_2$ mass measurements are
also necessary for estimates of star formation efficiency (SFE), which
have been determined in nearby galaxies \citep[e.g.,][]{young,rownd}.
In short, how do properties of the star-forming ISM in an evolved ring galaxy compare
with other systems?
\footnotetext[5]{CTIO is operated by AURA, Inc., under contract with the National Science Foundation.}
\footnotetext[6]{Spitzer Space Telescope is operated by JPL, California Institute of 
Technology for the National Aeronautics and Space Administration.}

This paper is organized as follows:
The observations and data reduction are outlined in Section 2. The results derived
using these observations are presented in Section 3, including the ring's metallicity, 
the distribution and kinematics of HI, the relative proportions of molecular 
and atomic gas, star formation efficiency, and their variations around the ring. 
We also consider the ring's gravitational stability, and whether or not star formation
follows a Schmidt law. In Section 4 we consider the role played by pressure and 
ambient FUV radiation field in determining the ring's molecular fraction.
We also discuss our results in the context of an alternate view of ring galaxies, in 
which the ring represents a self-sustaining propagating starburst. Our conclusions are
given in Section 5. Throughout this paper we adopt a flat $\Lambda$CDM cosmology based 
on the $7$ year WMAP results \citep{wmap7},
$\Omega_{\rm M}$=0.27, $\Omega_{\rm \Lambda}$=0.73, and a Hubble constant of 
70.3 km s$^{-1}$ Mpc$^{-1}$, giving AM0644-741 a luminosity distance of $96.9$ Mpc and
a linear scale of $0.45$ kpc arcsec$^{-1}$.

\section{Observations, Data Reduction, and Analysis}

	\subsection{GALEX Ultraviolet Imaging}
GALEX provides simultaneous FUV ($\lambda_{\rm eff} = 1516$ \AA, 
$\Delta\lambda_{\rm FWHM} = 269$ \AA) and NUV ($\lambda_{\rm eff} =
2267$ \AA, $\Delta\lambda_{\rm FWHM} = 616$ \AA) imaging capabilities. Details on
the telescope optics, instruments, calibration, and data reduction pipeline
can be found in \citet{martin} and \citet{morrissey05,morrissey07}. Figure 2
shows an archival FUV image of AM0644-741 taken on 20 September 2003
($t_{\rm int}=3562$ sec), which was processed nominally with the $GALEX$ pipeline (version 3).  
The angular resolution is $4.3\arcsec$ FWHM (1.9 kpc) and the point-source
sensitivity is estimated to be $7.2 \times 10^{-18}$ erg s$^{-1}$ cm$^{-2}$ \AA$^{-1}$
($3 \sigma$).

	\subsection{Optical Imaging and Spectroscopy}
AM0644-741 was imaged in $B$, $V$, and $I$-bands on the nights of
1991 February 18-19 using the CTIO 1.5 m telescope. The imaging array was a 
Tektronics 1024 CCD (TEK1) located at the f/13.6 Cassegrain focus, giving a pixel 
scale of $0.43$ arcsec pix$^{-1}$ and a $7.\arcmin4$ square field. Multiple 600-second 
integrations were made in each filter. The nights were photometric, and $UBVRI$ standards
\citep[E4][]{graham82} were used to calibrate the data. Reduction and calibration was routine 
and done using $IRAF$ and IDL.  The seeing was mediocre ($\approx2\arcsec$ FWHM)
but adequate given the angular resolution of the other data. Zero-points in \citet{bessel79}
were used to convert from magnitudes to flux units.

Star-forming complexes along the western half of AM0644-741's ring were observed with 
the EFOSC2 imaging spectrograph on the $ESO$ $3.6$ m telescope at La Silla on 
2002 December 6. Spectra covering $6200-7800$ \AA ~were obtained in multi-object mode (MOS) 
using aperture masks, together with the Loral CCD ($2048\times 2048$, $15$ ~$\mu$m pixels) 
and grism $\#10$ ($600$ ~mm$^{-1}$; $0.95$ \AA ~pix$^{-1}$ dispersion). 
Three slitlet positions are shown in Figure 1. Apertures situated
on blank sky regions were used for sky-subtraction. Blue spectra covering
$3500-5500$ \AA ~were also taken for these masks using grism $\#3$ ($400$ ~mm$^{-1}$; 
$1.5$ \AA ~pix$^{-1}$ dispersion). Flat-fielding problems rendered these data unusable, 
however, and only the red spectra are used in the following analysis.
Standard routines in $IRAF$ were used to extract ($apall$) and calibrate 
($identify$, $calibrate$) the red EFOSC2 spectra, which are shown in
Figure 3. Emission line fluxes and uncertainties were measured using $splot$, and 
are discussed below.

	\subsection{$Spitzer$ IRAC and MIPS Imaging}
AM0644-741 was imaged at 4.5 $\mu$m with IRAC on 2005 April 3 as part of our Guaranteed Time 
Program (PID 197). Preliminary data processing and calibration were carried out using
the standard IRAC pipeline (S12.0.2), producing a set of basic calibrated data products (BCDs).
Additional processing was done using the MOPEX software package \citep{makovoz}.
This consisted of ``self-calibration'' to improve flat-fielding and background subtraction
after the deletion of defective ``first scans'', and re-mosaicking. The resulting image is 
shown in Figure 2. At 4.5 $\mu$m the diffraction limited 
angular resolution is $3.0\arcsec$ (1.3 kpc). The $1 \sigma$ sensitivity of the final
image is 0.003 MJy sr$^{-1}$, though it should be noted that the photometric zero-point is 
accurate to $\pm 5\%$.

$Spitzer$ MIPS observations of AM0644-741 at 24 and 70 $\mu$m were also obtained on 2005 April 3 as
part of the same GT program. As with the IRAC data, preliminary data processing and calibration were 
carried out using the standard MIPS pipeline (version S13.2) as described in \citet{gordon},
with additional processing in MOPEX to improve the
flat-fielding, background removal, and final mosaic. The resulting 24 and 70 $\mu$m images are also 
shown in Figure 2. The diffraction limited resolution is $5.7\arcsec$ FWHM (2.5 kpc) at 24 $\mu$m and 
$16.6\arcsec$ FWHM (7.3 kpc) at 70 $\mu$m. Prominent Airy diffraction rings are visible around 
several of the brightest point-sources in the outer ring at 24 $\mu$m. The $1 \sigma$ sensitivity of
the MIPS images is 0.013 and 0.46 MJy sr$^{-1}$ at 24 and 70 $\mu$m respectively, 
or $3.1 \times 10^{-7}$ and $1.1 \times 10^{-5}$ Jy arcsec$^{-2}$, though again,
the zero-point is accurate to $\pm 5\%$ \citep[see][for a discussion of this and other calibration issues]{engelbracht}.

	\subsection{ATCA HI Interferometry}
AM0644-741 was observed using the ATCA between 1996 August to
1998 May with a variety of array configurations
(Table 2).  The baselines ranged from 30.6 m to 5.97 km with a total
of 114 hr on target.  The ATCA's correlator was configured
to provide a spectral line bandpass of 8 MHz subdivided into 512
channels separated by 3.3 km s$^{-1}$ and centered on a frequency
of 1390.0 MHz.  Simultaneous continuum observations were made with
a 128 MHz wide bandpass (32 channels) centered at 1380.0 MHz. The flux
scale and bandpass shape were calibrated using observations of radio source 
1934-638 (S$_{\rm 1390 MHz}$ = 14.95 Jy), while drifts in the antenna gain
coefficients were corrected using periodic observations of the
nearby radio source 0454-801 (S$_{\rm 1390 MHz}=0.9$ Jy).
 
The UV data from each observation was edited and calibrated 
separately using an {\em ATNF} implementation of $NRAO$'s  Astronomical Image 
Processing System (AIPS) package.  This included the correction for Doppler
shifts during each scan since the ATCA correlator does not frequency track.
Data from all configurations were concatenated into final
line and continuum data sets using the AIPS task DBCON, which were separately 
imaged and CLEANed using the task IMAGR.  The continuum was subtracted from the 
spectral line data set prior to imaging using a combination of the tasks UVSUB and UVLIN.  
HI line image cubes were made using both {\em natural} ($\theta_{\rm FWHM} = 12.\arcsec2
\times 9.\arcsec4$) and {\em robust} ({\em ROBUST = -0.5}; $\theta_{\rm FWHM} = 7.\arcsec9 
\times 6.\arcsec7$) weighting. Eight successive 3.3 km s$^{-1}$ channels
were averaged together during imaging, which yielded (after discarding
some data near the bandpass edges) 44 channels separated by 27.5 km s$^{-1}$.
The rms noise in the {\em natural} weighted channel maps was 0.38 mJy beam$^{-1}$,
while that in the {\em robust} weighted channel maps was 0.48 mJy beam$^{-1}$, which is close to
the expected sensitivity given the low elevation of AM0644-741 ($\la 40^{\circ}$)
during a significant fraction of the observations.

Integrated HI and flux-weighted velocity
images were made using the AIPS routine MOMNT and are shown in Figure 4,
where they have been superposed on optical $B$-band and $H\alpha$ gray scale images.
HI masses in solar units are calculated using
\begin{equation}
{\rm
M_{\rm HI} = 2.36 \times 10^{5}~~D^{2}_{\rm Mpc}~~\int{S_{\rm \nu}(v)~dv,}
}
\end{equation} 
where $S_{\rm \nu}(v)$ is the flux density in Janskys for a given
velocity channel and $dv$ represents the channel separation in km s$^{-1}$.
This equation is valid in the optically thin limit. We measured the run of 
radial velocity ($v_{\rm rad}$) around the ring in $10^{\circ}$ increments
using the flux-weighted velocity image shown in Figure 4. This will be used in 
the analysis of the ring's kinematics in Section $3$. The total HI
spectrum for AM0644-741 using the {\em robust} weighted data is presented
in Figure 11. All radial velocities reported
in this paper assume the optical/barycentric definition.

A {\em robust} weighted 20 cm continuum image 
($\theta_{\rm FWHM} = 7.\arcsec9  \times 6.\arcsec7$)  was made by averaging the 16 4-MHz 
continuum channels least affected by radio interference.  The $rms$ in the cleaned
image is 20 $\mu$Jy beam$^{-1}$, which is also near the expected
sensitivity.  This is shown contoured on the H$\alpha$ gray scale image in Figure 4.

	\subsection{SEST $^{12}$CO Mapping}

$^{12}$CO observations were carried out on the nights of 1999 July 18-22
and 2000 September 1-5 using the SEST. The telescope was operated in dual 115/230 GHz 
mode with SIS receivers.  Both back-ends were acoustic-optical 
spectrometers providing total band-passes of 1300 and 1600 km 
s$^{-1}$ in the redshifted $^{12}$CO(1-0) and $^{12}$CO(2-1) 
lines, with channel separations of 1.3 and 1.6 km s$^{-1}$ respectively.
The telescope FWHM beam widths at these frequencies
are 43$''$ (18.9 kpc) and 22$''$ (9.7 kpc).  Dual beam-switching mode
using a focal plane chopper wheel rotating at 6 Hz ($2.5-12\arcmin$ beam throw) 
was used to ensure flat baselines. Pointing was checked regularly
throughout the run using SiO masers, and is believed to be accurate to within 5$''$.
The highest quality data were obtained over the September 2000 run. System temperatures 
were exceptionally low on these nights, with T$_{\rm sys}$ = 230-300 K at 230 GHz.  

We searched the ring for $^{12}$CO(2-1) emission using 22$''$
beam separations, switching to 11$''$ spacing when line emission was
detected.  Altogether, we observed 14 positions in AM0644-741's
ring as well as single pointings at the optical nucleus and enclosed disk (Table 3).
At each observed position, the acoustic-optical spectrometer's central frequency was set
to correspond to the flux-weighted HI systemic velocity.
For the ring galaxy's nucleus and disk the central frequency corresponded
to the optical systemic velocity (6750 km s$^{-1}$; FMA).
The observed positions and beam FWHMs are shown in Figures 5 and 6 superposed on 
a H$\alpha$ gray scale image, along with the {\em robust} weighted HI contours. 
The 115 and 230 GHz spectra were reduced and analyzed 
in an identical manner using the CLASS software package.\footnotemark[7]
\footnotetext[7]{The CLASS analysis package is part of the GILDAS collection of software
produced and maintained by $IRAM$. See www.iram.fr/IRAMFR/GILDAS for further information.}
Scans showing curved baselines or ripples were discarded. The
remaining scans at each of the 14 observed positions
were then coadded, smoothed to a resolution of 30 km s$^{-1}$ and rebinned to
a 10 km s$^{-1}$ channel separation.  Linear baselines were then subtracted.
The final $^{12}$CO(1-0) and $^{12}$CO(2-1) spectra are shown in Figures 5 and 6. 
Typical {\em rms} for the final 115 and 230 GHz spectra are $\approx$1.5 mK and $\approx$1.0 mK
respectively. Integrated line intensities are defined as
\begin{equation}
{\rm
I_{1-0} = \int{T^{*}_{\rm A}(J=1-0)~dv} ~~~~~~and~~~~~~  I_{2-1} = \int{T^{*}_{\rm A}(J=2-1)~dv}
}
\end{equation}
for the 115 and 230 GHz lines. Where a single narrow line component was apparent, 
line intensities and uncertainties were derived by fitting single Gaussian profiles
using routines in CLASS. Otherwise, intensities were derived by summing the 
spectra over the full range of channels showing emission. Uncertainties in this case were 
estimated using $\sigma = \sigma_{\rm base} \sqrt{\Delta v_{\rm res} ~\Delta v_{\rm CO} }$,
where $\sigma_{\rm base}$ is the {\em rms} in the channels not showing line emission,
$\Delta v_{\rm res}$ is the velocity resolution and $\Delta v_{\rm CO}$ is the
line FWHM, both in km s$^{-1}$. Derived line intensities and uncertainties, peak T$_{\rm A}^{*}$, 
central velocities and widths (FWHM), and {\em rms} are listed in Table 4.

Because of the higher sensitivity and smaller beam diameter, we calculate H$_2$ masses
with the 230 GHz spectra using
\begin{equation}
{\rm
M_{\rm H_2} = {{ 2 m_{\rm p} I_{2-1} R({{1-0}\over{2-1}}) X_{\rm CO} \Delta\Omega_{\rm 230}}\over{\epsilon_{\rm mb}
 \epsilon_{\rm source}}},
}
\end{equation}
where $m_{\rm p}$ is the proton mass, R(${1-0}\over{2-1}$) is the 
locally determined $I_{\rm 1-0}/I_{\rm 2-1}$ ratio, $\epsilon_{\rm mb}$ is the SEST
main-beam efficiency ($\epsilon_{\rm mb}$= 0.70/0.50 at 115/230 GHz, 2002 $SEST ~Handbook$),
$X_{\rm CO}$ is the $I_{\rm CO}-N_{\rm H_2}$ conversion factor, and $\Delta\Omega_{\rm 230}$ 
is the beam area at $230$ GHz.  We further correct the main-beam efficiency by the 
source-coupling factor ($\epsilon_{\rm source}$) as well as the effect of overlapping beams. 
Both were calculated by numerically simulating the observations under the assumption that 
the molecular gas is distributed to first order like the  H$\alpha$ ring. This should be a 
reasonable approximation given the strong orbit crowding evident in the multiwavelength
images.\footnotemark[8] 
The H$_2$ surface density is defined $\Sigma_{\rm H_2} = M_{\rm H_2}/\Delta\Omega_{\rm ring}$, where 
$\Delta\Omega_{\rm ring}$ is the ring's solid angle, i.e., a $\approx$9$\arcsec \times$ 22$\arcsec$ 
rectangle, ignoring the slight curvature of the ring within a beam. Unless otherwise noted, we 
will quote HI and H$_2$ masses and surface densities without a correction for helium, which
entails scaling by a factor of $1.36$ \citep{kennicutt89}. This correction will
be necessary when we consider the ring's gravitational stability and pressure.

\footnotetext[8]{The assumed source geometry affects the value
of R(${1-0}\over{2-1}$). For sources small with respect to both the $115$ and $230$ GHz beams
R(${1-0}\over{2-1}$) is $({{I_{\rm 1-0}}\over{I_{\rm 2-1}}}) 
({{\Delta\Omega_{\rm 115}}\over{\Delta\Omega_{\rm 230}}}) = 4~({{I_{\rm 1-0}}\over{I_{\rm 2-1}}})$.
To account for the fact that the size of the CO emitting region (i.e., the ring) is 
larger in the $115$ GHz beam, R(${1-0}\over{2-1}$) becomes
$({{I_{\rm 1-0}}\over{I_{\rm 2-1}}})({{\Delta\Omega_{\rm 115}}\over{\Delta\Omega_{\rm 230}}})
({{\Delta\Omega^{\rm source}_{\rm 230}}\over{\Delta\Omega^{\rm source}_{\rm 115}}}) 
\approx 2~({{I_{\rm 1-0}}\over{I_{\rm 2-1}}})$.}

\section{Results}

	\subsection{Metal Abundances in the Ring and the CO-H$_2$ Conversion Factor}

Using $^{12}$CO intensity as a proxy for H$_2$ requires the correct $X_{\rm CO}$. 
Empirical calibrations for  molecular clouds in the Galactic plane 
using a variety of techniques find  
$X_{\rm CO} \approx 2 \times 10^{20}$ mol cm$^{-2}$ (K km s$^{-1}$)$^{-1}$ 
\citep[e.g.,][]{dickman,solomon,dame}, implying that physical conditions in the ISM (e.g.,
temperature, density, and ambient radiation field) average out to a 
great extent when integrated over a large number of clouds, i.e., for large 
beam sizes. One property that does not average out is the ISM's metal
abundance. Low metallicity reduces CO emission directly by
decreasing the available C and O needed to form CO, and indirectly
by reducing the dust columns required to shield the more fragile CO molecule 
from photodissociation within individual clouds. Since the formation and survival of 
H$_2$ is much less sensitive to metallicity, $X_{\rm CO}$ is expected
to depend on abundance \citep{maloney}. The precise nature of the
dependence is a matter of contention. Determinations of $X_{\rm CO}$ in other
galaxies that assume molecular clouds are in virial equilibrium find fairly
weak metallicity dependence \citep[e.g.,][]{wilson}, while studies using infrared
dust emission to estimate the total gas mass find strong metallicity
dependence \citep[e.g.,][]{israel,leroy09}, and in particular, $X_{\rm CO}$ that 
are $\approx3-10$ times larger than the Galactic value. It is thus important
to determine the metallicity in order to at least constrain the allowed range in
derived H$_2$ masses and surface densities.

We use H$\alpha$, [N~II] $\lambda 6584$ and [S~II] $\lambda\lambda 6717, 6731$ 
emission line fluxes extracted from the red EFOSC2 spectra in Figure 3 to determine oxygen 
abundances along the western half of AM0644-741's ring.
This is done following the empirical calibration given in Table 6 of \citet{nagao},
\begin{equation}
{\rm
log( F_{\rm [N~II]\lambda 6584}/F_{\rm H\alpha}) = 96.64 - 39.94y + 5.22y^{2} - 0.22y^{3}
}
\end{equation}
and 
\begin{equation}
{\rm
log( F_{\rm [N~II]\lambda 6584}/F_{\rm [S~II]}) = -80.62 + 31.32y - 4.10y^{2} + 0.18y^{3}
}
\end{equation}
where $y\equiv 12 +$ log(O/H) and where $F_{\rm [S~II]}$ is the sum of
the $6717$ and $6731$ \AA\ [S~II] lines. These relations were established using spectra 
of extra-galactic HII regions spanning $7.0 \la 12$+log(O/H)$ \la 9.4$   
taken from the literature and the Sloan Digital Sky Survey 
database. In each instance accurate oxygen 
abundances were determined either directly using the temperature sensitive 
[O~III]$\lambda4363$ line or through detailed photoionization modeling. 
The [N~II] $\lambda 6584$/H$\alpha$ and [N~II] $\lambda 6584$/[S~II] ratios have 
the added advantage of being insensitive to extinction. 
We find $F_{\rm [N~II] \lambda 6584}/F_{\rm H\alpha} = 0.36 \pm 0.02$, 
$0.38 \pm 0.01$, and $0.38 \pm 0.02$ at the three positions, which yields
$12$+log(O/H)=$9.04 \pm 0.15$, $9.09 \pm 0.14$, and $9.09 \pm 0.14$, respectively,
using Equation (4), with the uncertainty dominated by the dispersion in Nagao et al.'s 
graph (see their Figure 6). For the same three regions we measure 
$F_{\rm [N~II] \lambda 6584}/F_{\rm [S~II]} = 1.05 \pm 0.18$, $1.51 \pm 0.07$, and 
$1.51 \pm 0.12$, respectively, which gives using Equation (5)
$12$+log(O/H)=$8.96 \pm 0.10$, $9.11 \pm 0.06$, and $9.10 \pm 0.08$. The two sets of
oxygen abundances are in good agreement (imperfect night-sky line subtraction
may be partially responsible for the lower abundance at P5 using 
$F_{\rm [N~II] \lambda 6584}/F_{\rm [S~II]}$). Both line ratios 
give $12 +$ log(O/H) $\approx9.09$, which significantly exceed recent determinations 
of the Sun's oxygen abundance, e.g., $12$+log(O/H)$_{\odot} = 8.73-8.79$ \citep{caffau}. 
The western half of AM0644-741's ring thus exceeds solar metallicity and
shows no significant azimuthal variation despite (as we show below) large changes in 
$\Sigma_{\rm HI}$, $\Sigma_{\rm H\alpha}$, and molecular fraction. We 
therefore adopt the Galactic $I_{\rm CO}-N_{\rm H_2}$ determined by 
\citet[][$X_{\rm CO} = (2.3 \pm 0.3) \times 10^{20}$ mol cm$^{-2}$ (K~km~s$^{-1}$)$^{-1}$]{strong}
for AM0644-741's ring.

        \subsection{Star Formation in AM0644-741 Revisited}

For a distance of 96.9 Mpc the ring galaxy's H$\alpha$ luminosity is 
$(6.13 \pm 0.06) \times 10^{41}$ erg s$^{-1}$ (HW97) after correcting  
for Galactic extinction \citep[$A_{\rm V}=0.5$;][]{rc3}.
This corresponds to SFR$ = 4.9 \pm 0.1$ $M_{\odot}$ yr$^{-1}$ \citep{kennicutt98a}, 
giving AM0644-741 one of the highest SFRs among ring galaxies.\footnotemark[9] 
Internal extinction may cause the SFR to be significantly underestimated,
especially if the ISM is highly concentrated in the ring 
\citep[e.g., $A_{\rm V} \approx 2$ in the Cartwheel's ring,][]{fh}, representing a
$\approx 4$-fold increase in $L_{\rm H\alpha}$). We therefore consider three independent 
methods to determine internal $A_{\rm H\alpha}$ in the ring.
\footnotetext[9]{Excluding systems with obvious AGN, the median SFR in Marston \& Appleton's 
sample of ring galaxies is 2.8 $M_{\odot}$ yr$^{-1}$ after correcting for foreground 
extinction.}

		\subsubsection{Optical-Infrared Estimates of the SFR}
Emission at 24 and 70 $\mu$m from galaxies originates in dust illuminated either 
by an active nucleus or young stars. The H$\alpha$ 
and MIPS images clearly show that AM0644-741's mid- and far-infrared ($FIR$) emission
originates in the star-forming ring; the unresolved nuclear component makes a 
negligible contribution. This permits two independent approaches to
determine the extinction free SFR - calculating SFR directly
from the ring galaxy's infrared luminosity and by combining $L_{\rm H\alpha}$ and 
$L_{\rm 24 \mu m}$  to derive $A_{\rm H\alpha}$.

\citet{kennicutt98b} found SFR$ = 1.72 \times 10^{-10} ~L_{\rm IR}$ 
$M_{\odot}$ yr$^{-1}$, where $L_{\rm IR}$ is the $8-1000 ~\mu$m luminosity, which in units of
$L_{\odot}$  is $L_{\rm IR} = 1.8 \times 10^{12} (13.48~L'_{\rm 12} + 5.16~L'_{\rm 25} + 
2.58~L'_{\rm 60} + L'_{\rm 100})$ for $IRAS$ luminosity densities $L'_{\rm i} = 4 \pi D^{2}_{\rm L} F_{\rm i}$ 
\citep{sanders96}. Here $F_{\rm i}$ is the $IRAS$ flux density in Janskys and $D_{\rm L}$ is the source's 
luminosity distance. Infrared flux densities for AM0644-741 were taken from the {\em IRAS Faint Source Catalog}
\citep{moshir} and are listed in Table 1. We find
$L_{\rm IR} = (4.9 \pm 0.2)  \times 10^{10}$ $L_{\odot}$, giving 
SFR$ = 7.7 \pm 0.3$ $M_{\odot}$ yr$^{-1}$. For the above $L_{\rm H\alpha}$,
$A_{\rm H\alpha} \approx 1$ is implied.

We next use the tight correlation between 24 $\mu$m flux densities and Pa~$\alpha$ line 
fluxes in M51a's star-forming regions \citep[][hereafter K07]{kennicutt07} to 
derive $A_{\rm H\alpha}$. We first measure the 24 $\mu$m and 
H$\alpha$ luminosities ($L_{\rm 24\micron}$ and $L^{'}_{\rm H\alpha}$) within $22\arcsec$ FWHM apertures 
centered on the ring positions observed with the SEST. These are shown (after normalizing by
beam area) as a function of ring position angle in Figure 7. $L^{'}_{\rm H\alpha}$ are first corrected
for $0.33$ magnitudes of Galactic foreground extinction. Internal extinction (assuming a 
foreground screen geometry) is calculated using
\begin{equation}
{\rm
A_{\rm H\alpha} = 2.5~log(~ 1~+~{{a \nu L_{\rm 24 \mu m}}\over{L^{'}_{\rm H\alpha}}}   ~),
}
\end{equation}
with $a = 0.038 \pm 0.005$. \citet{calzetti} confirm this relation using
33 nearby galaxies from the SINGS sample \citep{kennicutt03} covering a wide range in 
Hubble type and metallicity, though they find a slightly smaller $a$ ($0.031 \pm 0.006$).
We will adopt the $a$ in K07 to allow a direct comparison with M51a's
star formation law in Section $3.6$.

Derived internal $A_{\rm H\alpha}$ are listed in Table 5, along with
corrected H$\alpha$ luminosities. $A_{\rm H\alpha} \approx0.9$ at all $14$
ring positions, which agrees with the $A_{\rm H\alpha}$ estimated
with $L_{\rm IR}$ above. We calculate a flux-weighted mean
$A_{\rm H\alpha}$ of $0.92 \pm 0.05$ for the entire ring, and an extinction corrected
$L_{\rm H\alpha} = (1.43 \pm 0.06) \times 10^{42}$ erg s$^{-1}$, giving
SFR$ = 11.2 \pm 0.4$ $M_{\odot}$ yr$^{-1}$  \citep{kennicutt98a}.
This is in reasonable accord with the SFR based on the galaxy's $8-1000 ~\mu$m luminosity, 
given the scatter in the $L_{\rm IR}-SFR$ relation.

		\subsubsection{Radio Continuum Estimates of SFR}

20 cm continuum emission in star-forming galaxies is dominated by synchrotron
emission from relativistic electrons spiraling around magnetic field lines. These electrons 
originate in supernovae, which provide the connecting link between star formation and radio 
emission. Radio continuum emission from the ring dominates
the weak ($\approx2\%$) nuclear source and enclosed disk, and resembles the galaxy's
morphology at H$\alpha$ and the infrared.\footnotemark[10] The galaxy's radio luminosity 
provides a second extinction free measure of the SFR, though the physics connecting 
massive star formation and the ensuing radio emission is complex (e.g., particle confinement, 
magnetic field strength, and even the radiative lifetime of radio supernova remnants) and 
still only partially understood.\footnotetext[10]{The elliptical intruder galaxy to the 
southeast is not detected in radio continuum (S$_{\rm 20 cm}$$<$ 0.14 mJy, 3$\sigma$), though 
we do detect ($S_{\nu} = 0.45 \pm 0.05$ mJy) emission from the nucleus of the face-on barred
spiral companion $\approx1.5\arcmin$ north of AM0644-741's nucleus.}

We calculate the ring's radio continuum density ($L_{\rm 20 cm}$, in W Hz$^{-1}$) with
\begin{equation}
{\rm
L_{\rm 20 cm} = 4 \pi D_{\rm L}^{2} (1 + z)^{\alpha - 1} F_{\nu}(20 cm),
}
\end{equation}
where $F_{\nu}(20 cm)$ is the 20 cm flux density in Janskys, $D_{\rm L}$ is the luminosity
distance, and $\alpha$ is the source's radio spectral index ($F_{\nu} \propto \nu^{-\alpha}$)
taken to be 0.7, a value appropriate for galaxies dominated by star formation. We find
$L_{\rm 20 cm} = (3.2 \pm 0.2) \times 10^{22}$ W Hz$^{-1}$, implying SFR$ = 17.6 \pm 0.9$
$M_{\odot}$ yr$^{-1}$ \citep{hopkins}. While significantly higher than the H$\alpha$ 
and infrared SFRs, this estimate is consistent with the H$\alpha$ and IR SFRs given the
factor of $\approx1.6$ scatter in the SFR$_{\rm 20 cm}-$SFR$_{\rm H\alpha}$ relation in
Hopkins et al.

We note that the connection between 20 cm emission and the SFR appears less direct 
than for other tracers, and indeed \citet{kennicutt09} find that the relation 
between 20 cm and the Balmer-corrected $L_{\rm H\alpha}$ in the SINGS sample is 
more nonlinear than for the IR-based tracers. We therefore adopt
$A_{\rm H\alpha}\approx0.9$ and SFR$=11$ $M_{\odot}$ yr$^{-1}$ throughout the 
remainder of this paper.

	\subsection{The Stellar Ring}

In the collisional picture, the stellar ring plays a role similar to that of a
spiral density wave. Regions of high stellar mass surface density ($\Sigma_{*}$, in
$M_{\odot}$ pc$^{-2}$) will correspond to regions of strongest orbit crowding, which is 
where the molecular cloud collision frequency and relative speed is expected to peak. 
The stellar component may also contribute
significantly to the ring's gravitational stability and pressure.

We derive $\Sigma_*$ in the ring by first calculating $M_{*}/L_{I, \odot}$,
the $I$-band stellar mass to luminosity ratio. This is done by fitting integrated FUV, optical, 
and 4.5 $\mu$m luminosity densities from each $22\arcsec$ diameter SEST aperture to Starburst99 
spectral energy distributions \citep[SEDs;][]{starburst99,vazquez}.
In modeling the ring we assume a constant 
SFR of $1$ $M_{\odot}$ yr$^{-1}$ over a 120 Myr lifetime and a Salpeter initial mass function 
($M_{\rm up} = 100$ $M_{\odot}$). The stellar tracks incorporate thermally 
pulsing asymptotic giant branch stars. The FUV-4.5 $\mu$m
luminosity densities are corrected for both internal and Galactic foreground extinction 
($R_{\rm V} = 3.1$). The stellar mass within a given
aperture was determined by scaling the SED by an amount $h$ until they matched the observed
data points (i.e., $M_{*}/M_{\odot} = h \times$ Age $\times$ SFR). We find a small 
range in stellar masses among the apertures, with $M_{*} = (1.2 - 2.2) \times 10^{9}$ $M_{\odot}$. 
The $I$-band mass to light ratio at each position follows after dividing $M_{*}$ by the corresponding
$I$-band luminosity in solar units, defined
\begin{equation}
{\rm
L_{I} = 4 \pi D_{\rm L}^{2} F_{I} \Delta\nu_{I}  L_{I,\odot}^{-1},
}
\end{equation}
where $F_{I}$ is the $I$-band flux density in Janskys, $\Delta\nu_{I} = 7.55 \times 10^{13}$ Hz
is the filter passband, and $L_{I,\odot}$ is the Sun's $I$-band luminosity 
\citep[$3.71 \times 10^{25}$ W;][]{bessel79}. We find $M_{*}/L_{\rm I,\odot} \approx 0.22$, 
with only small variations around the 
ring. No dependence on either $\Sigma_{\rm H\alpha}$ or optical color is evident. 
$\Sigma_*$ follows by multiplying $M_{*}/L_{I,\odot}$ by the ring's
average $I$-band surface brightness ($\Sigma_{I}$, in $L_{\odot}$ pc$^{-2}$) at each
position. The result is shown in Figure 7 and listed in Table 6. $\Sigma_*$ ranges 
from $21-51$ $M_{\odot}$ pc$^{-2}$, peaking
in the ring's southwest quadrant. $\Sigma_{\rm 24 \mu m}$, $\Sigma_{\rm H\alpha}$,
and $\Sigma_*$ show similar azimuthal variations.

     	\subsection{AM0644-741's Neutral ISM}

	     	\subsubsection{The Distribution of Atomic and Molecular Gas}

The {\em robust} and {\em natural} weighted map cubes yield nearly identical HI masses, and
we use the higher angular resolution data throughout this paper. We find
$M_{\rm HI}$ = (2.01 $\pm$ 0.06) $\times$ 10$^{10}$ $M_{\odot}$ for AM0644-741 using the
{\em robust} moment-0 map. We obtain a $\approx50\%$ higher HI mass 
($2.94 \pm 0.07 \times 10^{10} M_{\odot}$) using integrated HI profile in Figure 11. 
This wide discrepancy may arise from the ring galaxy's large total velocity width coupled
with (as we show) often wide and complicated HI line profiles which may be
left out in the moment analysis. We will adopt the larger $M_{\rm HI}$ for the determination
of global quantities.

The distribution of HI is shown in gray scale and as logarithmic contours of $\Sigma_{\rm HI}$
in Figure 4.  Atomic hydrogen is distributed non-uniformly around the ring,
as expected from an off-centered collision.  Roughly half of the galaxy's 
HI is concentrated in a 20 kpc long arc coincident with the optical ring's southwest 
quadrant, where the highest HI surface density ($\Sigma_{\rm HI}$ = 110 $M_{\odot}$ pc$^{-2}$)
is found. Another $1/3$ of the total $M_{\rm HI}$ is distributed along the
optical ring's southeastern quadrant, where the gas is clumpier, lower in
$\Sigma_{\rm HI}$ ($\approx 20-35$ $M_{\odot}$ pc$^{-2}$), and more
broadly distributed. The northern half of AM0644-741's ring is HI deficient by comparison.
Note that HI is detected only from the innermost of the twin H$\alpha$ rings in the northern quadrant
using the {\em robust} data. Patchy HI with $\Sigma_{\rm HI} = 3-4$ $M_{\odot}$ pc$^{-2}$ is
visible at the position of the outermost ring using the {\em natural} weighted data.

Molecular gas is detected at 9 of 14 ring positions observed with
the SEST at 230 GHz, as shown in Figure 5: five overlapping beams 
in the northern quadrant (P3-P7; P.A. $= 78^{\circ}-142^{\circ}$) and four in the southwest 
(P9-P12; P.A. $= 217^{\circ} - 266^{\circ}$). $^{12}$CO(1-0) was also detected at these positions
(Figure 6). Line intensities and limits are listed in Table 4. The latter 
are estimated assuming $\Delta v_{\rm HI} = \Delta v_{\rm CO}$.
Using Equation (3), we find $\Sigma_{\rm H_2} = 
5.0-24.1$ $M_{\odot}$ pc$^{-2}$ around the ring, peaking at P7 in 
the northern quadrant (Figure 7).  We derive integrated line intensities of 
$I_{\rm 2-1} = 9.50 \pm 0.29$ K km s$^{-1}$ and
$I_{\rm 1-0} = 3.08 \pm 0.12$ K km s$^{-1}$ in the northern quadrant, 
and $I_{\rm 2-1} = 4.02 \pm 0.24$ K km s$^{-1}$ and $I_{\rm 1-0} = 0.94 \pm 0.09$ 
K km s$^{-1}$ in the southwest.\footnotemark[11]\footnotetext[11]{To account
for beam overlap, the summed 115 GHz line intensities have been divided by
factors of 2.16 and 1.95 in the northern and southwest quadrants, while at
230 GHz the summed line intensities have been divided by 1.25 and 1.30 for the
same two regions.} 
The two regions show similar excitation, with $R({{1-0}\over{2-1}}) = 0.65 \pm 0.01$ 
in the northern and $0.47 \pm 0.01$ in the southwest quadrants. Much larger values
are typically found in the centers of spirals 
\citep[e.g., $R({{1-0}\over{2-1}}) = 1.12 \pm 0.08$;][]{braine93},
implying that CO in AM0644-741's ring is warm and possibly optically
thin.  We infer H$_2$ masses in the two
ring quadrants of $(1.76 \pm 0.10) \times$ 10$^{9}$ $M_{\odot}$ (north) and
$(7.70 \pm 0.08) \times$ 10$^{8}$ $M_{\odot}$ (southwest), for a total $M_{\rm H_2} =
(2.52 \pm 0.12) \times$ 10$^{9}$ $M_{\odot}$. The ring's {\em global}
molecular fraction ($f_{\rm mol} \equiv { {M_{\rm H_2}}\over{ M_{\rm HI} + 
M_{\rm H_2} }}$) is $0.079 \pm 0.005$. As a whole, AM0644-741's star-forming ring 
is poor in molecular gas.

Atomic and molecular gas follow very different distributions in
the ring (Figure 7). HI is strongly peaked in the southwest quadrant
($\Sigma_{\rm HI}=60-90$ $M_{\odot}$ pc$^{-2}$). While H$_2$ shows more
point to point variation, $\Sigma_{\rm H_2}$ is on average $\approx50\%$ higher
in the northern ring quadrant ($\overline{\Sigma}_{\rm H_2} =15 M_{\odot}$ pc$^{-2}$
vs. $11 M_{\odot}$ pc$^{-2}$). The northern quadrant is 
in fact the only part of AM0644-741 where $\Sigma_{\rm H_2}$ and $\Sigma_{\rm HI}$ 
are comparable. Note that there are {\em two} H$\alpha$
rings within most of the $230$ GHz SEST beams in the northern quadrant (see Figure 5), and we 
may be overestimating $\Sigma_{\rm H_2}$ at P4-P7 by assuming all the
CO to be confined to a single $9\arcsec \times 22\arcsec$ ring segment, possibly
by as much as $50\%$. This caveat should be kept in mind in the following discussion.

Figure 8 shows the azimuthal variations in $f_{\rm mol}$ around the ring.
Large systematic variations are evident. Perhaps the most surprising result
is that $f_{\rm mol}\ga0.5$ in only three ring positions, all in the
northern quadrant. Elsewhere, and notably throughout the southwest quadrant,
$f_{\rm mol}\approx0.1-0.3$. The ring's ISM appears to be overwhelmingly 
{\em atomic} rather than molecular where star formation is most intense.

HI is not detected from AM0644-741's nucleus ($\Sigma_{\rm HI} < 0.2$ $M_{\odot}$ pc$^{-2}, 3\sigma$) 
or from most of the region interior to the ring ($\Sigma_{\rm HI} < 0.1$ $M_{\odot}$ pc$^{-2}, 3\sigma$) 
assuming $\Delta v = 90$ km s$^{-1}$. Note that HI is considerably more abundant throughout 
the Cartwheel's interior, where a patchy $\Sigma_{\rm HI} \approx 2-6$ $M_{\odot}$ pc$^{-2}$ 
disk component exists \citep{wof2}. Similarly, no molecular gas was detected from AM0644-741's 
nucleus ($\Sigma_{\rm H_2} < 2.1$ $M_{\odot}$ pc$^{-2}$, $3\sigma$) or disk 
($\Sigma_{\rm H_2} < 1.8$ $M_{\odot}$ pc$^{-2}$, $3\sigma$) assuming $\Delta v = 90$ km s$^{-1}$
and $R({{1-0}\over{2-1}}) = 1.6$ \citep{braine93}. The confinement of AM0644-741's neutral atomic 
ISM to the 42 kpc ring applies equally well to its molecular component.

There is a sharp drop in $\Sigma_{\rm HI}$ at the optical ring's edge. 
Like the Cartwheel, either AM0644-741's ring has reached (or
exceeded) the original extent of the pre-collision HI disk, or beyond the optical
ring the system possesses extremely low HI surface densities \citep{wof2}.
No HI is detected from the elliptical intruder ($v_{\rm sys} = 6700$ km 
s$^{-1}$; FMA) or the faint optical bridge to AM0644-741, 
even after extra spatial and velocity smoothing. 
Likewise, HI was not detected from the face-on barred spiral to the north of 
AM0644-741 \citep[galaxy ``C'' in][$v_{\rm sys} = 7000$ km s$^{-1}$]{graham74}.

 	\subsubsection{The Relative Distributions of Gas and Massive Star Formation}

Figure 7 shows that HI and star formation tracers $\Sigma_{\rm 24 \mu m}$, $\Sigma_{\rm H\alpha}$,
and $\Sigma_{\rm 20 cm}$ have very similar distributions, even peaking at the same positions (P9-10).
$\Sigma_{\rm H_2}$ does not appear simply related with massive star formation 
in the ring. For example, where $\Sigma_{\rm H_{2}}$ peaks in the northern quadrant (P7), 
massive star formation is near its minimum (e.g., the local peak in $\Sigma_{\rm H\alpha}$ 
and $\Sigma_{\rm 24 \mu m}$ is at P3). This is in marked contrast with studies of nearby 
galaxies that find $\Sigma_{\rm H\alpha}$ and $\Sigma_{\rm H_{2}}$ to be 
well correlated \citep[e.g., K07,][]{rownd}. Figure 8 shows that the ring's 
molecular fraction is in fact anti-correlated with 
$\Sigma_{\rm H\alpha}$. (We will return to this result in Section $3.6$.) 
The neutral ISM in AM0644-741's ring appears largely 
atomic precisely where massive star formation is most intense.

The absence of measurable star formation in the interior disk follows
from the lack of a significant gas reservoir, either atomic or molecular. We infer 
a total $M_{\rm HI + H_{2}} \la 6 \times 10^{8}$ $M_{\odot}$
and a mean $\Sigma_{\rm HI + H_{2}} \la 2 $ $M_{\odot}$ pc$^{-2}$ over AM0644-741's
interior disk and nucleus. Both limits are $3 \sigma$ and include a mass contribution
from helium. The mass limit is not very stringent since giant molecular
associations (GMAs) in local spirals have $M_{\rm H_2}\approx3-7 \times 10^{8}$
$M_{\odot}$ \citep[e.g.,][]{rand93,rand95}. However, regions with such a low mean
surface density are not expected to provide adequate 
shielding from UV photons which retards the formation of molecular cloud complexes.

	\subsubsection{Kinematics and Velocity Structure of the Ring}

Differential rotation is apparent in the flux-weighted isovelocity contours
in Figure 4, with non-circular motions evident as kinks
in the velocity field. The distribution of HI interior to the ring is
not extensive enough to perform a detailed kinematic analysis (e.g., the
derivation of a rotation curve or measuring the radial infall of gas predicted
by ring galaxy models). The ring is sufficiently HI rich to
explore its kinematics by fitting inclined circular ring models, with and
without expansion, to the run of HI radial velocity ($v_{\rm rad}$) versus
position angle ($\psi$, measured counterclockwise from the northeast line of
nodes). These velocities are averages over $10^{\circ}$ 
sectors of a $6\arcsec$ wide annulus fit to the H$\alpha$ ring. Systemic ($v_{\rm sys}$), 
circular ($v_{\rm cir}$), and expansion ($v_{\rm exp}$) velocities 
are derived by minimizing the sums of the squares of the residuals between
model and observed $v_{\rm rad}$,
\begin{equation}
{\rm 
\chi^{2} = \sum_{\psi} w^{2}_{\psi}[v_{\rm rad,\psi} - v_{\rm sys} -
(v_{\rm cir} ~sin~i ~cos ~\psi ~ + ~v_{\rm exp} ~tan~i ~sin~\psi)/\rho_{\psi}]^{2},
}
\end{equation}
where
\begin{equation}
{\rm
\rho_{\psi} = \sqrt{~ sec^{2}~i ~-~ tan^{2}~i ~cos^{2}~\psi ~}.
}
\end{equation}
In the above, $w_{\psi}$ is the weight assigned to each measurement and
{\em i} is the inclination, which is $56.^{\circ}5 \pm 0.^{\circ}7$ 
from ellipse fitting to the (assumed) inclined circular H$\alpha$ ring.
For the non-expanding ring model, $v_{\rm exp}$ is set to zero in Equation (9).
Formal uncertainties in the velocities are calculated with
\begin{equation}
{\rm
\sigma^{2}(v_{x}) = \sum_{\psi} ~w_{\psi} ({{\partial v_{x}}\over{\partial v_{\psi}}})^{2} \delta v^{2},
}
\end{equation}
where $\partial v_{\psi} = v_{\rm model} - v_{\rm rad}$ at ring position angle $\psi$,
$\delta v$ is the velocity resolution in the HI data, taken to be the channel separation (27.5 km s$^{-1}$), 
and where $v_{\rm x}$ can be either $v_{\rm sys}$, $v_{\rm cir}$, or $v_{\rm exp}$. These
should be considered lower-limits, since systematic errors 
(e.g., the assumption of a circular ring) may contribute significantly.

Figure 9 shows the resulting $v_{\rm rad}-\psi$ diagram fits for 
inclined circular ring models with ({\em solid line}) and without 
({\em dashed line}) expansion. The error bars show the local velocity spread within
the region used to derive $v_{\rm rad}$. The expanding circular ring model is clearly the 
better fit to the data, with derived velocities and formal uncertainties of
$v_{\rm sys} = 6692 \pm 8$ km s$^{-1}$, $v_{\rm cir} = 357 \pm 13$ km s$^{-1}$, and
$v_{\rm exp} = 154 \pm 10$ km s$^{-1}$. FMA found significantly smaller 
$v_{\rm sys}$ ($6611 \pm 8$ km s$^{-1}$), $v_{\rm cir}$ ($311 \pm 19$ km s$^{-1}$), and 
$v_{\rm exp}$ ($128 \pm 14$ km s$^{-1}$) using optical emission lines. 
These discrepancies do not originate in the analysis routines, since we also derive
FMA's $v_{\rm sys}$, $v_{\rm cir}$, and $v_{\rm exp}$ using their ``blue'' velocities
(i.e., calculated using [O~II] $\lambda 3727$ \AA\ , [O~III] $\lambda\lambda 4959, 5007$ \AA\ , 
H$\beta$, H$\gamma$, and H$\delta$ lines), though with $\approx20\%$ larger uncertainties. 
A comparison of the two data sets
show the optical velocities to be systematically smaller by $\approx80$ km s$^{-1}$ at
all position angles, which accounts for the different $v_{\rm sys}$. The origin 
of such a large offset is unclear. However, Figure 10 shows that when HI and $^{12}$CO($J=2-1$)
peaks are strong and narrow (e.g., P10 and P11), they agree well with each other, 
with {\em both} $\approx60-90$ km s$^{-1}$ {\em higher} than FMA's velocities. 
This suggests that the discrepancy between the optical and radio velocities lies 
in a calibration issue with FMA's data.
We also note that FMA lack velocity measurements at position angles where expansion is most
pronounced (i.e., $120^{\circ} \le \psi \le 180^{\circ}$ and 
$260^{\circ} \le \psi \le 330^{\circ}$), which will lead to
$v_{\rm exp}$ being underestimated. For instance, including the 
measured velocity from knot ``k'' (not used by FMA) at $\psi = 144^{\circ}$ increases 
the derived $v_{\rm exp}$ from  
$128 \pm 14$ to $146 \pm 16$ km s$^{-1}$, which is consistent with our result.
Both optical and HI kinematic studies agree that the ring rotates counter-clockwise,
with the western half being the near side.

HI line profiles in AM0644-741's ring differ substantially from the Cartwheel's, which
are typically single component and narrow \citep[$< 50$ km s$^{-1}$ FWHM;][]{wof2}.
Figures 5 and 10 show that both $^{12}$CO(2-1) and HI can possess surprisingly large velocity
widths (e.g., $\Delta v_{\rm CO} = 558$ km s$^{-1}$ FWHM at P7). Figure 10 reveals a wide variety 
of profile shapes and  widths, from narrow single components (e.g., P2, $\Delta v_{\rm HI} = 104$ 
km s$^{-1}$ FWHM) to wider and more complicated spectra (e.g., P13, with 
$\Delta v_{\rm HI} = 261$ km s$^{-1}$ FWHM and at least two $\approx 100$ km s$^{-1}$ FWHM components). 
The former are more characteristic of profiles in the southwest ring quadrant. 
Where both HI and $^{12}$CO(2-1) are detected the line profiles are similar in shape 
and velocity, at least for the main component (e.g., P10 and P11). It may be significant that the widest 
HI and CO line-widths (P7) occur at the intersection of the two H$\alpha$ rings.
In terms of kinematics AM0644-741 is as rich as the Cartwheel. Whether this complicated 
velocity structure results from caustics in the gas or strong out-of-plane motions is not known.

The $^{12}$CO($J=2-1$) lines may be broadened significantly by rotation and expansion 
across the $22\arcsec$ SEST beam. To estimate this effect, we modeled the 230 GHz emission 
as a uniform annulus with inner and outer radii of $36\arcsec$ and $48\arcsec$, respectively, 
with inclination, rotation and expansion speeds as derived from the above kinematic analysis.  
We then convolved the model to the angular and spectral resolution of our SEST observations 
and derived corrections to $\Delta v_{\rm CO}$ (e.g., $\approx200$ km s$^{-1}$ near the minor 
axis from rotation, decreasing to $\approx 50$ km s$^{-1}$ near the major axis), which we subtract
in quadrature to derive $\Delta v_{\rm corr}$ (Table 4). We find that the smearing of the 
velocity field is significant but not dominant. In particular, it is not responsible for
the large $^{12}$CO($J=2-1$) line widths at P7 ($\Delta v_{\rm CO, corr} = 542$ km s$^{-1}$) or P6
($\Delta v_{\rm CO, corr} = 213$ km s$^{-1}$). Our model predicts wider $^{12}$CO($J=2-1$) line-widths
than we observe for P10, $\Delta v_{\rm CO, model} = 90$ km s$^{-1}$ FWHM versus the observed
$\Delta v_{\rm CO} = 55$ km s$^{-1}$ FWHM. This may be due to an underestimation of the width in the
relatively weak $230$ GHz line at this position or a breakdown in our assumption of a uniform
molecular ring. The broadening of the HI lines, estimated by the same method, is smaller because of the
smaller beam size, e.g., $\approx100$ km s$^{-1}$ near the major axis to about the channel width
near the major axis.

For a roughly constant $v_{\rm exp}$, the collision responsible for AM0644-741's
ring occurred $\approx rv_{\rm exp}^{-1} = 133$ Myr in the past.
Under the same assumption, the Cartwheel and its intruder collided $\approx400$ Myr ago.
We derive a dynamical mass $M_{\rm ind} = (r v^{2}_{\rm circ}/G) \approx 6 \times$ 10$^{11}$ $M_{\odot}$, 
which is comparable to the Cartwheel's \citep{wof2}. This is an approximate value since the 
ring is not likely to be in dynamical equilibrium. 
AM0644-741 appears to be a considerably younger system than the Cartwheel despite 
comparable linear diameter and total mass. 

As another illustration of  AM0644-741's unusually large velocity range, we show its total
HI spectrum in Figure 11, obtained by integrating the {\em robust} 
weighted data cube over the optical extent of the galaxy. We measure a FWZI velocity
width of $743 \pm 28$ km s$^{-1}$, with no correction for inclination. This range is 
extremely large, and is nearly twice the Cartwheel's \citep{wof2,mebold}.
The profile itself is noticeably asymmetric, reflecting the concentration of HI in the southwest and 
southeastern ring quadrants, with at least three peaks. By comparison, the Cartwheel's 
integrated HI spectrum shows a symmetric two-horned profile.

	\subsection{The Efficiency of Star Formation in the Ring}

We follow \citet{young} in parameterizing star formation efficiency,
the yield of massive stars per unit H$_2$ mass, as SFE = log$(L_{\rm H\alpha}/M_{\rm H_2})$,
with $L_{\rm H\alpha}$ and $M_{\rm H_2}$ in solar units. citet{rownd} derive global and
spatially resolved SFEs for 122 normal and interacting disk galaxies, and find little 
variation in mean SFE from Hubble types S0 to Scd, with  
SFE $\approx -1.8 \pm 0.2$. Later Hubble types showed progressively higher 
efficiencies, with mean SFE $= -1.3 \pm 0.2$ and $-0.8 \pm 0.4$ for Sd-Sm and Irr 
respectively. SFE in strongly interacting and 
merging systems was $\approx4$ times larger than in
isolated spirals, a result in agreement with earlier estimates using CO and $FIR$
data \citep[e.g.,][]{sage}. To draw a direct
comparison with their large galaxy sample, AM0644-741's $L_{\rm H\alpha}$ was
only corrected for Galactic extinction, and we scaled $M_{\rm H_2}$ upward by $20\%$
to match their larger $X_{\rm CO}$ \citep{bloemen}. Likewise, we did not factor in 
a helium contribution to the molecular mass (J. Young, private communication).
We find AM0644-741's global SFE $= -1.53 \pm 0.07$,
i.e., comparable to an Sd-Sm spiral. If the galaxy was originally an early
spiral, as suggested by its prominent bulge, this implies an increase in SFE by
a factor of $\approx2.5$ as a result of the collision, which is quite modest.

The $14$ individual SFE measurements in the ring are compared with
the ensemble of spatially resolved SFEs from Rownd \& Young's galaxy sample 
in Figure 12. The overall distribution reflects the modest increase in SFE.
The total range in measured SFEs around the ring, $\Delta$SFE$=1.49$,
is exceptional. Most of the galaxies in Rownd \& Young's sample
possess  $\Delta$SFE$<0.48$, i.e., a factor of $\approx3$, and only two galaxies
in their sample show a SFE dispersion wider than AM0644-741's. Rownd \& Young's
data show an abrupt decrease in the $\Delta$SFE within individual galaxies once the
$FCRAO$ beam exceeds $6$ kpc in linear resolution ($\Delta$SFE$<0.8$
or a factor of $6$) with little change in SFE, consistent with the
smoothing out of SFE variations for larger areas. The large $\Delta$SFE in 
AM0644-741's ring is therefore even more notable given the $\approx10$ kpc linear 
resolution of the $230$ GHz SEST beam.

SFE shows considerable point to point variation within the two ring
quadrants where $^{12}$CO is detected (Figure 13). Still, SFE is on average 
$4$ times higher in the starburst southwest quadrant 
(mean SFE$_{\rm north} = -1.72$ vs. SFE$_{\rm southwest}=-1.07$). 
SFE in AM0644-741 peaks at P10, where $\Sigma_{\rm HI}$ and $\Sigma_{\rm H\alpha}$ 
also reach maximum values. High SFEs limits are also indicated at P1, P2, P8, P9, P13 and P14.

Star formation efficiency might be expected to depend on the surface density of 
atomic plus molecular gas or stars if either provide a measure of orbit crowding in the
ring, analogous to the density wave amplitude in a spiral arm. We find no
correlation between log$(L_{\rm H\alpha}/M_{\rm H_2})$ and 
$\Sigma_{\rm HI + H_{2}}$ or $\Sigma_{*}$; the Pearson correlation coefficient $r$ is 
$\approx 0.1$ in both cases).

	\subsection{The Star Formation Law in AM0644-741's Ring}

The strong correlation between SFR density ($\Sigma_{\rm SFR}$)
and gas surface density in spiral galaxies is well established, and is observed 
to hold for disks and nuclear starburst regions as a whole, as well as in
spatially resolved star-forming complexes in nearby galaxies \citep[e.g.,][K07]{kennicutt98b}.
This corresponds to a range of scale sizes from $\sim0.5 - 10$'s of kiloparsecs and gas 
surface densities from $\la10$ to $10^{5}$ $M_{\odot}$ pc$^{-2}$. This correlation, usually
referred to as the Schmidt law \citep{schmidt}, can be expressed
$\Sigma_{\rm SFR} = A~\Sigma_{\rm gas}^{\rm N}$. For
$\Sigma_{\rm SFR}$ in units of $M_{\odot}$ yr$^{-1}$ kpc$^{-2}$ and $\Sigma_{\rm gas}$
(atomic, molecular, or both) in $M_{\odot}$ pc$^{-2}$, 
$N\approx1.4$.\footnotemark[12] The physical processes underlying the
Schmidt law are of considerable interest, as is its utility as a recipe for
star formation in numerical models of galaxy evolution \citep[e.g.,][]{richard}
and interacting systems, including ring 
galaxies \citep[e.g.,][]{weil,antunes}.\footnotetext[12]{Note that a Schmidt law with $N\approx1.5$ 
is to be expected from a disk of self-gravitating clouds with volume density $n_{\rm gas}$
that collapse and form stars on a free-fall timescale, i.e., 
$\Sigma_{\rm SFR} \sim n_{\rm gas}/t_{\rm ff}$ $\sim n_{\rm gas}/ n_{\rm gas}^{-0.5} 
\sim n_{\rm gas}^{1.5}$.}

The star formation law in the grand design spiral M51a was studied in detail by K07
using extinction corrected H$\alpha$ luminosities and interferometrically
derived HI and H$_2$ surface densities ($\approx 500$ pc linear resolution).
They find $\Sigma_{\rm SFR}$ and $\Sigma_{\rm HI + H_2}$ to be well
correlated ($r=0.74$), obeying
\begin{equation}
{\rm
log~\Sigma_{\rm SFR} = (1.56 \pm 0.04)~log ~\Sigma_{\rm HI + H_{2}} - (4.32 \pm 0.09).
}
\end{equation}
M51a has large amplitude spiral arms that are especially rich in
molecular gas, which follows a similar Schmidt law by itself ($r=0.73$), though with a 
significantly smaller exponent,
\begin{equation}
{\rm
log~\Sigma_{\rm SFR} = (1.37 \pm 0.05)~log ~\Sigma_{\rm H_2} - (4.36 \pm 0.09).
}
\end{equation}
The scatter in both star formation laws ($\approx0.4$ dex) is substantially larger 
than the individual measurement errors and the quoted uncertainties refer to the 
fit. The physical origin of this dispersion has yet to be firmly
identified, though candidates include local
variations in SFE, $X_{\rm CO}$, and age distribution of the individual 
star-forming regions. Given the similarities in Schmidt law parameters 
calculated for different linear resolutions (including M51a), K07
conclude that the power-law exponent and zero-point are relatively insensitive to 
resolution. We will use these results as a benchmark for AM0644-741.

The top panel of Figure 14 shows log$~\Sigma_{\rm SFR}$ plotted
against log$~\Sigma_{\rm HI + H_{2}}$ for the $14$ SEST ring 
positions. To be consistent with K07, SFR density is calculated 
using the extinction corrected H$\alpha$ luminosities with no helium correction for 
the neutral gas. While we find $\Sigma_{\rm SFR}$ and total hydrogen surface density to be
well correlated ($r=0.63$), a bivariate least-squares fit \citep{york}
yields significantly different Schmidt law coefficients,
\begin{equation}
{\rm 
log~\Sigma_{\rm SFR} = (1.29 \pm 0.06)~log~\Sigma_{\rm HI + H_{2}} - (3.67 \pm 0.16).
}
\end{equation}
This fit is shown as a solid line in the figure, with M51a's Schmidt law from
Equation (12) shown as a dashed line for comparison.
Again, the scatter is substantially larger than the individual measurement
uncertainties, though formally smaller ($\approx0.2$ dex) than that in M51a (K07). 

The star formation laws for atomic and molecular hydrogen are shown separately in the bottom
panel of Figure 14. AM0644-741's $\Sigma_{\rm SFR}$ is also well correlated with 
$\Sigma_{\rm HI}$ ($r=0.63$), but the relationship is essentially linear,
\begin{equation}
{\rm
log~\Sigma_{\rm SFR} = (0.96 \pm 0.04) ~log~\Sigma_{\rm HI} - (3.06 \pm 0.28).
}
\end{equation}
The real surprise is that $\Sigma_{\rm SFR}$ and $\Sigma_{\rm H_2}$ appear completely 
uncorrelated (ignoring the H$_2$ limits, they are actually weakly {\em anti-correlated}:
$r=-0.5$). This figure should be compared with Figure 5 in K07 where the situation is completely 
reversed: in M51a $\Sigma_{\rm H_2}$ correlates strongly with the $\Sigma_{\rm SFR}$ 
while it is $\Sigma_{\rm HI}$ that is uncorrelated. Given the distributions
of gas and massive stars in Figures 7 and 8, this result probably should not have been 
unexpected. Nevertheless, it is striking to see cold molecular gas, the star-forming 
component of the ISM, appear so disconnected from the process in the ring.

	\subsection{Gravitational Stability and Star Formation}

Groups of regularly spaced HII complexes are visible in the H$\alpha$ 
ring in Figure 1, with typical separations of $5-6$ kpc. This morphology,
reminiscent of ``beads on a string'', is also frequently observed in
spiral arms, tidal tails, nuclear starburst rings,
and other ring galaxies \citep[e.g.,][]{elmegreen2,wof2,mullan},
and is generally taken to indicate star formation
triggered through the action of large scale gravitational instabilities. 
In this section we use the kinematics and distribution of stellar,
HI, and H$_2$ mass to estimate the stability of AM0644-741's ring. 

A thin gaseous disk becomes gravitationally unstable wherever the $Q$ parameter, 
defined
\begin{equation}
{\rm
Q = {{\sigma_{\rm gas} \kappa}\over{\pi G \Sigma_{\rm gas}}}
}
\end{equation}
is less than unity \citep{safronov,toomre64}. Here $\sigma_{\rm gas}$ and 
$\Sigma_{\rm gas}$ are the gas disk's velocity dispersion and (atomic plus molecular) 
surface density, respectively, and $\kappa$ is the disk's epicyclic 
frequency, defined 
\begin{equation}
{\rm
\kappa = \sqrt{2}~{{v_{\rm circ}}\over{r}}~(1 + {{{\rm d}ln~v_{\rm circ}}\over{{\rm d}ln~r}})^{1/2}.
}
\end{equation}
For flat rotation curves the epicyclic frequency simplifies to 
$\kappa \approx \sqrt{2}~{v_{\rm circ} ~r^{-1}}$.
Since $\kappa$, $\sigma_{\rm gas}$, and $\Sigma_{\rm gas}$ can vary substantially over a
galaxy's disk, $Q$ will as well. When the above local condition is
satisfied, i.e., $Q < 1$, regions of the gas disk with scale size
$\lambda_{\rm J} \approx \sigma_{\rm gas}^{2}/(\pi G \Sigma_{\rm gas})$ 
become susceptible to gravitational collapse, and proceed to form 
molecular cloud complexes with typical 
$M \approx \pi \sigma_{\rm gas}^{4}/(G^{2} \Sigma_{\rm gas})$.\footnotemark[13]
These in turn become the sites of star formation. Observationally, $Q \approx 1-2$ throughout
the disks of spiral galaxies, which is attributed to negative feedback from
star formation acting to both increase $\sigma_{\rm gas}$ and decrease
$\Sigma_{\rm gas}$.\footnotetext[13]{At the Sun's position in the Galaxy, 
$\sigma_{\rm gas} \approx 5$ km s$^{-1}$, $\Sigma_{\rm gas} \approx 10$ $M_{\odot}$ pc$^{-2}$, 
and $\kappa \approx 1.2 \times 10^{-15}$ Hz \citep{sanders84,kerr}, giving local values 
of $\lambda_{\rm J} \approx 200$ pc and $M \approx 10^{7}$ $M_{\odot}$.}

\citet{kennicutt89} applied these ideas to a sample of spiral galaxies
and showed that the radial distribution of massive star formation, as traced by H$\alpha$
emission, could be understood in terms of a gravitational stability condition in the gas
disks through a critical gas surface density $\Sigma_{\rm crit}$ defined as
\begin{equation}
{\rm
\Sigma_{\rm crit} = {{\beta \sigma_{\rm gas} \kappa}\over{3.36 G}},
}
\end{equation}
where $\beta$ was determined empirically to be $0.7 \pm 0.2$. These two formulations are
equivalent since $\Sigma_{\rm crit} \propto Q \Sigma_{\rm gas}$. 
Sub-critical regions (i.e., $Q<1$) experience robust massive star formation and
obey a classic (i.e., $N\approx1.4$) Schmidt law. Conversely, star formation is greatly suppressed in regions where 
$Q$ substantially exceeds unity, even if the local gas surface density is high. 
$Q \approx 1$ identifies a critical state where additional factors like SNe shocks, 
magnetic field strength, and cloud-cloud collisions can tip the scales one way or the other. 

We calculate $Q$ for gas only ($Q_{\rm gas}$) at the $14$ ring positions observed 
with the SEST using the above equations. $\sigma_{\rm gas}$ is approximated by 
the observed line FWHM (corrected for velocity smearing within
the $230$ GHz beam) divided by 2.35, even though the line profiles are typically non-gaussian,
and regardless of the presence of multiple velocity components. Both HI and H$_2$ are
included in $Q_{\rm gas}$ and are scaled by $1.36$ to account for helium.
We assume a flat rotation curve and calculate $\kappa$ using the ring's
radius and $v_{\rm cir}$. The derived $Q_{\rm gas}$ are listed 
in Table 6 and plotted as a function of position angle in Figure 15.
AM0644-741's gas ring is decidedly {\em stable} gravitationally ($Q_{\rm gas}\approx2-7$),
apart from P9-11 in the starburst southwest quadrant where $Q_{\rm gas} = 0.3-0.5$. 
This is primarily due to the ring's typically large $\sigma_{\rm gas}$. However, 
this ignores the contribution of stars to the ring's gravitational stability. \citet{wang}
find that for a stellar component with surface density $\Sigma_{*}$ 
and velocity dispersion $\sigma_{*}$, a combined $Q$ parameter incorporating stars and gas 
can be written to good approximation as
\begin{equation}
{\rm
Q_{\rm tot} = {{\kappa}\over{\pi G}} ( { {\Sigma_{\rm gas}}\over{\sigma_{\rm gas}} } + 
{{\Sigma_{*}}\over{\sigma_{*}}} )^{-1}.
}
\end{equation}
\citet{leroy08} find that the additional self-gravity provided by stars can transform 
otherwise stable disks ($Q_{\rm gas}>2$) into critical or marginally stable ones ($Q_{\rm tot}\approx1-2$).
Locally, $Q_{\rm gas} \approx 2$ for the $\Sigma_{\rm gas}$ and $\sigma_{\rm gas}$ given in 
\citet{mckee} and \citet{sanders84}. When the local stellar component ($\Sigma_* \approx 35$
$M_{\odot}$ pc$^{-2}$ and $\sigma_* \approx 45$ km s$^{-1}$; \citet{kuijken} and
\citet{wielen} respectively) is included, Equation (19) gives $Q_{\rm tot} = 1.3$.
The stellar component can thus significantly affect gravitational stability.

We calculate $Q_{\rm tot}$ for AM0644-741's ring using the $\Sigma_{*}$ derived in
Section $3.2$. We have, unfortunately, no direct measure of $\sigma_*$, the ring's stellar velocity 
dispersion. A constraint on the radial component $\sigma_*$ can be
inferred from the narrow width of the H$\alpha$ ring, since 
B5 stars, the longest lived massive star capable of powering luminous HII regions,
will travel $d \approx \sigma_* \times \tau_{\rm B5}$. For the observed width
of the H$\alpha$ ring ($5\arcsec - 8\arcsec = 2-3.5$ kpc) and the main-sequence lifetime
of a B5 star ($40$ Myr), $\sigma_* \approx 50$ km s$^{-1}$. Comparable values are
found in spiral disks \citep[e.g.,][]{beltran} as well as the solar neighborhood \citep{wielen}. 
At the same time, it is difficult to reconcile such $\sigma_*$ with the significantly 
larger $\sigma_{\rm gas}$ implied by the HI and CO line profiles and the heating of the 
stellar disk that must have taken place during the collision. Given this inherent uncertainty, 
we evaluate $Q_{\rm tot}$ using $\sigma_* = 50$ km s$^{-1}$ and $\sigma_* = \sigma_{\rm gas}$ to 
bracket the correct value. The resulting azimuthal dependence of $Q_{\rm tot}$ is shown in Figure 15, 
with the hatched region reflecting $\sigma_*$'s expected range. The stellar component has significantly
lowered $Q$, but still not to the point that the entire ring is sub-critical. $Q_{\rm tot}
<1$ for only four positions, P9-12, which encompass the ring's starburst southwest quadrant.
However, over the remainder of the ring $Q_{\rm tot}\approx1.5-3$, including locations where
$\Sigma_{\rm SFR}$ is high, such as P1, P3, and P5. 
We are thus led to the unexpected conclusion that unless $\sigma_*<50$ km s$^{-1}$,
AM0644-741's star-forming ring is mostly stable against the growth of large scale density 
perturbations, despite indications that gravitational instabilities play a primary role 
in triggering star formation.

\section{Discussion}

	\subsection{Why is the Ring Rich in HI yet So Poor in H$_2$?}

Given AM0644-741's prominent bulge, both in terms of linear size and fraction of 
the galaxy's 4.5 $\mu$m emission, and the Hubble types of its companions, it was 
likely an early-type spiral prior to the collision. This makes the low molecular
content of its ring that more surprising, since the global $f_{\rm mol}$ for
Sa-Sb spirals is $\approx0.6$ \citep{young}. Just as striking, the inferred
H$_2$ mass for this large galaxy is only $\approx7$ times larger than a Magellanic
dwarf \citep{leroy05}. In trying to account for the ring's relatively small 
$M_{\rm H_2}$ and $f_{\rm mol}$ we discount the possibility 
that we are viewing AM0644-741 at a special time, e.g., when star formation has
exhausted the ring's molecular reservoir. The ring is rich in atomic gas, and
the timescale to convert HI into H$_2$ can be as short as $\approx2$ Myr in 
spiral arms \citep[e.g.,][]{hidaka}. Given the ring's large $\Sigma_{\rm gas}$ and 
$\ga$solar metallicity, HI should be quickly (and continuously) transformed into H$_2$. 
Likewise, metallicity variations cannot be the cause, since the spectra in 
Figure 3 indicate essentially constant abundances throughout the ring's western half,
over which $f_{\rm mol}$ varies by nearly an order of magnitude. Clearly, other 
factors are at work. We consider below the likelihood that insufficient ISM pressure, 
possibly in concert with high FUV radiation fields, are responsible for 
the observed global $f_{\rm mol}$ and its variations.

	\subsubsection{Insufficient ISM Pressure}
The gas phase pressure ($P_{\rm ISM}$) strongly influences the rate at
which HI is converted to H$_{2}$, since this process scales with the gas volume density as 
$n_{\rm gas}^2$, and $n_{\rm gas}\approx P_{\rm ISM}~\sigma_{\rm gas}^{-2}$.
In a spiral disk, the mid-plane gas pressure depends on both the mass density of stars and gas
as well as the respective velocity dispersions,
\begin{equation}
{\rm
P_{\rm ISM} \approx (\pi G/2 k_{\rm B}) 
~(\Sigma_{\rm gas}^2 + {{\sigma_{\rm gas}}\over{\sigma_{*}}}\Sigma_{\rm gas}\Sigma_{*}),
}
\end{equation}
where $k_{\rm B}$ is the Boltzmann constant, giving units of K~cm$^{-3}$ \citep{elmegreen93}.
The ISM pressure may in fact be the single most important parameter determining whether molecular 
or atomic gas dominates the ISM in the disks of spiral galaxies \citep{blitz}. 
\citet{elmegreen93} argued that the molecular fraction is a function of both $P_{\rm ISM}$ and 
the ambient FUV radiation field ($\chi_{\rm UV}$). We call this parameter $\phi_{mol}$,
which for solar abundances is
\begin{equation}
{\rm
  \phi_{\rm mol} \approx (P_{\rm ISM}/P_{\odot})^{2.2} (\chi_{\rm UV}/\chi_{\odot})^{-1},
}
\end{equation}
where the local ISM pressure ($P_{\odot}$) is $10^{4}$ K~cm$^{-3}$ 
and the local interstellar UV radiation field intensity
($\chi_{\odot}$) is $2.0 \times 10^{-4}$ erg s$^{-1}$  cm$^{-2}$ sr$^{-1}$ 
\citep{madden}. When $\phi_{\rm mol}$ approaches or exceeds unity, $f_{\rm mol}=1$.
\citet{hidaka} successfully modeled the abrupt HI to H$_2$ transition in M51a's 
disk and spiral arms using this criterion. To test the hypothesis that AM0644-741's
molecular content is primarily determined by the ISM pressure, we calculate $P_{\rm ISM}$ 
using $\Sigma_{\rm HI}$, $\Sigma_{\rm H_2}$, $\Sigma_*$, and velocity data in
Tables 5 and 6, and compare it with the observed $f_{\rm mol}$. In doing so we implicitly 
assume that the relationships expressed in the above two equations are valid in AM0644-741's 
ring.

We use Equation (20) to calculate the ISM pressure in each of the 14 ring positions
observed with the SEST. As we did for $Q_{\rm tot}$, we bracket
$\sigma_*$ with  $50$ km s$^{-1}$ and $\sigma_{\rm gas}$. The top panel in Figure 16 shows 
that $P_{\rm ISM}\approx6-80 P_{\odot}$ everywhere in the outer ring regardless of
$\sigma_*$. $P_{\rm ISM}$ is high where $f_{\rm mol}$ is small, and actually peaks
in the southwest quadrant where the measured $f_{\rm mol}$ reaches a minimum.
The lack of correspondence between azimuthal variations of 
$P_{\rm ISM}$ and $f_{\rm mol}$ is compelling evidence that pressure is 
not the primary factor determining the ring's molecular content. If it were, 
the entire ring, and in particular, the southwest quadrant, would be 
dominated by H$_2$.

	\subsubsection{Photodissociation of Molecular Gas}
Whether the ISM is largely atomic or molecular depends on the eventual 
equilibrium between  the formation of H$_{2}$ out of HI on dust grains
and its destruction by ultraviolet photons with wavelengths between $0.05$
and $0.22 ~\mu$m. Can the ambient FUV radiation field intensity (and
variations) explain the relatively H$_2$ poor ring, or for that matter, the 
H$_2$ dominated northern quadrant?  We estimate $\chi_{\rm UV}$ 
in a two step process. First, we calculate the ring's integrated $40-500 ~\mu$m luminosity, 
$L_{\rm FIR, tot} = 2.02 \times 10^{12}(2.58L'_{\rm 60 \mu m} ~+~ L'_{\rm 100 \mu m})$,
using data tabulated in the $IRAS$ Point Source Catalog \citep{moshir}. At $24$ and $70$ 
$\mu$m the nucleus contributes $<5\%$ of the total emission (Figure 2), so we assign all of
the $IRAS$ flux to the ring. Second, we assume that $L_{\rm FIR}$ is traced reasonably
well by the 70 $\mu$m image, such that within each of the 14 SEST positions,
$L_{\rm FIR}(i) = (L_{\rm FIR, tot}/F_{\rm 70 \mu m, tot}) F_{\rm 70 \mu m}(i)$.
We then follow \citet{stacey} and calculate the ambient FUV radiation field at each
position in units of $\chi_{\odot}$ with
\begin{equation}
{\rm
\chi_{\rm UV}(i) = {{L_{\rm FIR}(i)}\over{4\pi D^2_{\rm L} \Delta\Omega (2.0 \times 10^{-4})}},
}
\end{equation}
where $D_{\rm L}$ is the luminosity distance (96.9 Mpc), $\Delta\Omega$ is the source solid angle, 
($4.66 \times 10^{-9}$ sr). These are listed in Table 7. The ring's FUV radiation field is 
moderate in strength ($\chi_{\rm UV}/\chi_{\odot} \approx 7-25$, peaking in the southwest 
quadrant) and much less extreme than a typical nuclear starburst, where 
$\chi_{\rm UV}/\chi_{\odot} \approx 10^{2}-10^{4}$ is typical \citep{stacey}.

Values of $\phi_{\rm mol}$ derived using Equation (21) are shown in the bottom panel 
of Figure 16. We find $\phi_{\rm mol}\gg1$ everywhere in AM0644-741's ring
($\sigma_*$ again has a relatively small effect). This is especially true in the 
starburst southwest quadrant, where $\phi_{\rm mol} \approx 25-1200$. Clearly, by
this criterion the ring should be primarily molecular in composition. 
Only if $\chi_{\rm UV}/\chi_{\odot}$ were $50-1000$ larger would we recover the 
observed $f_{\rm mol}$ (Figure 8). Again, other factors must ultimately be determining 
$f_{\rm mol}$.

As an alternative to the semi-analytic approach of Elmegreen, we consider the
photodissociation region (PDR) models in \citet{kaufman99}. These take a simple
1-dimensional semi-infinite slab of gas with constant volume density of hydrogen 
($n = n_{\rm HI} + 2n_{\rm H_{2}}$) and gas to dust ratio, and 
irradiate one side with a constant (one-dimensional) flux of
FUV photons ($h\nu = 6-13.6$ eV). Depending on $n$ and ambient 
FUV radiation field $G_{\circ}$ ($\equiv 1.71\chi_{\rm UV}/\chi_{\odot}$), 
the models yield HI column density ($N_{\rm HI}$) 
and $^{12}$CO($J=1-0$) intensity ($I_{\rm CO}$). For high $n$ and weak $G_{\circ}$, 
HI is produced via photodissociation in relatively small amounts as a thin layer on the 
molecular cloud surface, giving low $N_{\rm HI}$ and high $I_{\rm CO}$. At lower
volume densities and higher $G_{\circ}$, the FUV photons penetrate deeper into
the cloud, destroying CO molecules and producing more HI by photodissociation, which
gives rise to higher $N_{\rm HI}$ and lower $I_{\rm CO}$.
Given $G_{\circ}$ and $N_{\rm HI}$, $n$ can be inferred to constrain the ISM's state.

We compare the ring's $N_{\rm HI}$ and $G_{\circ}$ with the PDR model grid shown
in Figure 3 of \citet{allen}. Throughout the northern ring quadrant (P2-7) where
$G_{\circ} \approx 10$ and $N_{\rm HI} \approx 10^{21}$ cm$^{-2}$, the models give
$n \approx 100-300$ cm$^{-3}$. This is 
characteristic of the cold neutral medium (CNM;~$T\approx50-100$ K), 
which  makes up a large fraction of the ISM in ordinary spirals \citep{madden}.
The CNM is generally considered to be the precursor of
cold molecular clouds and cloud complexes. In M33, for example, this component comprises 
roughly $50\%$ of the neutral atomic ISM \citep{braun}. The situation is very 
different in the southwest quadrant, where the higher FUV radiation field 
($G_{\circ} = 15-25$) and HI column densities
($N_{\rm HI} \ga 10^{22}$ cm$^{-2}$) imply $n \approx 1-5$
cm$^{-3}$. This density range is characteristic of the warm neutral
medium (WNM; ~$T = 5000-10,000$K). Intermediate $n$ are found outside the northern and 
southwest ring quadrants (Table 7). While the implied $n$ are ensemble averages, it 
nevertheless appears that the ISM in the southwest ring quadrant is highly deficient in
cold dense gas, even compared with the ring's northern half.  

This raises the obvious question of how can AM0644-741's ring sustain the observed 
high $\Sigma_{\rm SFR}$ in an ISM dominated by the WNM? One possibility is that most of the HI in the
southwest quadrant is not directly participating in star formation, and for all
practical purposes, {\em is} the WNM. Suppose $75\%$ of the HI is a photodissociation
product that simply ``hangs around'' because it is confined to the ring. If only
$\approx25\%$ of the $N_{\rm HI}$ is really appropriate for the PDR models, then the same
range of $G_{\circ}$ gives $n \approx 20-80$ cm$^{-3}$, which is still
low for the CNM. If $90\%$ of the observed HI in the ring's southwest
quadrant is a photodissociated {\em background} component, then $n \ga 100$ cm$^{-3}$, 
and the star-forming ISM begins to more closely resemble the CNM. If so, this leaves the 
highly unusual situation where most of the ISM in AM0644-741's ring {\em traces} rather than 
{\em sustains} star formation. How is this possible?

\subsection{The Star-forming Environment in AM0644-741's Ring}

AM0644-741 presents a number of puzzles. Despite physical conditions that
clearly favor the dominance of molecular over atomic gas (e.g., $\ga$solar metallicity,
large $P_{\rm ISM}$, and only moderate $\chi_{\rm UV}$), three-quarters 
of the ring is characterized by low $f_{\rm mol}$.
Only in the northern quadrant does atomic and molecular gas appear on equal footing.
$\Sigma_{\rm HI}$ in the starburst southwest quadrant, moreover, is 
far in excess of what is observed in spiral arms and dwarf galaxies, and is 
in fact WNM-like in terms of $n$. How can such an 
ISM support efficient and robust star formation? The galaxy's star formation law,
in which $\Sigma_{\rm SFR}$ appears coupled with HI but not H$_2$, is highly
peculiar. And finally, despite indications of gravitational instabilities
operating on large scales, we find the ring to be largely stable (or at best critical) 
even after including the stellar mass (i.e., $Q_{\rm tot}\approx 1.3-5$ for plausible 
values of $\sigma_*$). We propose that all of these peculiarities are a 
direct consequence of the ISM's unusually long confinement time in the star-forming ring, 
which greatly increases its exposure to the destructive effects of massive stars.

We illustrate schematically in Figure 17 the differences between the ``over-cooked'' ISM 
within the ring of a large and evolved ring galaxy like AM0644-741 and a spiral arm. 
Unlike the situation in a spiral galaxy, where a giant molecular 
cloud (GMC) may spend $\approx15$ Myr passing through an arm, and not 
encountering another arm for $\approx100$ Myr \citep[e.g.,][]{kimostriker}, 
GMCs in AM0644-741 and the Cartwheel are confined to the dense star-forming rings 
for $\ga100-300$ Myr. GMCs in both environments
are disrupted after $\approx10$ Myr, primarily due to expanding HII 
regions powered by the young massive stars embedded within them \citep{bally,maddalena}, 
since recent work implies that most of the damage
occurs before the first supernovae \citep[e.g.,][]{murray}. 
Unlike their spiral arm counterparts, GMCs confined within the star-forming rings accrue
still further damage from the moderately high $\chi_{\rm UV}$ as well as shocks 
as the OB stars proceed to supernova and deposit still more energy into the ISM. 
Enhanced cloud-cloud collisions in the crowded rings are an additional source of 
damage \citep[e.g.,][]{asm87}. With $\ga90\%$ of AM0644-741's 
atomic and molecular ISM collected in the ring (Figure 4 and 5), 
molecular cloud fragments do not coalesce to re-form GMCs in a 
low $\chi_{\rm UV}$ environment: the Oort cycle \citep{oort} has been 
essentially short-circuited. As these destructive effects accumulate, the ring 
ISM evolves into a mixture dominated by GMC-fragments and
photodissociated HI, with embedded massive stars and SNe remnants.
An equilibrium is eventually reached where GMC growth in the gas rich ring is balanced by 
their destruction. The balance will depend on the ring's $\chi_{\rm UV}$, 
$\Sigma_{\rm gas}$, metallicity, and age (i.e., the confinement time).

For the moderate $\chi_{\rm UV}$ we infer in AM0644-741's ring, CO is expected to 
survive only in the innermost cloud cores where dust columns are the highest 
\citep[e.g.,][]{maloney}. In an ``over-cooked'' ISM that is dominated by small 
H$_2$ clouds and GMC fragments, little CO may exist at all. As a result,
$^{12}$CO emission will substantially underestimate H$_2$ in the ring, 
particularly where $\Sigma_{\rm SFR}$ is high. This provides a 
simple explanation for the ring's apparent low $f_{\rm mol}$ and $M_{\rm H_2}$,
as well as the anti-correlation between $f_{\rm mol}$ and $\Sigma_{\rm H\alpha}$
in Figure 8. Similarly, if the $^{12}$CO lines are significantly underestimating the 
H$_2$ mass, then log$(L_{\rm H\alpha}/M_{\rm H_2})$ will {\em overestimate} SFE,
particularly where $\Sigma_{\rm SFR}$ is high, accounting for the observed
wide SFE range. If the underlying $I_{\rm CO}-N_{\rm H_2}$ relation is a function of the
local $\Sigma_{\rm SFR}$, the correlation between $\Sigma_{\rm H_2}$ and $\Sigma_{\rm SFR}$ 
would be expected to break down as well, giving rise to the peculiar
star formation law in Figure 14.

In the Oort cycle, molecular clouds are photodissociated by embedded
OB stars downstream from the arm (Figure 17 upper panel). Such offsets between $^{12}$CO 
and HI emission are observed in galaxies with large
amplitude spiral arms like M51 \citep[e.g.,][]{vogel}. 
For AM0644-741's ``over-cooked'' ring, moderately high $\chi_{\rm UV}$ combined
with fragmented clouds will produce a large photodissociated HI component, 
giving rise to unusually high $\Sigma_{\rm HI}$. This process should scale with 
the local $\Sigma_{\rm SFR}$, accounting for to the observed linear relation between 
$\Sigma_{\rm HI}$ and $\Sigma_{\rm H\alpha}$ in
Figure 14. Photodissociation of cloud fragments would be expected to produce low volume density
HI, and account for the dominant WNM-like ISM in the ring's southwest quadrant. 
The low volume density of photodissociated HI will retard the conversion back 
to H$_2$ and contribute to the large $\Sigma_{\rm HI}$ and low $f_{\rm mol}$ 
despite the high $P_{\rm ISM}$.
This provides a simple explanation for the observed variations 
in $f_{\rm mol}$ around the ring in Figure 8. In the southwest quadrant, 
$^{12}$CO emission is suppressed due to small cloud sizes and peak 
$\chi_{\rm UV}$, while photodissociated HI is produced in abundance. $f_{\rm mol}$ 
will appear small as a result. The opposite situation exists in the northern 
ring quadrant (P3-7), where $\Sigma_{\rm SFR}$ and $\chi_{\rm UV}$ are lower and 
fragmentation should be less severe. Consequently, $^{12}$CO emission will better 
trace H$_2$ and photodissociated HI will exist in smaller proportions, accounting for
the observed $f_{\rm mol}\approx0.5$.

We have argued that $^{12}$CO emission will substantially underestimate
the ring's molecular mass. But by how much?  Suppose that the molecular ISM 
obeyed M51's Schmidt law (K07). We can rewrite Equation (13) to estimate 
$\Sigma_{\rm H_2}$ directly from the observed $\Sigma_{\rm SFR}$, i.e.,
$\Sigma_{\rm H_2} = 1522~\Sigma_{\rm SFR}^{0.72}$. The predicted $\Sigma_{\rm H_2}$
are $\approx3-20\times$ higher than what we infer from $^{12}$CO in the northern quadrant. 
In the southwest quadrant however the discrepancy is much greater, with the predicted
$\Sigma_{\rm H_2}\approx 150-250 ~M_{\odot}$ pc$^{-2}$, implying that H$_2$ is
underestimated by factors of $\approx10-20$. The overall effect on the ring's molecular
fraction is dramatic: $f_{\rm mol}\approx0.7-0.9$ around the ring. The resulting
$Q$ from such a modification is shown in Figure 18, where the ring becomes mostly
sub-critical, even when gas alone is considered. $Q_{\rm tot}<1$ everywhere 
except for P7 and P14, where little star formation occurs.
Regions of high $\Sigma_{\rm SFR}$ that are not detected in
$^{12}$CO emission are now sub-critical.
We emphasize that while this thought experiment is not intended as an
accurate measure of the ring's molecular mass, it does suggest large amounts
of unseen H$_2$ are present in AM0644-741's ring, sufficient to render most 
of the ring gravitationally unstable. Alternate assays of the ring's molecular 
content are clearly needed.

The above effects will be heightened in low metallicity environments such as the
Cartwheel's ring, as CO will be further hard pressed to survive the lowered
dust columns, even less so in cloud fragments. However, not every ring galaxy's ISM is
expected to be ``over-cooked''. Systems that are either young (i.e., 
$Age \approx R/v_{\rm exp} \la 30$ Myr) or which possess low 
$\Sigma_{\rm SFR}$ should possess more typical $n$, SFE, $f_{\rm mol}$,
and star formation laws.  Arp 147, with a $\approx33$ Myr old ring 
\citep[$Age=4.5$ kpc/$136$ km s$^{-1}$;][]{jeske} is an example of the former. 
It possesses a robustly star-forming
ring, with a globally averaged $\Sigma_{\rm SFR}$ ($0.06$ $M_{\odot}$ yr$^{-1}$ 
kpc$^{-2}$) nearly identical to the Cartwheel's and AM0644-741's. The ring is 
also gas rich, with M$_{\rm HI}=(3.80 \pm 0.05) \times 10^{9}$ $M_{\odot}$ and 
M$_{\rm H_2} = 2.5 \times 10^{9}$ $M_{\odot}$ with $X_{\rm Gal}$ 
\citep{jeske,hor95}, giving a more 
normal average $f_{\rm mol} = 0.4$. Other examples
include NGC 2793 \citep[SFR$\approx 0.3$ $M_{\odot}$ yr$^{-1}$;][]{higdon01} and the 
``Sacred Mushroom'' \citep[SFR$< 0.05$ $M_{\odot}$ yr$^{-1}$ ($3 \sigma$);][]{higdon01}.
High angular resolution CO, infrared and submillimeter wave observations will be 
needed to test this prediction.

Are the rings of AM0644-741 and the Cartwheel unique star-forming environments, 
or do analogs exist in other systems? Nuclear starburst rings,
resulting from bar induced resonances, would at first glance appear to be good candidates. 
Gas is confined to the rings for $\approx100$ Myr timescales, and they are often sites of 
intense star formation. Examples include NGC~1097 \citep{gerin88} and NGC~4314 \citep{fritz92}.
One might therefore expect large amounts of photodissociated HI and low $n$ and $f_{\rm mol}$.
Not so. The ISM of nuclear rings appear dominated by molecular rather than atomic gas 
\citep[e.g.,][]{gerin91,fritz96,curran,hsieh}. This is likely a consequence of both the high metallicity
and large $\Sigma_*$ and $\Sigma_{\rm gas}$ in the ring and nuclear regions promoting
rapid conversion of HI to H$_2$, and keeping $f_{\rm mol} \approx 1$. Polar ring galaxies, 
such as NGC~4650~A and NGC~660, where a close passage 
has apparently transferred gas and dust into a 
polar orbit around an S0 galaxy may be the closest analogs to ring galaxies. Polar 
rings appear to persist for $>100$ Myr timescales, and several show high $\Sigma_{\rm SFR}$ 
and $\Sigma_{\rm HI}$ \citep[e.g.,][]{mahon,arnaboldi97}. These objects are 
still relatively unexplored observationally, and would be good targets for comparison 
studies at radio, millimeter, and infrared wavelengths. 
The rings in evolved starburst ring galaxies like AM0644-741 and the Cartwheel
appear to represent rather unique star-forming environments. They also further demonstrate 
the hazards of trying to infer basic properties of the molecular ISM using rotational transitions of
$^{12}$CO where the physical conditions differ markedly from the Milky Way's disk.

\subsection{Follow-up Observations}

Current and next-generation infrared and sub/millimeter wave facilities 
will allow the ideas outlined in the previous sections to be explored observationally.

$Herschel$ makes available a suite of infrared emission lines that provide
important diagnostics of PDR regions (e.g., [C~II] $\lambda 157.7 ~\mu$m
and [O~I] $\lambda\lambda 63.2, 145.5 ~\mu$m with PACS, and 
[C~I] $\lambda\lambda 369.9, 609.1 ~\mu$m with SPIRE), which should be prominent
in the rings of AM0644-741 and the Cartwheel. Just as important, an independent 
estimate of $\Sigma_{\rm H_2}$ can be obtained by combining MIPS$ ~24 \mu$m
and $70 \mu$m with PACS $100$ and $160$ $\mu$m images to estimate the
optical depth of dust ($\tau_{\rm dust}$). When multiplied by the gas-to-dust ratio
($\delta$), the H$_2$ surface density can be estimated with
$\Sigma_{\rm H_2} = \case{1}{2}( {{\tau_{\rm dust}}\over{\delta}} - \Sigma_{\rm HI})$
\citep[e.g.,][]{israel,leroy09}. $\delta$ cannot be measured directly, 
but can be reasonably constrained using measured obscuration and metal abundances.

The Atacama Large Milimeter Array (ALMA) will revolutionize astronomy in the
$84-950$ GHz range due to its high angular resolution ($\la1\arcsec$),
wide range of spectral resolution, and superb sensitivity. These gains are the result 
of the large number of array elements ($>60$ 12 m telescopes), state of the art receivers 
and correlator, and a very dry site. ALMA observations of AM0644-741
and the Cartwheel would immediately impact the question of their gravitational
stability and star formation law by spatially resolving the $^{12}$CO($J=1-0$) and
$^{12}$CO($J=2-1$) emission in the rings. This will allow the precise determination 
of the molecular ISM's surface density, kinematics, and velocity dispersion, which
are quantities we can only estimate with our single-dish data (particularly so
in AM0644-741's double-ringed northern quadrant).

We have invoked an ``over-cooked'' ISM to explain AM0644-741's (and by inference, 
the Cartwheel's) peculiar star formation law, $n$, SFE, and $f_{\rm mol}$.
The dramatically improved sensitivity and angular resolution provided by ALMA will
permit the detection of individual molecular cloud complexes, allowing us to 
constrain their size and mass distributions.
Interferometric $^{12}$CO($J=1-0$) observations of spiral arms in nearby galaxies 
have identified discrete objects termed giant molecular associations (GMAs), with 
$M_{\rm H_2} \approx 3-7 \times 10^{7}$ $M_{\odot}$, velocity widths of $\approx10$ km s$^{-1}$,
and $300-450$ pc diameters \citep[e.g.,][]{rand93,rand95}. For a Galactic 
$X_{\rm CO}$ we would expect the H$_2$ mass equivalent of $\approx100$ $5 \times 10^{7}$
$M_{\odot}$ GMAs throughout AM0644-741's ring. 
Each would subtend $\approx1\arcsec$ if similar to those in M51,
and emit a 115 GHz line flux of 0.36 Jy km s$^{-1}$. This would be detected
at the $5 \sigma$ level in $^{12}$CO($J=1-0$) in 2.5 hr per 
array pointing ($\approx55\arcsec$
primary beam) with angular and velocity resolution of $0.5\arcsec$ and $2$ km s$^{-1}$
for the sensitivity estimates in \citet{wootten}. 
An obvious first question would be whether the molecular ISM in AM0644-741's
ring is dominated by a small number of GMAs, or by many smaller lower mass
complexes. Resolved studies of molecular cloud associations in the rings of AM0644-741
and the Cartwheel will allow us to test the expectation of smaller or less massive GMAs, 
and make important comparisons with the size and mass distributions of GMAs in other systems
(e.g., spiral arms with similar $\Sigma_{\rm SFR}$), and how their properties vary with local
conditions. Observations in higher $^{12}$CO transitions
will help constrain the excitation dependence of $X_{\rm CO}$. 
Additionally, observations of other molecules that may serve as 
more reliable tracers of the dense star-forming ISM (e.g., ammonia or CS) can be 
carried out to see if the densest gas correlates with local $\Sigma_{\rm SFR}$.
In other words, can a standard Schmidt law be recovered by using CS instead of
$^{12}$CO?

Recent advances in implementing a multi-phase ISM in galaxy evolution
codes \citep[e.g.,][]{saitoh} additionally open the prospect of modeling the effects 
of confining an otherwise normal molecular ISM to a dense star-forming ring for long
timescales. Ring galaxy models with more realistic star formation recipes
would be important adjuncts to observational studies, and we encourage such
studies to be done.

 	\subsection{Density Wave or ``Forest Fire''?}
\citet{korchagin98,korchagin99,korchagin01} account for radial color gradients
interior to the ring and slight offsets between the H$\alpha$ and continuum rings by invoking
a self-sustaining radially propagating starburst. This model dispenses with the
orbit crowded ring altogether and is reminiscent of the explanation of flocculent spiral
arms by \citet{gerola}.
The central passage of a companion galaxy through a spiral disk is still required,  
but now acts as a ``detonator'' that  initiates  a nuclear starburst. Once started, 
the nuclear starburst self-propagates radially 
through the disk in a ring shaped distribution of HII complexes. 
The orbits of stars are left largely undisturbed, while the ISM experiences
local accelerations from SNe shocks, stellar winds, and ionization fronts 
that trigger new star formation at larger radii. Post-starburst populations 
left in the wake of this rampaging ``forest fire'' give rise to radial color and
abundance gradients on galaxy-wide scales. \citet{korchagin98} successfully
accounted for the color gradients in the Cartwheel's disk \citep{marcum}.
Such a novel (and radical) interpretation 
bears consideration in light of published studies of the gaseous 
\citep{wof2,higdonrandlord} and stellar \citep[][HW97]{applemarston97,wof1}
components of ring galaxies, including this one. 
Does the ``forest fire'' hypothesis hold up to observational scrutiny? 
There are in fact major problems with this interpretation. 

First, a Gerola \& Seiden-esque ``forest fire'' cannot produce the HI distributions
of ring galaxies. $\Sigma_{\rm HI}$ in rings of AM0644-741 and Cartwheel
greatly exceed that found in outer spiral disks. Gas must have traveled from their interiors. 
However, shocks and winds from star-forming regions are ill suited to produce such 
coherent and large scale gas outflows, even ignoring inhomogeneities in the disk 
(e.g., spiral arms and HI holes) that would work against the formation of narrow HI rings.
The low $\Sigma_{\rm gas}$ encircled disks present another problem: where did all
the gas go? Conversion into stars would require essentially $100\%$ ~SFE on
$\approx100-300$ Myr timescales, which is clearly unfeasible.
 Second, analysis of the Cartwheel's HI kinematics found large-scale
gas inflows plus an $\approx$solid-body rotation curve \citep{wof2}.
While the forest fire interpretation makes no clear kinematic predictions,
it is not obvious how a propagating starburst would result in these bulk motions.
Third, differential rotation will distort and eventually destroy the symmetry
of a self-propagating starburst unless it was perfectly
centered, i.e., unless the intruder passed through the exact center of the target
disk. An analogy can be made with the evolution of HI holes in spiral disks,
created by expanding superbubbles of hot gas from SNe blasts
\citep[e.g.,][]{deul}. As the holes grow
they are invariably distorted by shear.  One would therefore
predict that large off-centered ring galaxies would have highly distorted rings. The
opposite is true: the three largest ring galaxies - AM0644-741, the Cartwheel, and the ``Sacred
Mushroom'' - all have highly symmetric  rings, even the two formed in a markedly off-centered 
collision. If the rings are just propagating HII region fronts, one would expect 
that preexisting stellar structures like spiral arms and bars to survive, and in fact, 
made more obvious in optical images. On the contrary, apart from the Cartwheel's spokes, 
no other ring galaxy shows obvious  spiral arms. 
Finally, roughly $10\%$ of spirals in the local universe with 
$9.7 \le$ log$(L_{\rm FIR}/L_{\odot}) \le 10.7$
possess starburst nuclei, representing a space density
$\rho_{\rm SB}\approx 10^{-4}$ Mpc$^{-3}$ \citep[e.g.,][]{dickhead,derobertis}.
This is $50$-times larger than the estimated space density of ring galaxies,
$\rho_{\rm RG}\approx 2 \times 10^{-6}$ Mpc$^{-3}$ \citep{fewmadore}.
If ring galaxies are a consequence of a nuclear starburst, then they would be
expected to be much more common, even considering the transient nature of both the
rings and nuclear starbursts. 

The collisional interpretation has no such problems. Numerical models since
the classic work of \citet{lynds} reproduce the narrow and high $\Sigma_{\rm gas}$
rings and gas poor interiors through orbit crowding. SPH models by \citet{struckhigdon}
successfully accounted for the Cartwheel's large-scale kinematics. Stellar structures in
the disk (e.g., pre-collision spiral arms) are effectively erased as their orbits are 
crowded and phase mixed in the expanding ring. Ring galaxies are 
rare simply because they require a low probability event - a low impact-parameter 
collision with a companion. Stochastic effects undoubtedly play a role in these systems.
However, the weight of evidence argues that they must be secondary to the ability of 
the ring density wave to concentrate a disk's stellar and gaseous components into a dense 
environment for the action of gravitational instabilities.

\section{Conclusions}

Ring galaxies are unique systems in which to study the triggering
and regulation of massive star formation on large scales. In this
paper we examine one of the best examples, AM0644-741, which is 
notable for its large diameter ring ($42$ kpc), inferred age ($133$
Myr), and global SFR. From the analysis of H$\alpha$, FUV, 
infrared, and radio continuum images we conclude that SFR$ = 11$ $M_{\odot}$
yr$^{-1}$, with relatively low internal extinction ($A_{\rm H\alpha}\approx1$),
and no significant star formation within the encircled disk or nucleus.
Analysis of optical spectra show that AM0644-741's ring possesses
$\ga$solar metal abundances (in marked contrast with its ``twin'' the Cartwheel), 
opening the possibility of characterizing its molecular ISM using rotational
transitions of $^{12}$CO. The distribution of HII complexes gives the appearance 
of the classic ``beads on a string'' morphology, which is a hallmark of star 
formation triggered by the action of large scale gravitational instabilities. 
One therefore expects $Q<1$ throughout the ring. 

Observations using ATCA and SEST show that
essentially all of AM0644-741's neutral ISM is concentrated in 
the expanding $42$ kpc diameter ring, which agrees both with collisional 
models and high resolution observations of other ring galaxies (e.g.,
Arp~143 and the Cartwheel). Contrary to our expectations, we find 
the ring's global molecular fraction to be surprisingly small ($f_{\rm mol} = 
0.078 \pm 0.003$), particularly given that Sa-Sb galaxies, the likely progenitor 
of AM0644-741, normally possess comparable amounts of HI and H$_2$.
It is this apparent deficiency of H$_2$ that is responsible for
the ring's other peculiar properties: elevated SFE,
a star formation law where atomic gas - but not molecular gas -
correlates with star formation, and typically $Q_{\rm tot}\ga1$.
The concentration of the disk's ISM into the dense ring appears to 
have not generated a molecular rich environment. Quite to the contrary, 
applying PDR models to the observed $\Sigma_{\rm HI}$, $\Sigma_{\rm H_2}$,
and $\chi_{\rm UV}$ yield WNM-like total hydrogen volume densities 
($n\approx2$ cm$^{-3}$) throughout the ring quadrant experiencing
peak SFR. 

Our data allow us to reject the possibility that the ring's apparently
small molecular component is the result of low and/or variable metallicity, 
unusually high $\chi_{\rm UV}$, or insufficient pressure (which 
regulates the conversion of HI to H$_2$). We find the opposite to be
true: AM0644-741's ring would have been expected to be dominated by 
molecular gas given our estimates of its $P_{\rm ISM}$ and 
$\chi_{\rm UV}$. To explain how AM0644-741 can form stars in a WNM 
dominated ISM, we propose that the rings in evolved and
robustly star-forming ring galaxies like AM0644-741 and the Cartwheel
possess an ``over-cooked'' ISM, where GMCs are fragmented and 
photodissociated by prolonged exposure to SNe shocks, expanding HII 
regions, and moderately high $\chi_{\rm UV}$ ($\approx 7-25 \chi_{\odot}$)
during their $>100$ Myr confinement to the ring. Consequently, CO 
survives only in the small innermost cloud cores due to reduced 
shielding available in the cloud fragments, and traces H$_2$ differently 
than for a Galactic cloud. In particular, $^{12}$CO transitions 
will significantly {\em underestimate} the amount of H$_2$ 
in the ring in a way that depends on the local SFR density. An
``over-cooked'' ISM can thus account for the unusually high SFE, the 
small $f_{\rm mol}$ and its variations around the ring, as well as the 
peculiar star formation law, where CO appears to be 
decoupled from local star formation. This also explains the observed linear
relationship between $\Sigma_{\rm SFR}$ and $\Sigma_{\rm HI}$: the 
abundant atomic gas we detect with the ATCA is primarily a 
photodissociation product, which traces rather than fuels star 
formation. It is no longer surprising, then, that this component 
will resemble the WNM. For all practical purposes, it is
the WNM. 

If, as we suspect, $^{12}$CO significantly underestimates the
ring's molecular component, then $Q_{\rm tot}$ may be significantly
overestimated. If the H$_2$ obeys an M51-like Schmidt Law (e.g., K07),
then we find that $Q_{\rm tot}<1$ is satisfied for nearly all of the ring,
and in particular, where massive star formation is observed to be robust.
Gravitational instabilities then become the dominant star
formation trigger. Independent estimates of the ring's molecular component,
as well as better constraints on the stellar velocity dispersion, are
clearly needed to conclusively settle this point. 

Another peculiar aspect of AM0644-741's ISM, and one not evident in the Cartwheel, 
is the wide and multi-component line profiles, both in the atomic and molecular gas. 
It is not clear at present if this reflects caustics in the gas flow or out-of-plane
motions, though preliminary $N$-body simulations appear to favor the former interpretation
(J. Wallin, private communication). 

Our results imply that the rings of systems like AM0644-741 and the Cartwheel represent
substantially different star-forming environments than that typically found in
spiral arms due to the sustained accumulation of damage to the ISM from concentrated
populations of OB stars. At the same time, the modest FUV radiation fields
mean that they are still different from intense nuclear starbursts and ULIRGs.
Ring galaxies remain interesting and valuable systems to study the interaction between
the ISM and intense star formation, and follow-up observations with $Herschel$ and ALMA,
in addition to models incorporating more sophisticated treatments of the ISM, are 
clearly warranted.

Finally, we examined the ``forest fire'' interpretation of ring galaxies 
in light of these (and previous) observational results. We argue that a 
propagating starburst \citep[e.g.,][]{gerola} cannot reproduce the observed gas
distribution or large-scale kinematics of ring galaxies. Likewise, the space density 
of nuclear starburst galaxies, i.e., systems possessing the forest fire ``detonator'', 
is $\approx50$ times larger than that of ring galaxies, implying that if ring 
galaxies arise this way, they should be much more common. The collisional 
interpretation has no such problems.

\acknowledgements
We thank the staff of the SEST, in particular Philipe McAuliffe, for their 
hospitality and assistance with the observations. We also thank Michael Nord
for his assistance with the CO observations. Discussions with Mark Hancock  
were helpful in our estimates of the ring's stellar mass.  J.L.H. also thanks the
Kapteyn Astronomical Institute, especially Thijs van der Hulst, for supporting
this and related research, and David McConnell of the {\em ATNF} for
the generous allotment of telescope time. We acknowledge
the anonymous referee for valuable comments and suggestions, and Jonathan Braine for
alerting us to an error in our H$_2$ mass estimation. This research has made 
use of the NASA/IPAC Extra-galactic Database (NED), which is operated by the Jet Propulsion 
Laboratory, California Institute of Technology, under contract with the National 
Aeronautics and Space Administration. J.L.H. and S.J.U.H. acknowledge support from NASA/$Spitzer$ 
grant 1346930.

%............................
\clearpage
\begin{table}
\caption{AM0644-741 Global Properties}
\begin{tabular}{lll}
\tableline \tableline

R.A. \tablenotemark{(a)} 			&&06$^{\rm h}$ 43$^{\rm m}$ 06.$^{\rm s}$18 \\
Dec. \tablenotemark{(a)}	       	 	&&-74$^{\circ}$ 14$\arcmin$ 10$.\arcsec7$ \\
{\it l, b} 					&&	285.26, -26.59\\
$v_{\rm sys}$ \tablenotemark{(b)}   		&&6692 $\pm$ 8 km s$^{-1}$ \\
Luminosity distance                             &&96.9 Mpc \\
$IRAS$ flux densities \tablenotemark{(c)}	&& 0.073, 0.152, 1.41, 3.89 Jy\\
Ring diameter 				       	&&95$''$$\times$ 52$''$  (42 $\times$ 23 kpc)\\
$L_{\rm IR}$\tablenotemark{(d)}			&&(4.7 $\pm$ 0.2) $\times$ 10$^{10}$ $L_{\odot}$ \\
$I_{1-0}$\tablenotemark{(e)} 			&& 4.02 $\pm$ 0.15 K km s$^{-1}$\\
$I_{2-1}$\tablenotemark{(e)} 			&& 13.52 $\pm$ 0.38 K km s$^{-1}$\\
$R({{1-0}\over{2-1}})$\tablenotemark{(f)}    	&& 0.59 $\pm$ 0.03\\
$L_{\rm 20 cm}$\tablenotemark{(g)}              && (3.2 $\pm$ 0.2) $\times$ 10$^{22}$ W Hz$^{-1}$ \\
$L_{\rm H\alpha}$\tablenotemark{(h)} 	 	&& 1.43 $\pm$ 0.06 $\times$ 10$^{42}$ erg s$^{-1}$\\
SFR\tablenotemark{(i)}                          && 11.0 $\pm$ 0.4 $M_{\odot}$ yr$^{-1}$\\
SFE$_{\rm global}$\tablenotemark{(j)}  		&& -1.53 $\pm$ 0.07\\
$M_{\rm HI}$\tablenotemark{(k)}  		&&(2.94 $\pm$ 0.07) $\times$ 10$^{10}$ $M_{\odot}$\\
$M_{\rm H_2}$\tablenotemark{(l)}   		&&(2.52 $\pm$ 0.12) $\times$ 10$^{9}$ $M_{\odot}$\\
$f_{\rm mol}$\tablenotemark{(m)}		&&0.079 $\pm$ 0.005\\
\end{tabular}
\tablenotetext{} {Notes - (a) J2000 coordinates of nucleus. (b) Systemic velocity of ring derived
from HI kinematics (optical/heliocentric definition). (c) $IRAS$ 12, 25, 60 and 100 ~$\mu$m point source
flux densities \citep{moshir}. (d) Integrated $8-1000 ~\mu$m luminosity derived using $IRAS$
data \citep{sanders96}. (e) Integrated $^{12}$CO line fluxes measured in this paper. Corrections
for main-beam efficiency, beam-source coupling, and over sampling are described in Section $2.5$. (f) The ring's
global $I_{\rm 1-0}/I_{\rm 2-1}$ ratio. See footnote 8 in Section 2.5. (g) Integrated 20 cm radio continuum
luminosity using the {\em robust} weighted data. (h) Integrated H$\alpha$ luminosity of the ring 
corrected for Galactic \citep[$A_{\rm H\alpha} = 0.33$;][]{rc3} and internal extinction
($A_{\rm H\alpha} = 0.92$) as described in Section3.2.1. (i) SFR derived from extinction corrected 
$L_{\rm H\alpha}$ after \citet{kennicutt98a}. (j) Global star formation efficiency (Section $3.5$) following
\citet{young}. We have corrected for Galactic extinction but not for helium mass. (k) Total HI mass using
{\em robust} weighted ATCA data. No helium mass correction. (l) Total H$_2$ mass assuming $X_{Gal}$. 
No helium mass correction. (m) Global molecular mass fraction, defined 
$f_{\rm mol} = M_{\rm H_2}/(M_{\rm HI} + M_{\rm H_2})$.}
\end{table}

\clearpage

\begin{table}
\caption{ATCA Observations}
\begin{tabular}{lcc}
\tableline \tableline
Array configurations:             & 6A,6B,6C,6D,1.5A,1.5B,750B,375 &\\
Minimum/maximum baseline (m)      & 30.6, 5972 & \\
R.A. pointing center (J2000)      & 06$^{\rm h}$ 43$^{\rm m}$ 06.$^{\rm s}$66 &\\
Dec. pointing center (J2000)      &-74$^{\circ}$ 16$'$ 16$''$.0 &\\
Integration time (hr)          & 114 &\\
Primary beam FWHM (arcmin)        & 31.5 &\\
Correlator configuration          & full\_8\_512-128 & \\
Center frequency (MHz)            & 1390.0 &\\
Number of channels                &44&\\
Channel separation (km s$^{-1}$)  &27.3&\\
Primary flux calibrator           &1934-638&\\
Phase calibrator                  &0454-801&\\
Bandpass calibrator               &0407-658 &\\
Synthesized beam FWHM (arcsec)    &  7.9 $\times$ 6.7 ({\em robust} weight)& \\
Continuum bandwidth (MHz)         &128&\\
Channel map rms (mJy beam$^{-1}$)  &0.48 ({\rm robust} weight)&\\
Continuum image rms (mJy beam$^{-1}$) &0.08 ({\em robust} weight)&\\
\\

\end{tabular}
\end{table}

\newpage
\clearpage
\begin{deluxetable}{lccc}
\tablenum{3}
\tablecaption{Positions Observed with the SEST}
\tablewidth{0pc}
\tabletypesize{\footnotesize}
\tablehead{
\colhead{Position} & \colhead{P.A.\tablenotemark{(a)}} & \colhead{R.A.} & \colhead{ Decl.} \\
\colhead{        } & \colhead{(deg)}                    & \colhead{(J2000)}                      & \colhead{(J2000)} }
\startdata
P1      &   6 & 06:43:10.79 & -74:14:34.0 \\
P2      &  49 & 06:43:11.00 & -74:14:12.0 \\
P3      &  78 & 06:43:08.94 & -74:13:52.0 \\
P4      &  96 & 06:43:06.19 & -74:13:49.0 \\
P5      & 110 & 06:43:03.55 & -74:13:51.0 \\
P6      & 125 & 06:43:01.67 & -74:13:57.0 \\
P7      & 142 & 06:43:00.25 & -74:14:06.0 \\
P8      & 195 & 06:42:58.62 & -74:14:32.0 \\
P9      & 217 & 06:42:58.45 & -74:14:43.0 \\
P10     & 237 & 06:42:59.31 & -74:14:53.0 \\
P11     & 252 & 06:43:00.34 & -74:15:05.0 \\
P12     & 266 & 06:43:02.20 & -74:15:13.0 \\
P13     & 281 & 06:43:05.00 & -74:15:15.0 \\
P14     & 313 & 06:43:08.57 & -74:14:55.0 \\
        &     &             &             \\
Disk    &...  & 06:43:04.54 & -74:14:33.0 \\
Nucleus &...  & 06:43:06.32 & -74:14:09.0 \\
\enddata
\tablenotetext{a}{The SEST beam's position angle on the ring, measured 
clockwise in degrees from the western minor axis line of nodes, as illustrated
in Figures 4 and 5. }
\end{deluxetable}

\newpage
\clearpage
\begin{deluxetable}{lccccccccccc}
\tablenum{4}
\tablecaption{Results of the SEST Observations}
\tablewidth{0pc}
\rotate
\tabletypesize{\footnotesize}
\tablehead{
\colhead{Pos.} & \colhead{$I_{\rm 1-0}$\tablenotemark{(a)}} &
\colhead{Velocity\tablenotemark{(b)}} & \colhead{$\Delta v$} & \colhead{T$_{\rm A}^{*}$(1-0)} & 
\colhead{$\sigma_{1-0}$} & \colhead{$I_{\rm 2-1}$\tablenotemark{(a)}} & \colhead{Velocity} & 
\colhead{$\Delta v$\tablenotemark{(b)}} & \colhead{$\Delta v_{\rm cor}$\tablenotemark{(c)}} & \colhead{T$_{\rm A}^{*}$(2-1)}  &
\colhead{$\sigma_{2-1}$} \\
\colhead{} &  \colhead{(K~km~s$^{-1}$)} & 
\colhead{(km s$^{-1}$)} & \colhead{(km~s$^{-1}$)} & \colhead{(mK)} & \colhead{(mK)} & 
\colhead{(K~km~s$^{-1}$)} & \colhead{(km~s$^{-1}$)} &  \colhead{{\small (km~s$^{-1}$)}} &  
\colhead{(km~s$^{-1}$)} & \colhead{(mK)} & \colhead{(mK)} }
\startdata
P1  & $<$0.22     & ...      & ...     & ... & 1.8 & $<$0.21     &...       & 242~(20) &230& ... & 0.8 \\
P2  & $<$0.11     & ...      & ...     & ... & 0.9 & $<$0.14     &...       & 140~(11) &125& ... & 0.8 \\
P3  & 0.91~(0.05) & 6949~(~6)& 107~(~3)& 8.3 & 0.9 & 0.48~(0.04) & 6941~(~4)& ~85~(~8) &130& 5.3 & 0.8\\
P4  & 1.57~(0.18) & 6814~(13)& 232~(24)& 9.3 & 1.6 & 1.25~(0.08) & 6903~(12)& 195~(21) &175& 7.3 & 0.7\\
P5  & 1.71~(0.19) & 6887~(19)& 220~(37)& 7.7 & 1.4 & 1.01~(0.10) & 6869~(17)& 208~(18) &183& 5.2 & 0.9\\
P6  & 1.10~(0.19) & 6887~(21)& 264~(40)& 6.6 & 1.8 & 0.74~(0.10) & 6827~(21)& 247~(49) &213& 2.9 & 0.9\\
P7  & 2.02~(0.18) & 6686~(28)& 642~(40)& 4.2 & 1.5 & 1.68~(0.09) & 6771~(18)& 558~(31) &542& 4.2 & 0.7\\
P8  & $<$0.17     & ...      & ...     & ... & 1.4 & $<$0.28     & ...      & 193~(14) &172& ... & 1.2\\
P9  & 0.88~(0.14) & 6472~(10)& 100~(21)& 7.3 & 1.2 & 1.20~(0.09) & 6512~(12)& ~89~(10) &~89& 9.6 & 0.9\\
    &             &          &         &     &     & 0.41~(0.06) & 6201~(16)& ~87~(16) &~87& 4.4 & 0.9\\
P10 & 1.00~(0.12) & 6445~(~6)& 103~(12)& 8.4 & 1.3 & 0.48~(0.08) & 6439~(15)& ~55~(11) &~55& 5.5 & 1.1\\
P11 & 0.79~(0.17) & 6433~(11)& ~98~(22)& 7.8 & 2.8 & 0.93~(0.13) & 6411~(10)& ~97~(12) &~66& 9.6 & 1.5\\
P12 & 0.70~(0.13) & 6459~(~7)& ~78~(15)& 8.5 & 1.8 & 1.04~(0.09) & 6525~(20)& 247~(40) &230& 6.2 & 1.0\\
P13 & $<$0.25     & ...      & ...     & ... & 2.1 & $<$0.27     & ...      & 240~(20) &233& ... & 1.0\\
P14 & $<$0.25     & ...      & ...     & ... & 2.1 & $<$0.23     & ...      & 420~(30) &410& ... & 0.8\\
    &             &          &         &     &     &             &          &          &   &     &   \\
Disk&$<$0.25      & ...      & ...     & ... & 2.2 & $<$0.16     & ...      & ...      &...& ... & 1.0\\
Nucleus &$<$0.29  & ...      & ...     & ... & 2.4 & $<$0.14     & ...      & ...      &...& ... & 0.9\\  
\enddata
\tablenotetext{(a)}{Integrated CO line fluxes as defined in Section 2.2. The upper-limits are $3\sigma$ and assume 
                   $\Delta v_{\rm CO} = \Delta v_{\rm HI}$.}
\tablenotetext{(b)}{For ring positions with no CO detections (P1, P2, P8, P13, and P14) we list the corresponding
                    HI linewidths (FWHM) in km s$^{-1}$.}
\tablenotetext{(c)}{ Linewidths after correcting for effects of rotation and expansion within the $22\arcsec$ beam. The
corrected widths are calculated with $\Delta v^{2}_{\rm cor}
= \Delta v^{2}_{\rm CO} - \Delta v^{2}_{\rm rot} - \Delta v^{2}_{\rm exp}$, as described in Section $3.4.3$. Where
$^{12}$CO($J=2-1$) emission is not detected we list the corrected HI widths.}

\end{deluxetable}

\newpage
\clearpage
\begin{deluxetable}{lccccccccc}
\rotate
\tablenum{5}
\tablecaption{Derived Mean Surface Densities, Masses, and Luminosities}
\tablewidth{0pc}
\tabletypesize{\footnotesize}
\tablehead{
\colhead{Position} & \colhead{$\Sigma_{\rm H_2}$\tablenotemark{(a)} } & \colhead{$\Sigma_{\rm HI}$\tablenotemark{(b)}} & 
\colhead{$\Sigma_{\rm H\alpha}$\tablenotemark{(c)}} & \colhead{$\Sigma_{\rm 24 \mu m}$\tablenotemark{(d)}}  & 
\colhead{$A_{\rm H\alpha}$\tablenotemark{(e)}} & \colhead{M$_{\rm H_2}$\tablenotemark{(f)}} & 
\colhead{M$_{\rm HI}$\tablenotemark{(f)}} & \colhead{$L_{\rm H\alpha}$\tablenotemark{(g)}} & \colhead{SFE\tablenotemark{(h)}} \\
\colhead{}  & \colhead{{\small (M$_\odot$ pc$^{-2}$)}} & \colhead{\small (M$_\odot$ pc$^{-2}$)} & 
\colhead{\small ($L_{\odot}$ pc$^{-2}$) } &  \colhead{\small (L$_\odot$ pc$^{-2}$)} & \colhead{(mag)} & 
\colhead{\small (10$^{8}$ M$_\odot$)} & \colhead{\small (10$^{8}$ $M_{\odot}$)} & 
\colhead{\small (10$^{7}$~$L_{\odot}$)} & \colhead{ } }
\startdata
P1     &$<$3.0         &  37.7~(3.2) &  0.33~(.01)  &11.5~(0.1) & 0.80~(0.14) & $<$1.2     &  15.2~(1.1) &  ~3.72~(.05) & $>$-0.91\\
P2     &$<$2.0         &  21.6~(1.8) &  0.19~(.01)  &~5.4~(0.1) & 0.98~(0.14) & $<$0.8     &  ~8.7~(0.8) &  ~2.54~(.04) & $>$-0.97\\
P3     &  ~6.9~(0.6)   &  13.4~(1.4) &  0.26~(.01)  &13.7~(0.1) & 1.03~(0.14) & ~2.8~(0.3) &  ~5.4~(0.6) &  ~3.73~(.06) & ~-1.37~(0.08)\\
P4     &  18.0~(1.2)   &  12.7~(2.8) &  0.23~(.01)  &~5.7~(0.1) & 0.97~(0.14) & ~7.3~(0.4) &  ~5.1~(0.8) &  ~3.06~(.05) & ~-1.84~(0.05)\\
P5     &  14.6~(1.5)   &  13.0~(2.6) &  0.20~(.01)  &~4.3~(0.1) & 0.92~(0.14) & ~5.9~(0.6) &  ~5.2~(0.7) &  ~2.49~(.04) & ~-1.82~(0.09)\\
P6     &  10.7~(0.3)   &  12.7~(0.8) &  0.15~(.01)  &~3.6~(0.1) & 0.90~(0.14) & ~4.3~(0.1) &  ~5.1~(0.3) &  ~1.98~(.04) & ~-1.78~(0.03)\\
P7     &  24.1~(1.3)   &  14.4~(1.9) &  0.13~(.01)  &~2.8~(0.1) & 0.86~(0.14) & ~9.7~(0.6) &  ~5.8~(0.2) &  ~1.59~(.03) & ~-2.21~(0.04)\\ 
P8     &$<$4.0         &  34.5~(0.8) &  0.52~(.01)  &19.3~(0.1) & 0.83~(0.14) & $<$1.6     &  13.9~(0.5) &  ~6.06~(.06) & $>$-0.84\\
P9     &  16.7~(1.2)   &  65.6~(4.1) &  0.94~(.01)  &35.2~(0.2) & 0.84~(0.14) & 6.7~(0.5) &  26.3~(0.8) &  11.09~(.09) & ~-1.20~(0.04)\\
P10    &  ~5.0~(0.9)   &  84.8~(5.9) &  0.83~(.01)  &36.3~(0.2) & 0.92~(0.14) & ~2.0~(0.4) &  34.1~(1.3) &  10.62~(.09) & ~-0.72~(0.12)\\
P11    &  ~9.7~(1.4)   &  89.9~(1.9) &  0.45~(.01)  &20.8~(0.2) & 0.96~(0.14) & ~3.9~(0.6) &  35.6~(0.6) &  ~5.96~(.07) & ~-1.28~(0.11)\\
P12    &  10.8~(1.0)   &  77.8~(3.1) &  0.30~(.01)  &11.6~(0.1) & 0.85~(0.14) & ~4.4~(0.4) &  31.9~(1.1) &  ~3.56~(.05) & ~-1.51~(0.08)\\
P13    &$<$2.8         &  57.0~(2.1) &  0.16~(.01)  &~6.8~(0.1) & 0.92~(0.14) & $<$1.2     &  22.9~(0.8) &  ~1.99~(.04) & $>$-1.23\\
P14    &$<$2.4         &  25.2~(2.4) &  0.12~(.01)  &~4.7~(0.1) & 0.84~(0.14) & $<$1.0     &  20.4~(0.8) &  ~1.49~(.03) & $>$-1.24\\
       &               &             &	            &           &             &            &             &              &     \\
Nucleus&$<$2.1         &$<$0.2       &  0.03~(.01)  &10.4~(0.1) & 2.17~(0.15) & $<$2.2     &  $<$0.93    &  ~1.73~(.06) & ... \\
Disk   &$<$1.8         &$<$0.1       &  $<$0.02     &~3.5~(0.1) & 1.74~(0.14) & $<$2.0     &  $<$0.52    &  $<$0.78     & ... \\
\enddata

\tablenotetext{a} {H$_{2}$ surface density, defined $\Sigma_{\rm H_2} = M_{\rm H_2}/\Delta\Omega_{\rm ring}$. 
No helium correction applied.}
\tablenotetext{b}{HI surface density, defined $\Sigma_{\rm HI} = M_{\rm HI}/\Delta\Omega_{\rm ring}$. No helium correction applied.}
\tablenotetext{c}{The H$\alpha$ luminosity density derived by dividing $L_{\rm H\alpha}$ by $\Delta\Omega_{\rm ring}$. No extinction correction applied.}
\tablenotetext{d}{Average 24 $\mu$m luminosity density, defined $\Sigma_{\rm 24 \mu m} = \nu L_{\nu}(24 \mu m)/\Delta\Omega_{\rm ring}$.}
\tablenotetext{e}{Extinction correction for the H$\alpha$ line calculated after \citet{kennicutt07}. See discussion in Section $3.1.1$.}
\tablenotetext{f}{Total H$_2$ mass within each SEST beam using Equation (3) in Section $2.5$. No helium correction applied.}
\tablenotetext{g}{Integrated H$\alpha$ luminosity within each SEST beam after correcting for both internal and foreground extinction. See Section $3.2.1$.}
\tablenotetext{h}{Star formation efficiency, defined SFE $= $log$( L_{\rm H\alpha} / M{_{H_2}} )$ following \citet{young},
with $L_{\rm H\alpha}$ and $M_{\rm H_2}$ in units of $L_{\odot}$ and $M_{\odot}$ respectively. See Section $3.5$.}
\end{deluxetable}

\newpage
\clearpage
\begin{deluxetable}{lccccc}
\tablenum{6}
\tablecaption{Gravitational Stability in AM0644-74's Ring}
\tablewidth{0pc}
\tabletypesize{\footnotesize}
\tablehead{
\colhead{Position} & \colhead{$\Sigma_{*}$\tablenotemark{(a)}} & \colhead{$\Sigma_{\rm gas}$\tablenotemark{(b)}} & 
\colhead{$Q_{\rm gas}$\tablenotemark{(c)}} & \colhead{$Q_{*}$\tablenotemark{(d)}} & 
\colhead{$Q_{\rm tot}$\tablenotemark{(e)}} \\
\colhead{} & \colhead{($M_{\odot}$ pc$^{-2}$)} & \colhead{($M_{\odot}$ pc$^{-2}$)} & 
\colhead{} & \colhead{} & \colhead{} }
\startdata
P1      & 37.1 & ~55.4~(4.4) & 2.9~(0.5) & 1.7$-$~4.3 & 1.1$-$1.7 \\ 
P2      & 29.7 & ~32.1~(2.4) & 2.7~(0.5) & 2.2$-$~2.9 & 1.2$-$1.4 \\   
P3      & 37.9 & ~27.6~(2.1) & 3.3~(0.6) & 1.7$-$~2.4 & 1.1$-$1.4 \\ 
P4      & 32.6 & ~41.8~(4.1) & 2.9~(0.6) & 2.0$-$~3.7 & 1.2$-$1.6 \\   
P5      & 29.5 & ~37.5~(4.1) & 3.4~(0.7) & 2.2$-$~4.3 & 1.3$-$1.9 \\   
P6      & 23.5 & ~31.8~(1.2) & 4.6~(0.8) & 2.8$-$~6.3 & 1.7$-$2.7 \\   
P7      & 22.4 & ~52.4~(3.1) & 7.1~(1.3) & 2.9$-$16.7 & 2.1$-$5.0 \\   
P8      & 39.4 & ~52.4~(1.1) & 2.4~(0.4) & 1.6$-$~3.1 & 1.0$-$1.3 \\   
P9      & 49.6 & 111.93~(5.8) & 0.6~(0.1) & 1.1$-$~1.3 & 0.4   \\   
P10     & 50.2 & 122.1~(8.1) & 0.3~(0.1) & 0.8$-$~1.3 & 0.3  \\   
P11     & 45.6 & 135.5~(3.2) & 0.3~(0.1) & 1.0$-$~1.4 & 0.2$-$0.3 \\   
P12     & 37.0 & 120.5~(4.4) & 1.3~(0.2) & 1.8$-$~4.3 & 0.8$-$1.0 \\   
P13     & 29.7 & ~81.3~(2.9) & 2.0~(0.3) & 2.2$-$~5.4 & 1.0$-$1.4 \\   
P14     & 21.1 & ~37.5~(3.3) & 7.5~(1.4) & 3.1$-$13.4 & 2.2$-$4.8 \\   
%        & & & & &	\\ 
%Disk    &  6.0  & $<$6  &  ...   &  ...  &  ...  \\
\enddata
\tablenotetext{(a)}{Stellar mass surface density estimated by multiplying the $I$-band luminosity
density ($L_{\rm I}$) by stellar $M_{*}/L_{\rm I,\odot}$ derived by fitting the FUV-4.5 $\mu$m
flux densities to Starburst99 SED. Uncertainties are $\approx10\%$ (Section $3.3$).}
\tablenotetext{(b)}{Gas surface density, $\Sigma_{\rm gas} = ( \Sigma_{\rm HI}
  + \Sigma_{\rm H_{2}})$. A helium mass correction is included.}
\tablenotetext{(c)}{Toomre $Q$ parameter for HI$+$H$_2$. A correction for helium mass is included.
Quoted uncertainties assume a flat rotation curve and Galactic $X_{\rm CO}$.}
\tablenotetext{(d)}{Toomre $Q$ parameter for the stellar ring. The two values correspond to
different stellar velocity dispersions: $\sigma_{*}=\sigma_{\rm gas}$ (larger) and
$\sigma_{*}= d_{\rm ring}/\tau_{\rm B5}$ (smaller), where $d_{\rm ring}$ is the thickness of the
H$\alpha$ ring and $\tau_{\rm B5}$ is the main-sequence lifetime of a B5 star.}
\tablenotetext{(e)}{The combined stellar and gaseous Toomre $Q_{tot}$ parameter, defined
   $Q_{\rm tot} = {{\kappa}\over{\pi G}} ( { {\Sigma_{\rm gas}}\over{\sigma_{\rm gas}} } + 
{{\Sigma_{*}}\over{\sigma_{*}}} )^{-1}$ \citep{wang}. The two values listed again correspond to
$\sigma_{*}=\sigma_{\rm gas}$ and $\sigma_{*}= d_{\rm ring}/\tau_{\rm B5}$, with formal
uncertainties in the range $0.1-0.4$. }
\end{deluxetable}

\newpage
\clearpage
\begin{deluxetable}{lccccccc}
\tablenum{7}
\tablecaption{Derived Physical Conditions in AM0644-741's ISM}
\tablewidth{0pc}
\tabletypesize{\footnotesize}
\tablehead{
\colhead{Position} & \colhead{$P_{\rm ISM}/P_{\odot}$\tablenotemark{(a)}} & \colhead{$\phi_{mol}$\tablenotemark{(b)}} &
\colhead{$f_{\rm mol}$\tablenotemark{(c)}} & 
\colhead{$\chi_{\rm UV}/\chi_{\odot}$\tablenotemark{(d)}} & 
\colhead{$\chi_{\rm FUV}/\chi_{\odot}$\tablenotemark{(e)}}  &  
\colhead{$\Phi_{\rm UV}$\tablenotemark{(f)}} & \colhead{$n$\tablenotemark{(g)}} \\
\colhead{} & \colhead{} & \colhead{} & \colhead{} & \colhead{} & \colhead{} & \colhead{} & \colhead{(cm$^{-3}$)}}
\startdata
P1      & ~17.0$-$~26.8   & ~~47.9$-$~131.2   & $<$0.07     & 10.6~(0.5) & 1.16~(0.02) &  ~9.1~(0.6) & ~20  \\ 
P2      & ~~6.6$-$~~7.6   & ~~~7.1$-$~~~9.8   & $<$0.09     & ~8.9~(0.3) & 0.59~(0.02) &  15.1~(0.9) & 125  \\   
P3      & ~~6.0$-$~~7.3   & ~~~5.1$-$~~~8.0   & 0.34~(0.05) & 10.0~(0.5) & 0.70~(0.02) &  14.3~(1.0) & 320  \\ 
P4      & ~10.3$-$~14.2   & ~~17.2$-$~~34.9   & 0.59~(0.08) & ~9.8~(0.4) & 0.69~(0.03) &  14.2~(1.0) & 100  \\   
P5      & ~~8.3$-$~11.8   & ~~12.5$-$~~26.9   & 0.53~(0.10) & ~8.5~(0.4) & 0.53~(0.01) &  16.0~(0.9) & 170  \\   
P6      & ~~5.8$-$~~9.0   & ~~~6.1$-$~~15.8   & 0.46~(0.02) & ~7.9~(0.3) & 0.36~(0.02) &  21.9~(1.4) & 275  \\   
P7      & ~13.0$-$~31.5   & ~~34.3$-$~241.5   & 0.63~(0.06) & ~8.2~(0.3) & 0.27~(0.02) &  30.4~(1.9) & 125  \\   
P8      & ~15.9$-$~22.1   & ~~25.2$-$~~51.9   & $<$0.10     & 17.5~(0.5) & 0.42~(0.02) &  41.7~(1.8) & 175  \\   
P9      & ~59.9$-$~76.8   & ~352.7$-$~608.1   & 0.20~(0.02) & 23.1~(0.5) & 0.69~(0.01) &  33.5~(1.0) & ~~2  \\   
P10     & ~69.1$-$~69.8   & ~446.0$-$~455.3   & 0.06~(0.01) & 25.0~(0.5) & 0.89~(0.01) &  28.1~(0.8) & ~~1  \\   
P11     & ~75.2$-$~81.3   & ~710.2$-$~843.2   & 0.10~(0.02) & 18.9~(0.4) & 0.81~(0.02) &  23.3~(0.9) & ~~2  \\   
P12     & ~62.9$-$~84.3   & ~633.9$-$1206.7   & 0.12~(0.01) & 14.3~(0.5) & 0.76~(0.03) &  18.8~(1.1) & ~~5  \\   
P13     & ~29.9$-$~41.8   & ~192.3$-$~400.3   & $<$0.05     & ~9.2~(0.4) & 0.62~(0.02) &  14.8~(0.9) & ~20  \\   
P14     & ~~7.3$-$~16.1   & ~~11.3$-$~~64.8   & $<$0.09     & ~7.0~(0.4) & 0.67~(0.02) &  10.4~(0.8) & ~30 \\   
\enddata
\tablenotetext{(a)}{ISM pressure from 
$P_{\rm ISM} = (\pi G/2k_{\rm B})~(\Sigma_{\rm gas}^2 + {{\sigma_{\rm gas}}\over{\sigma_{*}}}\Sigma_{\rm gas}\Sigma_{*})$
expressed in units of the local ISM pressure \citep[$P_{\odot} = 10^{4}$ K~cm$^{-3}$,][]{elmegreen93}. The range of values follows
from the uncertainty in the stellar velocity dispersion: $\sigma_{*}=d_{\rm ring}/\tau_{\rm B5}\approx 50$ km s$^{-1}$ to
$\sigma_{*}=\sigma_{\rm gas}$.}
\tablenotetext{(b)} {Molecular fraction parameter, defined $\phi_{mol} = (P_{\rm ISM}/P_{\odot})^{2.2} 
(\chi_{\rm UV}/\chi_{\odot})^{-1}$ \citep{elmegreen93}. The ISM is primarily molecular when $\phi_{mol}\ga1$.}
\tablenotetext{(c)} {The molecular gas fraction at each SEST beam position, 
defined $f_{\rm mol} = M_{\rm H_2}/(M_{\rm HI} + M_{\rm H_2})$.}
\tablenotetext{(d)} {UV radiation field intensity in the ring in units of the
local value \citep[$\chi_{\odot} = 2.0 \times 10^{-4}$ erg s$^{-1}$ cm$^{-2}$ sr$^{-1}$;][]{draine},
defined $\chi_{\rm UV} = L_{\rm FIR}/4 \pi D_{\rm L}^{2} \Delta\Omega_{\rm ring} (2.0 \times 10^{-4})$,
$L_{\rm FIR}$ has been estimated by scaling the {\em IRAS} $40-500 ~\mu$m luminosity 
by the fraction of the total {\em Spitzer} $70 ~\mu$m emission contained in the $22''$ beam. $\Delta\Omega_{\rm ring}$ is the ring's solid angle.}
\tablenotetext{(e)} {Normalized FUV radiation field intensity in the ring, defined
$\chi_{\rm FUV}/\chi_{\odot} = F_{\rm FUV}/\Delta\Omega_{\rm ring} (2.0 \times 10^{-4})$, with $F_{\rm FUV}$ derived 
from the {\em GALEX} {\em FUV} image.}
\tablenotetext{(f)}{Average cloud filling factor, defined $\Phi = \chi_{\rm UV}/\chi_{\rm FUV}$.}
\tablenotetext{(g)}{Total gas volume density ($n = n_{\rm HI} + 2n_{\rm H_2}$ cm$^{-3}$) 
inferred from PDF model of \citet{allen}.}
\end{deluxetable}

\clearpage
\begin{figure}
\figurenum{1}
\epsscale{1.1}
\plotone{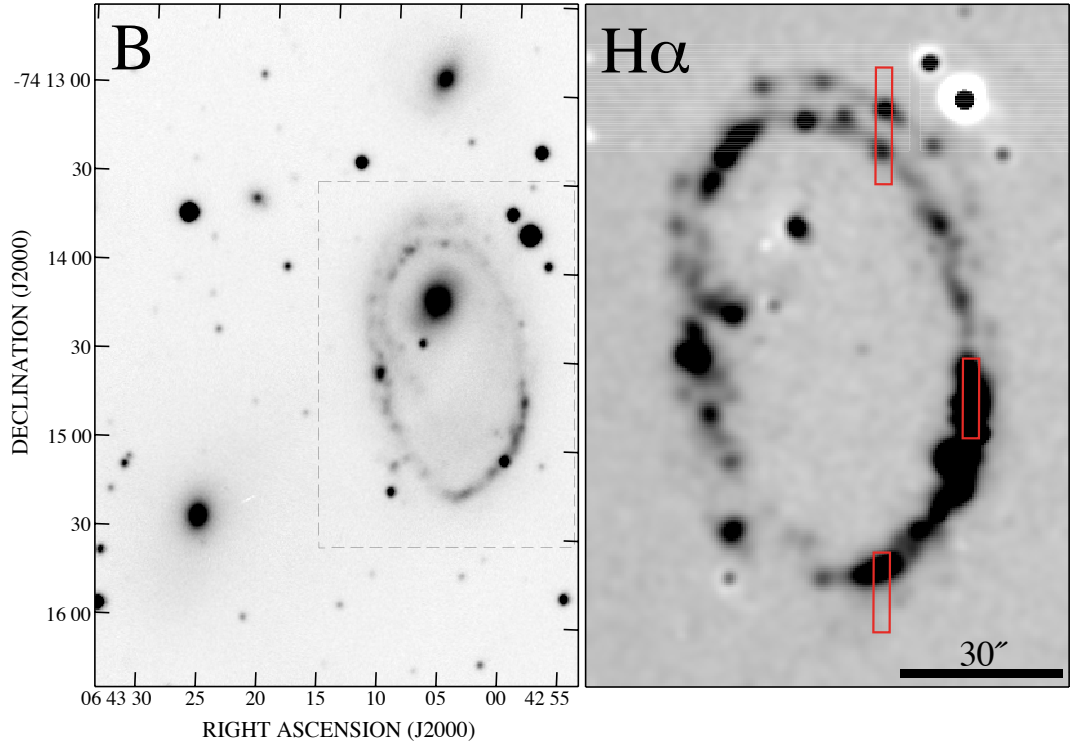}  
\caption{\small Left: optical $B$-band image of AM0644-741 and its two closest
companion galaxies. Right: continuum subtracted H$\alpha$ 
image of the ring galaxy corresponding to the dashed region at left. A linear
stretch is used in both. The rectangles (red in the online version)
mark the slitlets  used for the EFOSC2 spectra shown in Figure 3. Both images 
were obtained at CTIO. The star-forming complexes visible in the H$\alpha$ ring show 
a pronounced ``beads on a string'' morphology, which is indicative of large scale
gravitational instabilities providing the triggering mechanism.
The elliptical galaxy $1.8\arcmin$ to the southeast of the ring's nucleus is the ``intruder''.
A faint stellar bridge connects it to the ring's center (see Figure 1 in HW97).}
\end{figure}

\clearpage
\begin{figure}
\figurenum{2}
\epsscale{1.1}
\plotone{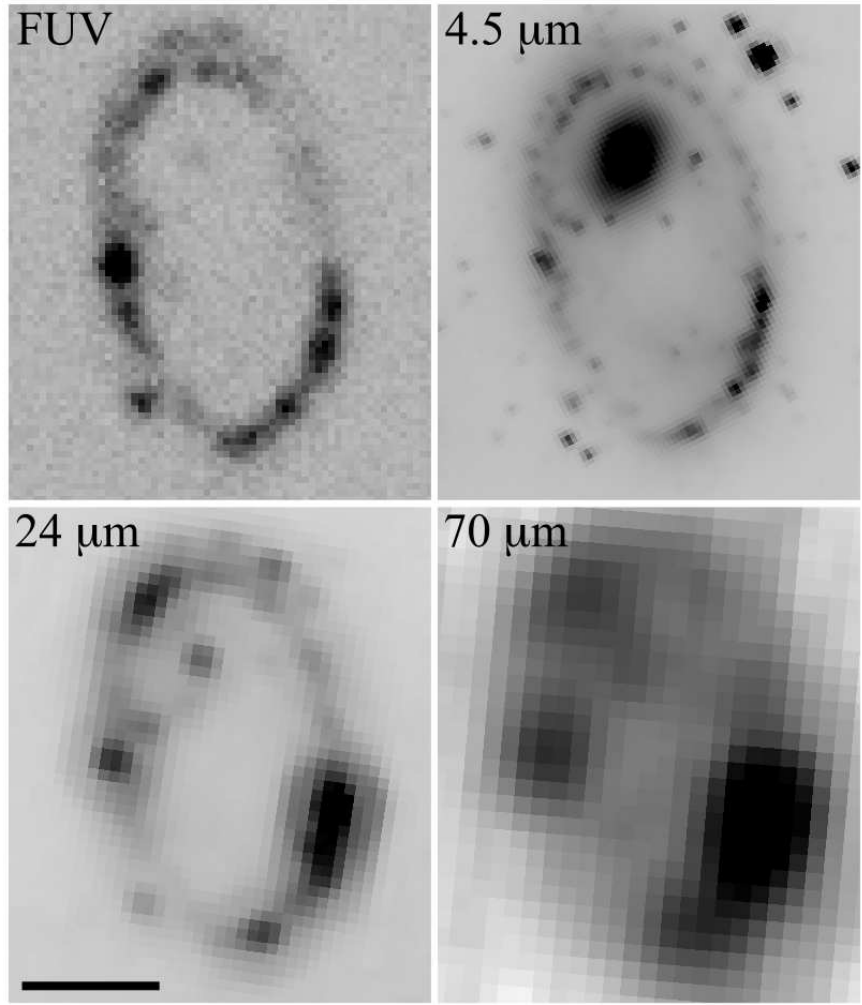} 
\caption{\small Emission from AM0644-741 spanning far-ultraviolet to 
far-infrared wavelengths. Upper left:
FUV obtained with $GALEX$ ($\lambda_{\rm eff} = 0.153 ~\mu m$); upper right:
$4.5$ $\mu$m obtained with IRAC; lower left: $24$ $\mu$m with MIPS;
lower right: $70$ $\mu$m with MIPS. All are shown using a logarithmic stretch.
Emission in the FUV, $24$ and $70$ $\mu$m wavebands is dominated
by the $42$ kpc diameter ring, with only a negligible contribution from the nucleus
and enclosed disk, in agreement with the H$\alpha$ image. Emission from evolved
stars dominates the $IRAC ~4.5$ $\mu$m image. A $30\arcsec$ scale bar is shown in the bottom left panel.}
\end{figure}

\clearpage
\begin{figure}
\figurenum{3}
\epsscale{.70}
\plotone{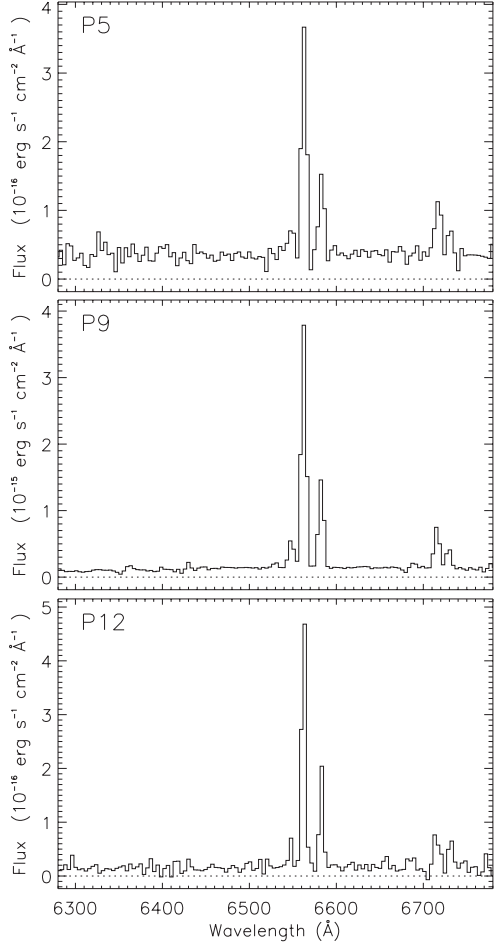}
\caption{EFOSC2 red spectra for three HII complexes in AM0644-741's ring.
The slitlet positions are shown in Figure 1, and correspond closely to the
centers of SEST positions P5, P9, and P12 in Figures 5 and 6. Emission lines
of H$\alpha$, [N~II] $~\lambda\lambda 6548, 6584$, and [S~II] $\lambda\lambda 6716, 6731$ 
~are evident. The $F_{\rm [N II] 6584}/F_{\rm H\alpha}$ and 
$F_{\rm [N~II] 6584}/F_{\rm [S~II] 6716 + 6731}$  emission line ratios
yield $12$+log(O/H)$ \approx9.1$ using the empirical
calibration of \citet{nagao}, with no significant azimuthal variations. 
This is substantially in excess of solar metallicity \citep[$12$+log(O/H)$_{\odot}=8.73-8.79$;][]{caffau}.}
\end{figure}

\clearpage
\begin{figure}
\figurenum{4}
\epsscale{0.8}
\plotone{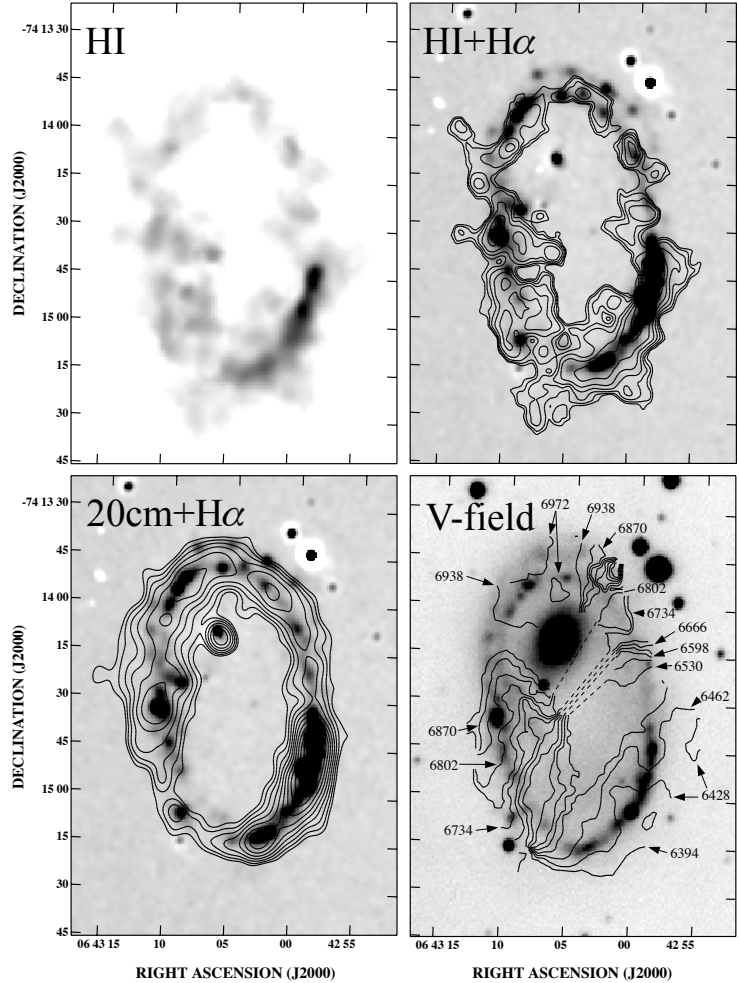} 
\caption{\small ATCA observations of AM0644-741.  Upper left: {\em robust}
weighted total HI distribution in gray scale with a linear stretch. The synthesized beam
is $7.\arcsec9 \times 6.\arcsec7$ FWHM. Upper right: {\em robust} weighted
HI surface density contours ($\Sigma_{\rm HI}$) on the H$\alpha$ image.
The logarithmic contours correspond to $\Sigma_{\rm HI}$ of
$5.0, 7.7, 12.0, 18.6, 28.8, 44.6, 69.0, 103.0$ $M_{\odot}$ pc$^{-2}$. 
Lower left: Contours of  20 cm radio continuum using {\em robust} weighting
on the H$\alpha$ image.  The contours correspond to flux densities of
$0.11, 0.15, 0.20, 0.27, 0.36, 0.49, 0.66, 0.89, 1.20, 1.60$ mJy beam$^{-1}$.
Lower right: HI isovelocity contours (km s$^{-1}$) superposed on the
$B$-band gray scale image. }
\end{figure}

\clearpage
\begin{figure}
\figurenum{5}
\plotone{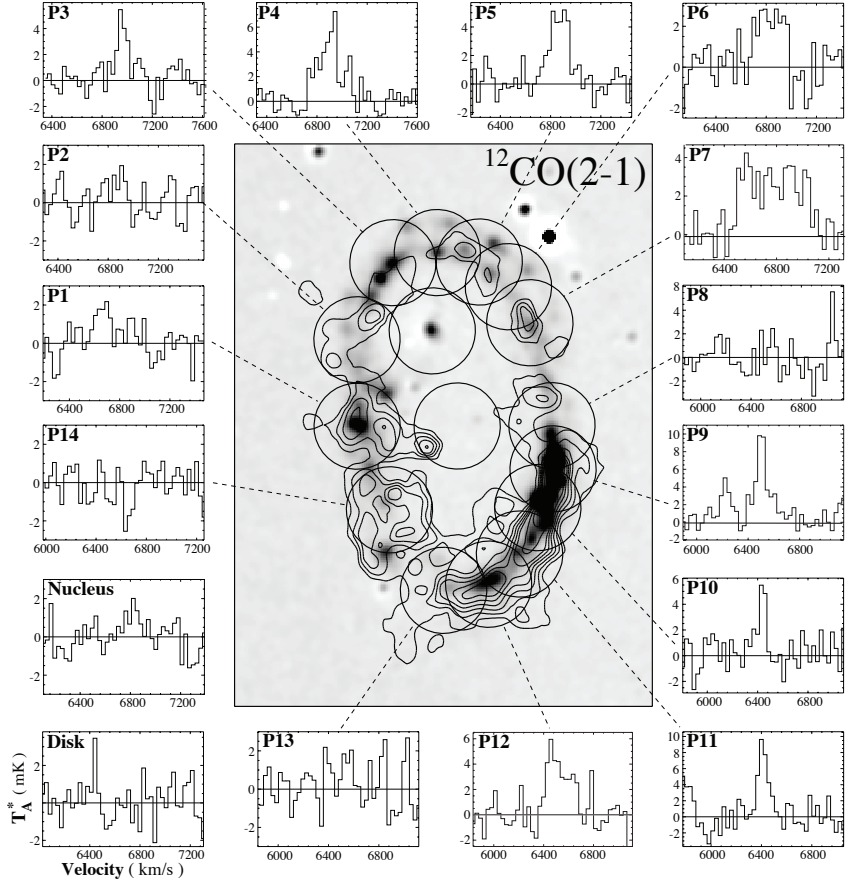}  
\caption{\small Individual $^{12}$CO($J=2-1$) spectra measured for AM0644-741's
ring, disk and nucleus. The circles represent the SEST $230$ GHz beam ($22\arcsec$ FWHM)
positions, and are shown superposed on a composite H$\alpha$ (gray scale)
and HI (contours) image. Molecular gas is detected in two sections of the ring, the
northern arc (P3-7) and the southwest quadrant (P9-12). A variety of profile shapes and widths
are visible. No line emission is detected from the nucleus or encircled disk.}
\end{figure}

\clearpage
\begin{figure}
\figurenum{6}
\plotone{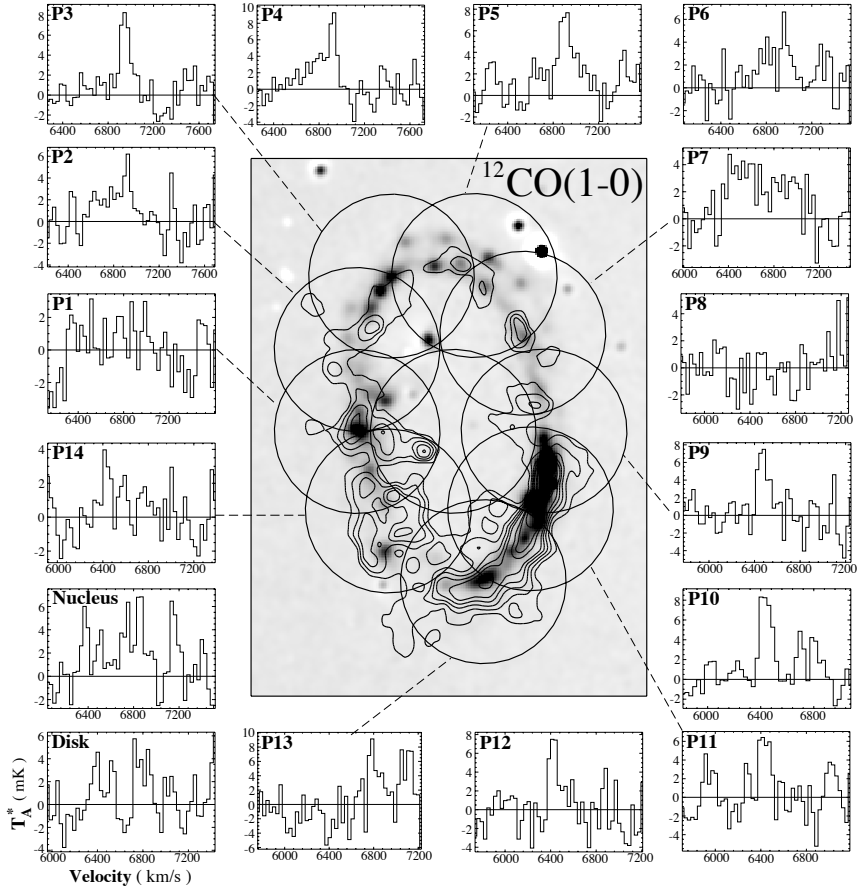}
\caption{\small Individual $^{12}$CO($J=1-0$) spectra measured for AM0644-741's
ring, disk and nucleus.  The circles represent alternating SEST $115$ GHz beam ($44\arcsec$ FWHM)
positions, and are shown superposed on a composite H$\alpha$ (gray scale)
and HI (contours) image. Only 10 of the 16 observed positions are shown
for clarity due to the large and overlapping beams. Note that $^{12}$CO($J=1-0$) and $^{12}$CO($J=2-1$) emission
is detected at the same positions (P3-7 and P9-12) and show similar profiles, despite the very different
beam widths.}
\end{figure}

\clearpage
\begin{figure}
\figurenum{7}
\epsscale{0.8}
\plotone{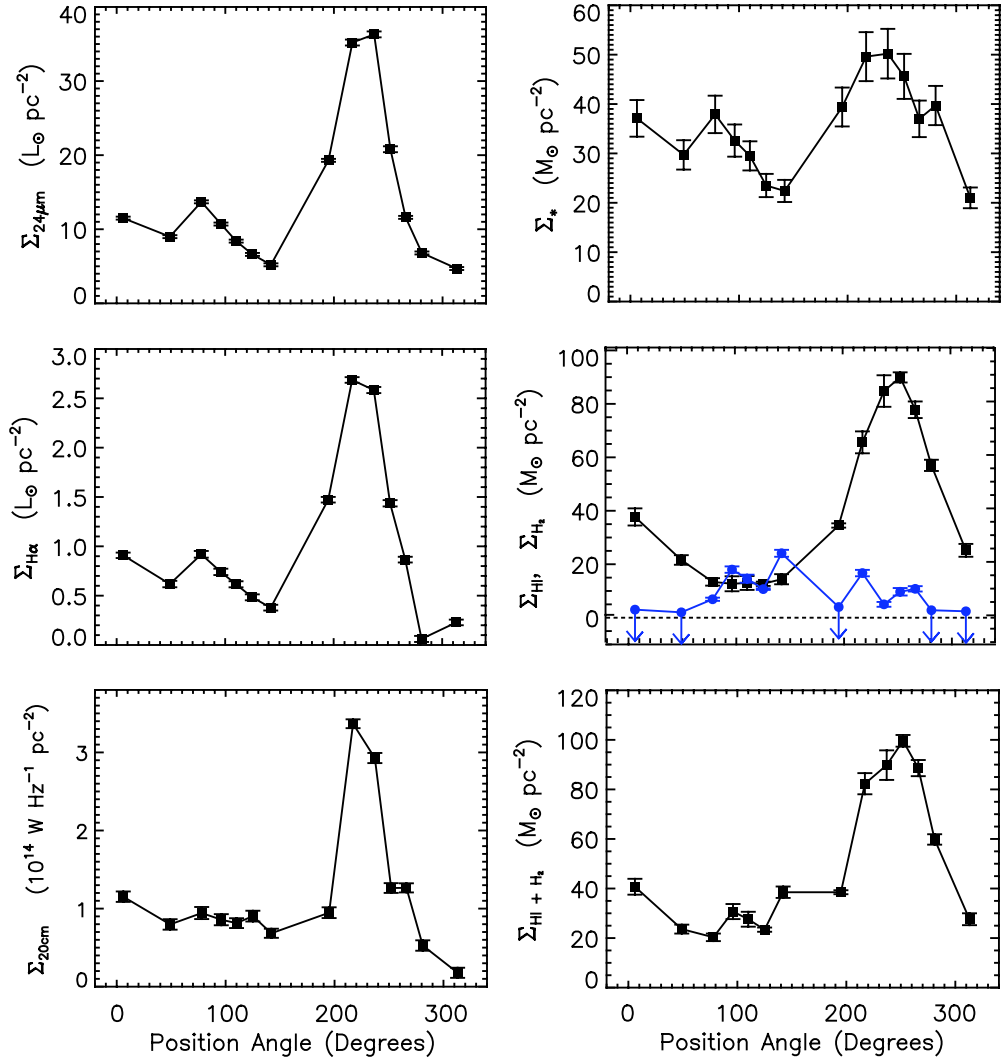}
\caption{Azimuthal variations in emission around AM0644-741's ring. The left
column shows star formation indicators --  upper left: warm dust as
traced by MIPS at $24 ~\mu$m; middle left: massive stars as traced
by H$\alpha$ line emission (corrected for internal and foreground extinction); 
bottom left: non-thermal radio continuum 
luminosity density as traced by ATCA $1.4$ GHz interferometery. The right column
shows the distribution of gas and stars --
upper right: surface mass density of stars derived by fitting FUV-$4.5~\mu$m
flux densities with Starburst99 SEDs; middle right: HI (black squares) and H$_2$
(blue circles) surface density plotted separately. No helium
correction has been applied. $\Sigma_{\rm H_2}$ may be overestimated
where two H$\alpha$ rings exist within the $230$ GHz beam (i.e., P4-7); 
bottom right: the total HI and H$_2$ surface density.
In each panel the positions run from P1 (far left) to P14 (far right).)
}
\end{figure}

\clearpage
\begin{figure}
\figurenum{8}
\epsscale{0.9}
\plotone{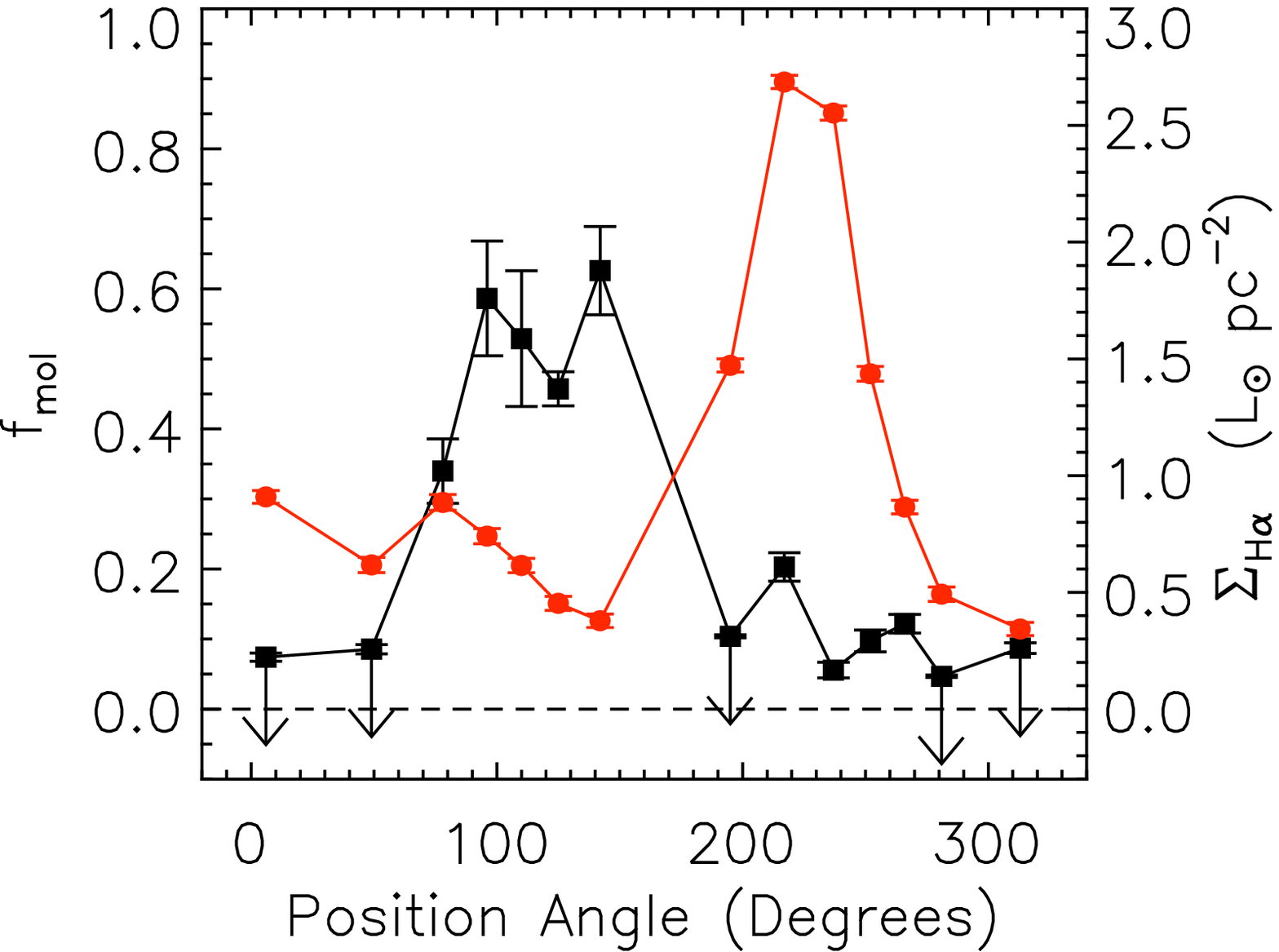}   
\caption{\small Azimuthal variations in the ring's molecular fraction (black squares)
and H$\alpha$ luminosity density (red circles). The molecular 
fraction ($f_{\rm mol}$) is defined $f_{\rm mol} = M_{\rm H_{2}} / (M_{\rm HI} + M_{\rm H_{2}})$. 
The ring is poor in molecular gas ($f_{\rm mol}\la0.2$) over $\approx3/4$ 
of its circumference. Only in the northern quadrant is $f_{\rm mol}\approx0.5$.
By comparison, $f_{\rm mol} \ga 0.75$ for M51's inner ($r \le 9$ kpc) spiral arms 
\citep{hidaka}. The figure reveals a clear 
anti-correlation between massive star formation and the cold molecular component 
of the ISM. $\Sigma_{\rm H\alpha}$ has been corrected for internal ($A_{\rm H\alpha}\approx1$)
and foreground extinction. The data points run from P1 (far left) to P14 (far right).}
\end{figure}

\clearpage
\begin{figure}
\figurenum{9}
\epsscale{0.9}
\plotone{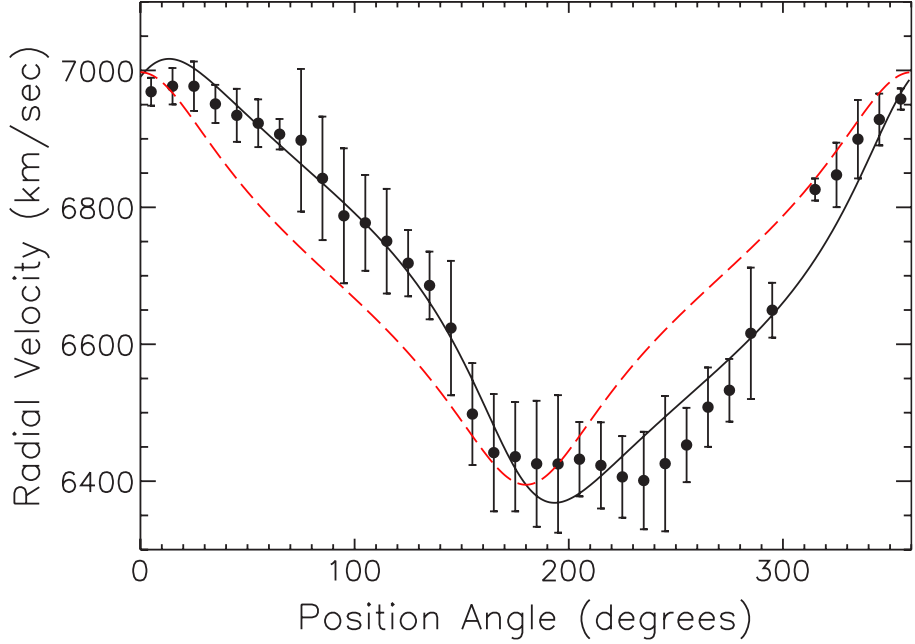}   
\caption{\small Radial velocity-position angle diagram for AM0644-741's ring using
the {\em robust} weighted HI data in Figure 4. The red dashed line
shows the least-squares fit to the data for an inclined 
($i = 56.5^{\circ}$) circular ring  that is rotating but not 
expanding, while the solid line shows the fit when the ring is also allowed to expand. 
Error bars represent the velocity spread local to where $v_{\rm rad}$ is defined. The
rotating/expanding ring model is a significantly better fit to the data, and we
find $v_{\rm sys} = 6692 \pm 8$ km s$^{-1}$, $v_{\rm cir} = 357 \pm 13$ km s$^{-1}$, 
and $v_{\rm exp} = 154 \pm 10$ km s$^{-1}$, with the quoted errors representing formal 
uncertainties of the fit. Position angle is measured counterclockwise from the northern
major-axis line of nodes.}
\end{figure}

\clearpage
\begin{figure}
\figurenum{10}
\epsscale{0.9}
\plotone{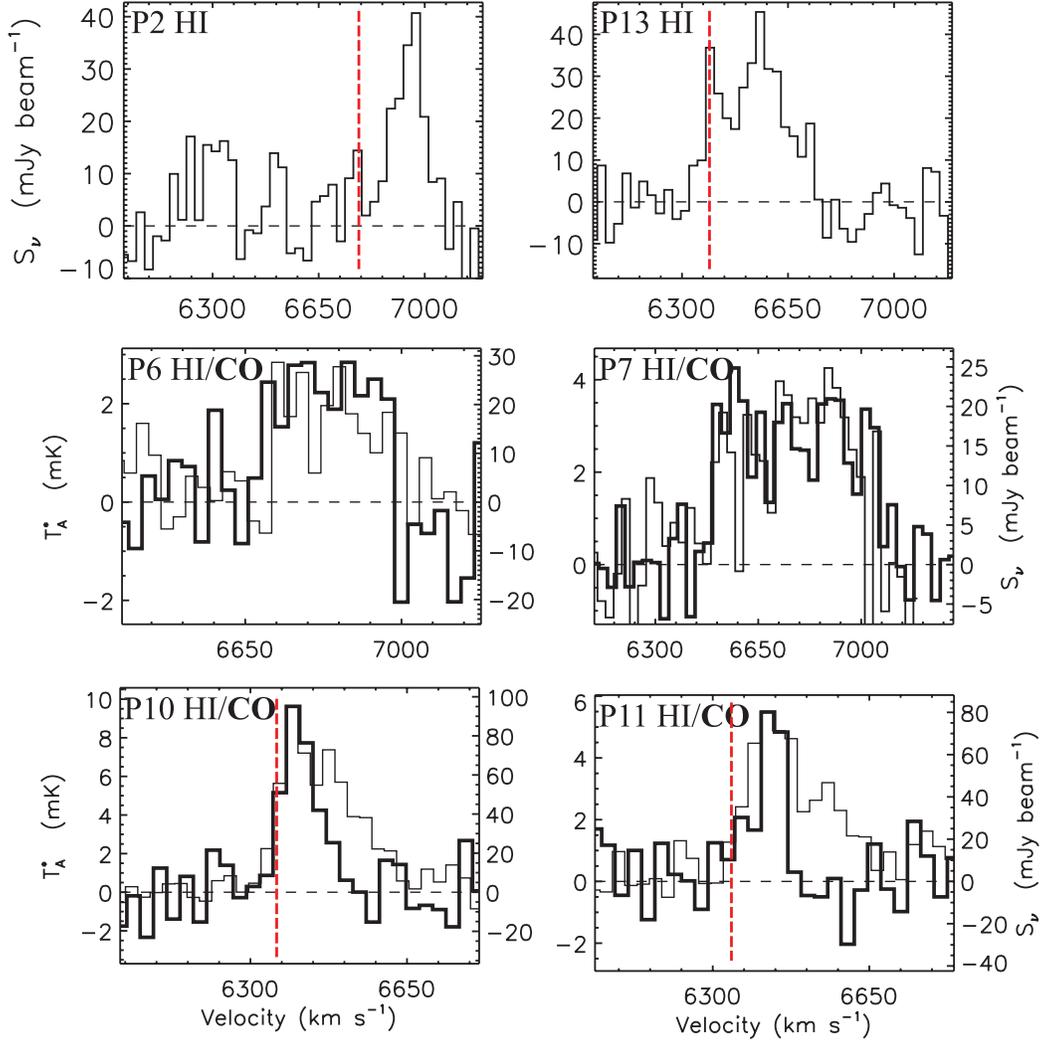}   
\caption{\small HI and $^{12}$CO($J=2-1$) spectra extracted from identical $22\arcsec$ diameter 
regions in AM0644-741's ring. Only HI is detected in the upper two panels (P2 \& 13). 
Elsewhere, the CO profiles are shown using thicker lines. Multiple velocity components are
sometimes seen in HI (e.g., P13), and both HI and CO lines can be very broad (P7).
HI and CO peaks agree well,
though HI sometimes shows a broad high-velocity wing not present in CO. The red vertical
dashed lines mark the corresponding optical velocities 
determined by FMA. Where the HI and/or CO
lines are sufficiently narrow and strong, the optical velocities are $\approx60-90$ km s$^{-1}$
smaller.}
\end{figure}

\clearpage
\begin{figure}
\figurenum{11}
\epsscale{1.0}
\plotone{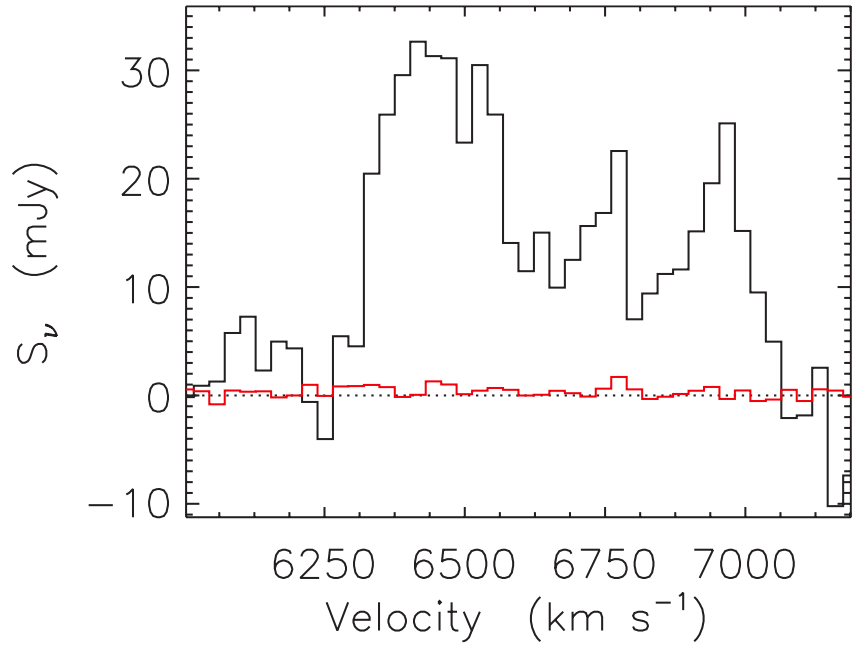}  
\caption{\small Total HI spectrum for AM0644-741 obtained by integrating
the {\em robust} weighted map cube over the optical extent of the ring
galaxy. An extremely wide profile ($\Delta v_{\rm FWZI} = 743 \pm 28$ km s$^{-1}$) with 
multiple peaks is evident. Also shown in red
is the integrated HI spectrum from a 320 arcsec$^{-2}$ ($62$ kpc$^{-2}$) region overlapping
the gas poor nucleus and interior disk (i.e., the HI ``hole'' in Figure 4), giving an
HI surface density upper limit of 1.9 $M_{\odot}$ pc$^{-2}$ ($3 \sigma$).}
\end{figure}

\clearpage
\begin{figure}
\figurenum{12}
\epsscale{1.0}
\plotone{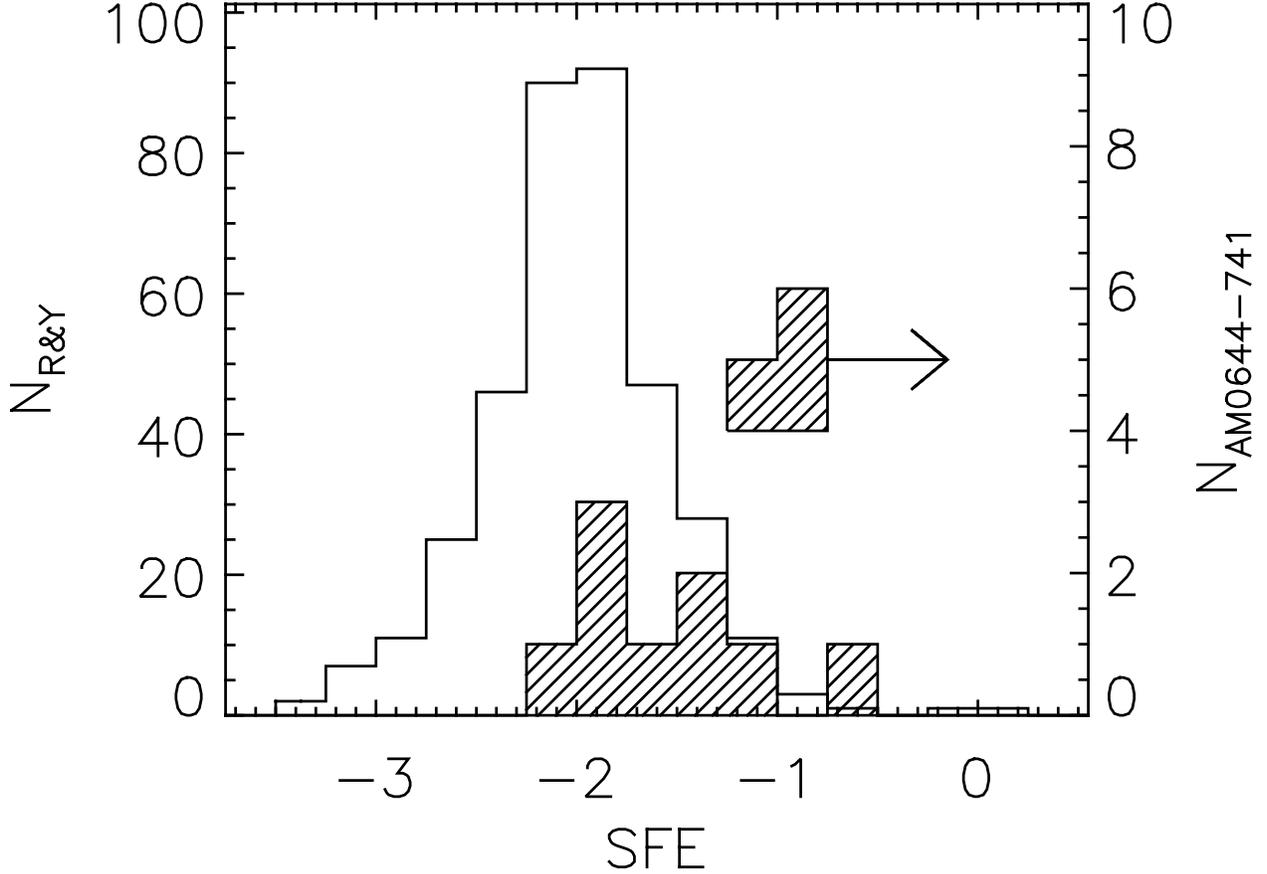}  
\caption{\small Star formation efficiencies, defined SFE
= log$( L_{\rm H\alpha}/M_{\rm H_2} )$, in AM0644-741's ring compared with
spatially resolved regions within a sample of 122 normal and interacting disk galaxies from
\citet{rownd}. The latter are represented by the unfilled histogram. For AM0644-741's
ring, SFE (and upper-limits) are shown using hatched histograms. 
$L_{\rm H\alpha}$ and $M_{\rm H_2}$ are in solar units in the above equation. To be consistent 
with Rownd \& Young we use their $X_{\rm CO}$, correct $L_{\rm H\alpha}$ for foreground
extinction only, and do not correct $M_{\rm H_2}$ for helium. SFE in the ring
appears somewhat elevated relative to the galaxies in Rownd \& Young's sample,
though with an exceptionally wide (factor of $31$) dispersion.}
\end{figure}

\begin{figure}
\figurenum{13}
\epsscale{0.9}
\plotone{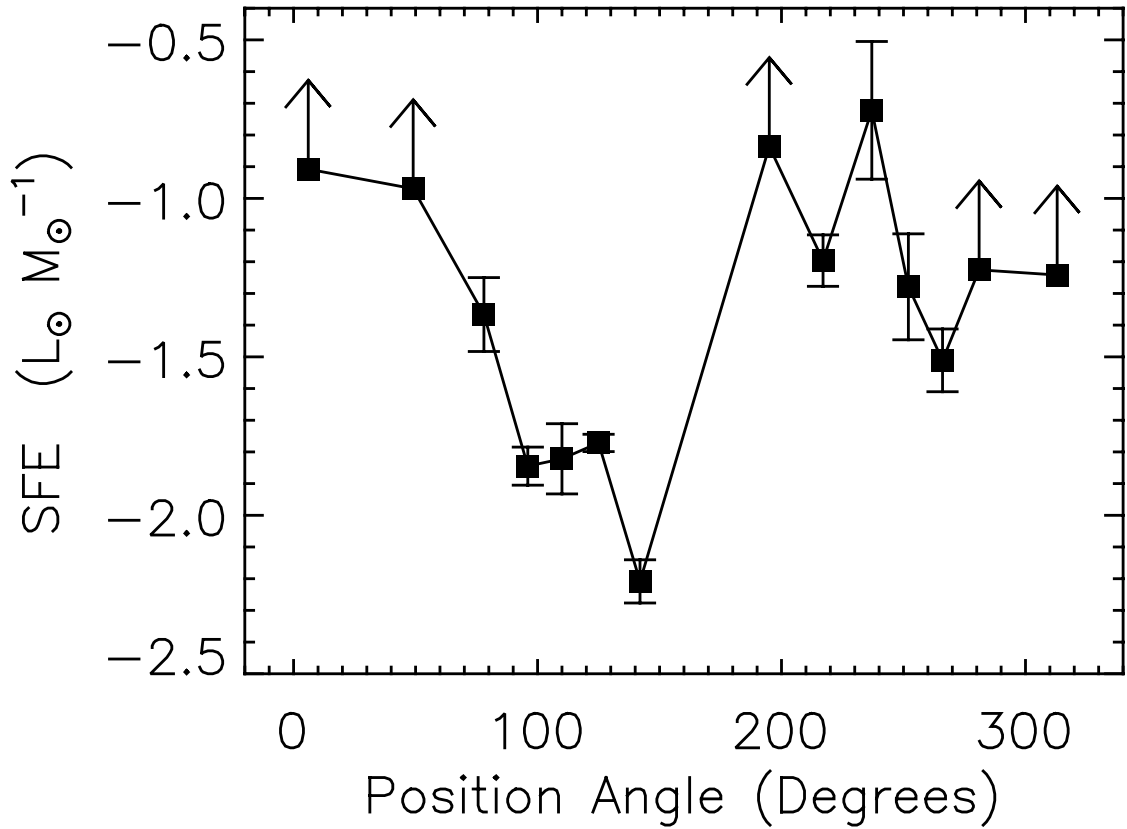}  
\caption{\small Variations in star formation efficiency around AM0644-741's ring. SFE
appears systematically higher in the starburst southwest quadrant relative to the
northern quadrant. As in all the azimuthal plots, the plotted points run from P1
(far left) to P14 (far right).}
\end{figure}

\begin{figure}
\figurenum{14}
\epsscale{0.5}
\plotone{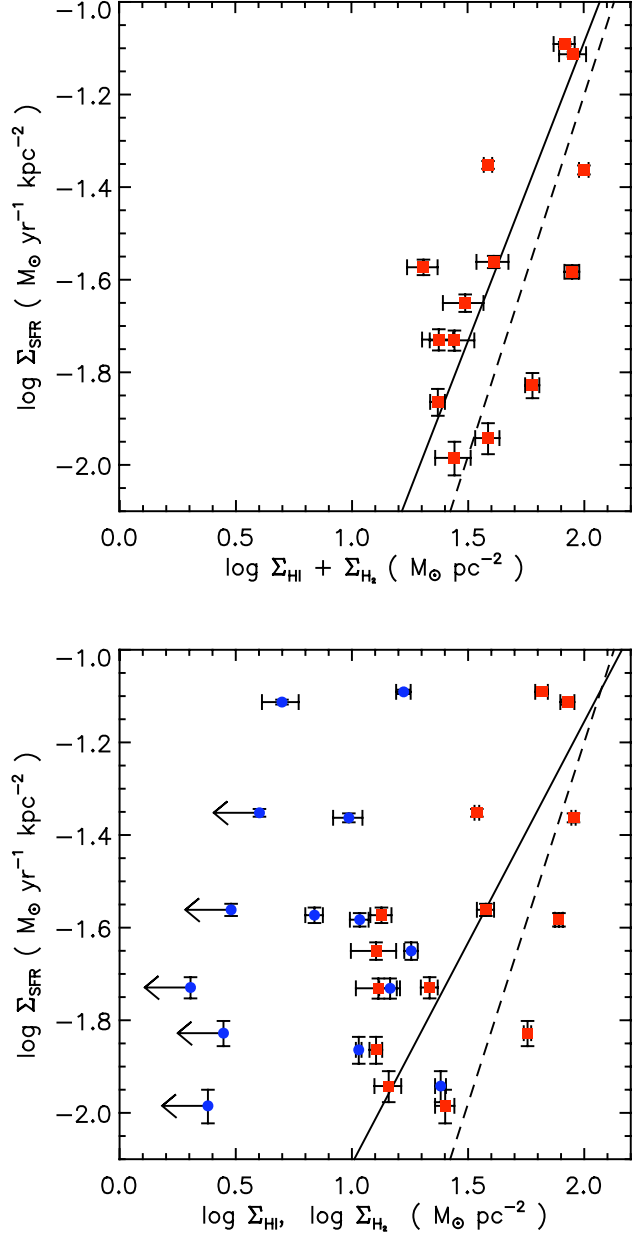}
\caption{\small The star formation law in AM0644-741's ring. Top: the correlation between
SFR density and atomic plus molecular hydrogen surface densities is shown with 
red squares. The solid line
shows a bivariate least-squares fit to the data for a Schmidt law expressed in power-law form:
log~$\Sigma_{\rm SFR} = (1.29 \pm 0.06)$~log$~\Sigma_{\rm HI + H_{2}} - (3.67 \pm 0.02)$,
with a correlation coefficient $r=0.63$. The dashed line shows 
the fit for the spatially resolved star-forming regions in the inner
14 kpc of M51a, log~$\Sigma_{\rm SFR} = (1.56 \pm 0.04)~$log$~\Sigma_{\rm HI + H_{2}} - 
(4.32 \pm 0.09)$ (K07) for comparison. Bottom: the relation between SFR density
and HI (red squares) and H$_2$ (blue circles) surface densities shown separately. Molecular hydrogen
and SFR density are uncorrelated in AM0644-741's ring (ignoring the limits, actually
{\em anti}correlated; $r=-0.48$), while $\Sigma_{\rm SFR}$ scales linearly with
HI surface density: log~$\Sigma_{\rm SFR} = 
(0.96 \pm 0.04)$~log~$\Sigma_{\rm HI} - (3.06 \pm 0.01)$, with
$r=0.63$. $\Sigma_{\rm SFR}$ has been corrected for extinction as described in  
Section $3.1.1$. We have not included a helium contribution to
enable direct comparisons with K07.}
\end{figure}

\clearpage
\begin{figure}
\figurenum{15}
\epsscale{0.8}
\plotone{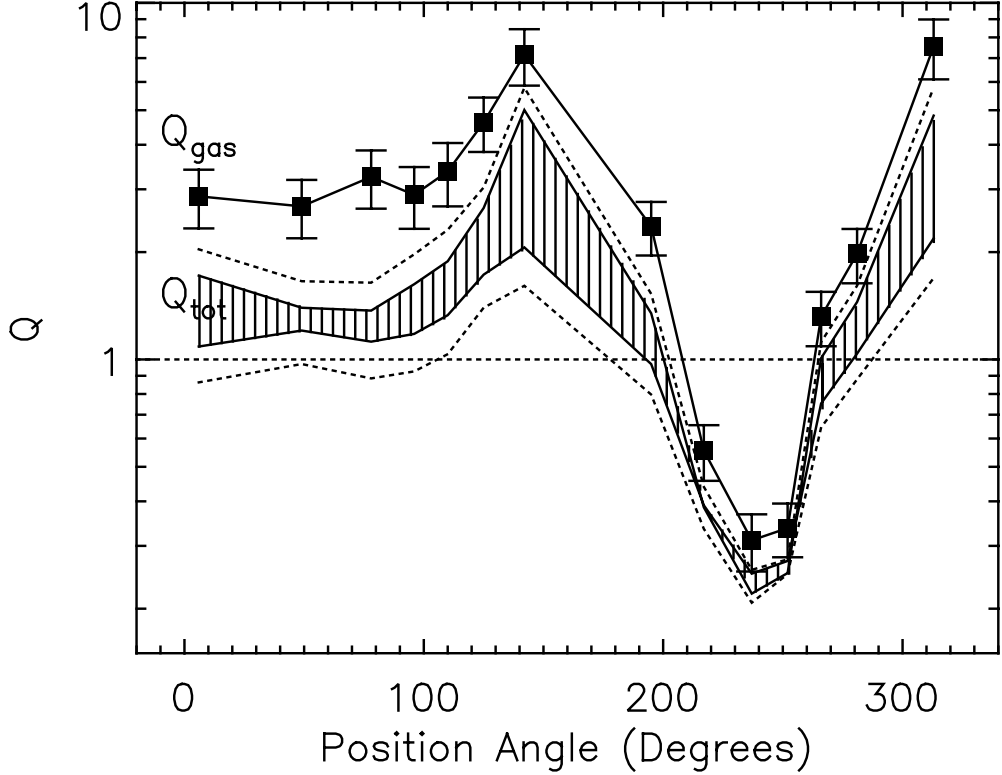}   
\caption{\small Toomre stability parameter $Q$ for AM0644-741's ring, starting at P1 (far left)
and ending at P14 (far right). The ring will be susceptible
to the growth of gravitational instabilities wherever $Q < 1$. The filled squares show the
$Q$ parameter for gas only ($Q_{\rm gas}$) using both atomic and molecular components (including
helium), and velocity dispersions derived from the line widths after
correcting for rotation/expansion smearing within the $22\arcsec$ diameter beams. Only three
ring positions (P9 - 11) possess $Q_{\rm gas}\la1$. Elsewhere the gas ring is stable. 
The hatched region shows how the ring's gravitational stability changes when the 
stellar component is included, using $Q_{\rm tot} = {{\kappa}\over{\pi G}} ( {{\Sigma_{\rm gas}}\over{\sigma_{\rm gas}}} + {{\Sigma_*}\over{\sigma_*}} )^{-1}$ \citep{wang}. Its width
reflects the range of likely stellar velocity dispersions: $\sigma_* = 
\sigma_{\rm gas}$ (upper-range) or $\sigma_* \la d_{\rm ring}/\tau_{\rm B5} \approx 50$ km s$^{-1}$
(lower-range) as implied by the narrow H$\alpha$ rings, where 
$\tau_{\rm B5}=40$ Myr is the main-sequence lifetime of a B5 star. The dotted lines 
represent the formal uncertainty in $Q_{\rm tot}$.
The ring remains largely {\em stable} ($Q_{\rm tot}\approx 1.5-4$) after including
stars except for P9-12 in the starburst southwest quadrant, where $Q_{\rm tot}<1$. Unless we
are underestimating $\Sigma_{\rm H_2}$ (or overestimating $\sigma_*$), star formation 
in the ring is not primarily triggered by large-scale gravitational instabilities.}
\end{figure}

\begin{figure}
\figurenum{16}
\epsscale{0.6}
\plotone{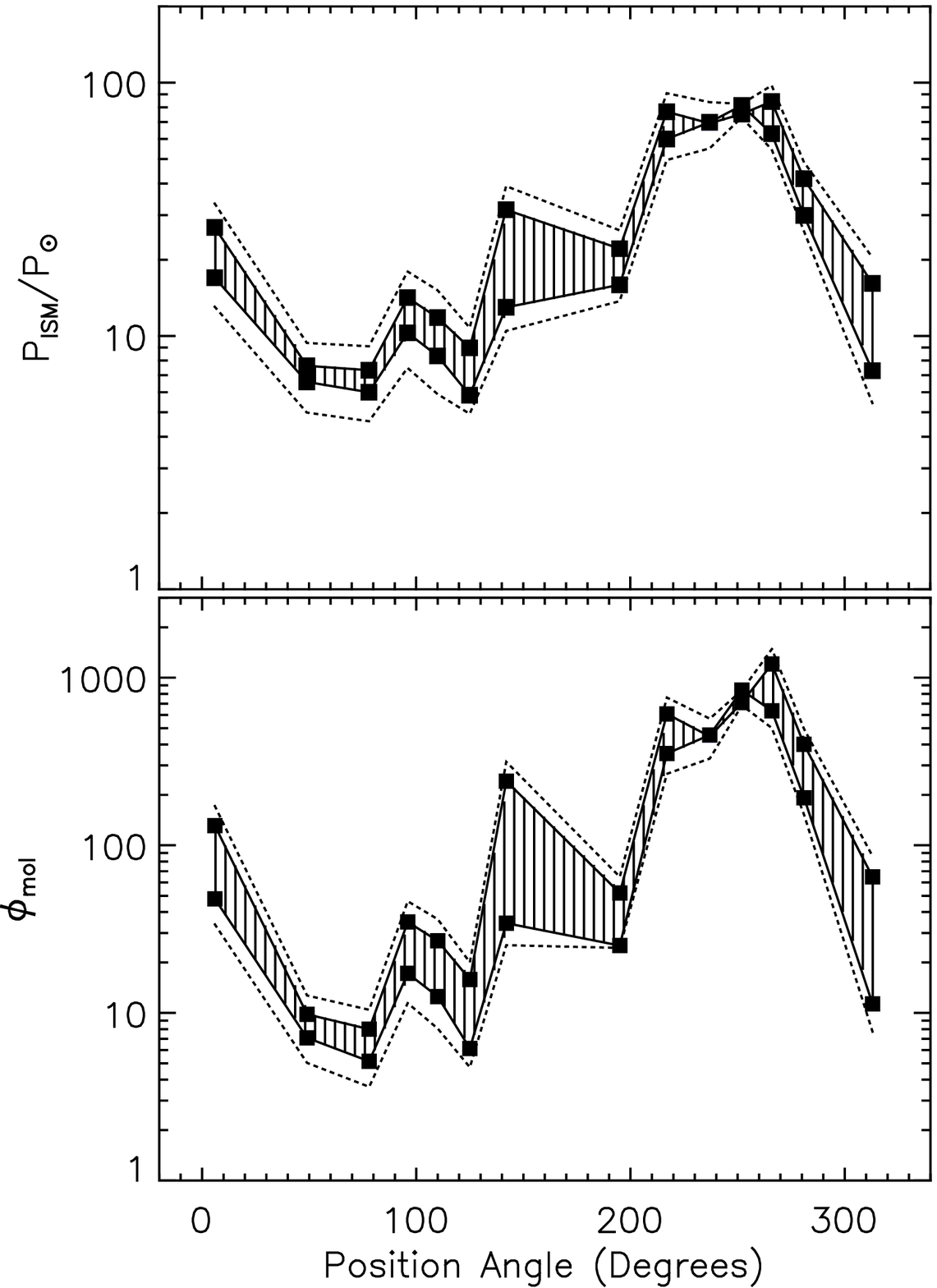}    
\caption{\small Top: azimuthal variations in the ISM pressure in AM0644-741's ring, defined
$P_{\rm ISM} = {{\pi G}\over{2 k_{\rm B}}} ( \Sigma_{\rm gas}^2  + 
{{\sigma_{\rm gas}}\over{\sigma_*}}\Sigma_{\rm gas}\Sigma_*)$ \citep{elmegreen93}, where
we use measured $\Sigma_*$, $\Sigma_{\rm gas}$, and $\sigma_{\rm gas}$. The hatched 
region reflects the uncertainty in the stellar velocity dispersion, which we bracket 
with $\sigma_* = \sigma_{\rm gas}$ and $\sigma_* \la d_{\rm ring}/\tau_{\rm B5} \approx 50$ km s$^{-1}$, 
the latter derived from the narrow width of the H$\alpha$ rings and the main-sequence
lifetime of a B5 star. $P_{\rm ISM}$ is normalized by the local ISM pressure
($P_{\odot} = 10^{4}$ K cm$^{-3}$). High $P_{\rm ISM}$ promotes the prompt
conversion of HI to H$_2$. The trends evident in this plot do not match
the observed azimuthal variations of $f_{\rm mol}$ in Figure 8, indicating that the ISM
pressure is not the dominant factor determining the ratio of atomic to molecular gas mass.
Bottom: molecular fraction parameter $\phi_{mol} \equiv
(P_{\rm ISM}/P_{\odot})^{2.2} (\chi_{\rm UV}/\chi_{\odot})^{-1}$ \citep{elmegreen93} 
for the same 14 positions. $\chi_{\rm UV}$ is estimated using $IRAS$ and MIPS
70 $\mu$m data as described in Section $4.1.2$. The hatched region reflects the uncertainty 
in the stellar velocity dispersion. The ISM will be primarily molecular ($f_{\rm mol} \approx 1$) 
wherever $\phi_{mol}\ga 1$. This condition is satisfied everywhere in AM0644-741's ring, 
leading to the expectation of a primarily molecular ISM, not the molecular poor ISM inferred 
from $^{12}$CO measurements. The ring positions run from P1 (far left) to
P14 (far right).}
\end{figure}

\begin{figure}
\figurenum{17}
\epsscale{1.0}
\plotone{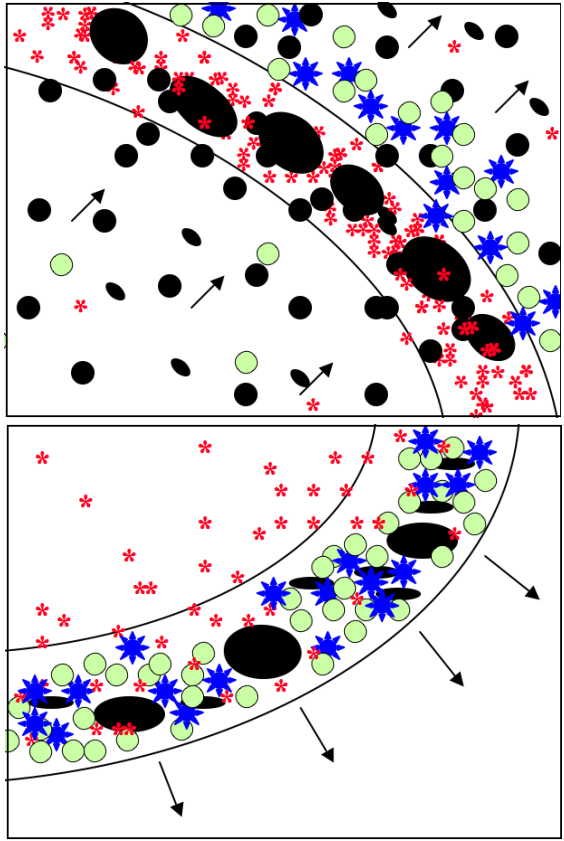}   
\caption{\small Schematic to illustrate differences between the conditions and
evolution of the ISM for a grand design spiral arm as in M~51 (top) and the
ring of a large and evolved ring galaxy such as AM0644-741 or the Cartwheel (bottom).
In a spiral arm molecular (black) and atomic (green) clouds flow into the large-amplitide
density wave, where giant molecular cloud complexes (GMCs) are formed and star formation 
initiated. The GMCs eventually flow out of the spiral density wave after $\tau \approx 15$ 
Myr but by now feel the effects of the embedded star formation, which results in young
blue star clusters, HII complexes, and photodissociated HI being found downstream from
the molecular dominated arm \citep[e.g.,][]{vogel}. The escaping molecular cloud fragments
will not experience such high ISM density, pressure, and FUV fields until an arm is again
encountered $\approx 100$ Myr later. By contrast, in an evolved and robustly star-forming
ring galaxy the molecular clouds are confined to the
expanding ring for $>100$ Myr, where they are subjected to high FUV
radiation fields for much longer periods. The result is a ring dominated by 
photodissociated HI and (on average) smaller molecular clouds, where CO will only be adequately 
shielded in the innermost cores (if at all), leading to H$_2$ being substantially underestimated by
$^{12}$CO emission even for solar abundances. $\Sigma_{\rm H\alpha}$ would also
be expected to correlate with atomic but not molecular gas, accounting for AM0644-741's
peculiar star formation law. (The arrows in the bottom panel represent the ring's radial flow only).}
\end{figure}

\begin{figure}
\figurenum{18}
\epsscale{1.0}
\plotone{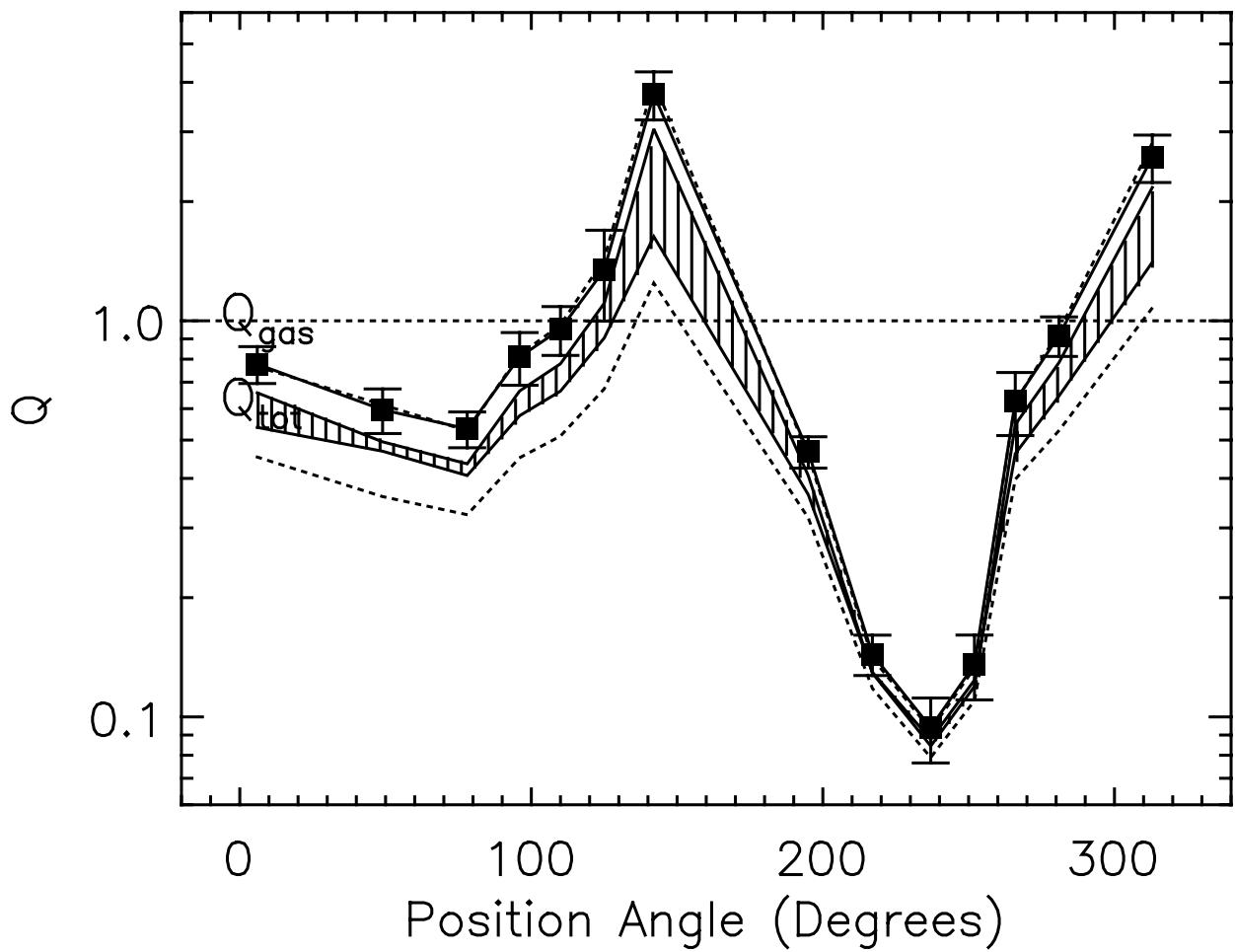}
\caption{\small The resulting Toomre stability parameter for atomic plus molecular gas ($Q_{\rm gas}$) 
and gas plus stars ($Q_{\rm tot}$) in the ring of AM0644-741 after $\Sigma_{\rm H_2}$ is made to 
obey M51a's molecular Schmidt law (i.e., $\Sigma_{\rm H_2} = 1522~\Sigma^{0.73}_{\rm SFR}$). As
before, the plotted points run from P1 (far left) to P14 (far right).
Compared to Figure 15, there is a much better agreement between $Q_{\rm tot}$ 
and the location of intense star formation, in particular P1, P3, and P8-12. 
The point of this exercise is to show that if a Schmidt law holds in AM0644-741's ring, then
large quantities of H$_2$ not traced by $^{12}$CO sufficient to make large stretches gravitationally
unstable likely exist.}
\end{figure}


\begin{thebibliography}

\bibitem[Allen et al.(2004)]{allen}
Allen, R. J., Heaton, H. I., \& Kaufman, M. J. 2004, \apj, 608, 314

\bibitem[Antunes \& Wallin(2007)]{antunes}
Antunes, A. \& Wallin, J. 2007, \apj, 670, 261

\bibitem[Appleton \& Marston(1997)]{applemarston97}
Appleton, P. \& Marston, A. 1997, \aj, 113, 201

\bibitem[Appleton \& Struck(1987)]{asm87}
Appleton, P. \& Struck-Marcell, C. 1987, \apj, 318, 103

\bibitem[Arnaboldi et al.(1997)]{arnaboldi97}
Arnaboldi, M., Oosterloo, T., Combes, F., Freeman, K. C., \& Koribalski, B. 1997, \aj, 113, 585

\bibitem[Arp \& Madore(1987)]{arp}
Arp, H. J. \& Madore, B. F. 1987, A Catalog of Southern Peculiar
Galaxies and Associations (Cambridge: Cambridge Univ. Press)

\bibitem[Bally \& Scoville(1980)]{bally}
Bally, J. \& Scoville, N. 1980, \apj, 239, 121

\bibitem[Bell et al.(2006)]{bell06}
Bell, E. F., Phelps, S., Somerville, R., et al. 2006, \apj, 652, 270

\bibitem[Beltran et al.(2001)]{beltran}
Beltran, J. C., Pizzella, A., Corsini, E. M., et al. 2001, \aap, 374, 394

\bibitem[Benedict et al.(1992)]{fritz92}
Benedict, G. F., Higdon, J. L., Tollestrup, E., Hahn, J., \& Harvey, P. 1992, \aj, 103, 757

\bibitem[Benedict et al.(1996)]{fritz96}
Benedict, G. F., Smith, B. J., \& Kenney, J. 1996, \aj, 112, 1318

\bibitem[Bessel(1979)]{bessel79}
Bessel, M. S. 1979, \pasp, 91, 589

\bibitem[Blitz \& Rosolowsky(2006)]{blitz}
Blitz, L. \& Rosolowsky, E. 2006, \apj, 650, 933

\bibitem[Bloemen et al.(1986)]{bloemen}
Bloemen, J. G., Strong, A. W., Mayer-Hasselwander, H. A. et al. 1986, \aap, 154, 25

\bibitem[Braine et al.(1993)]{braine93}
Braine, J., Combes, F., Casoli, F. et al. 1993, \aaps, 97, 887

\bibitem[Braun \& Thilker(2004)]{braun}
Braun, R., \& Thilker, D. A. 2004, \aap, 417, 421

\bibitem[Caffau \& Ludwig(2008)]{caffau}
Caffau, E., \& Ludwig, H.-G.\ 2008, in IAU Symposium, 252, The Art of Modeling Stars
in the 21st Century (Cambridge: Cambridge Univ. Press), 35

\bibitem[Calzetti et al.(2007)]{calzetti}
Calzetti, D., Kennicutt, R. C., Engelbracht, C. W., et al. 2007, \apj, 666, 870

\bibitem[Conselice et al.(2003)]{conselice}
Conselice, C. J., Bershady, M., Dickinson, M., \& Papovich, C. 2003, \aj, 126, 1183

\bibitem[Corwin, Buta \& de~Vaucouleurs(1994)]{rc3}
Corwin, H. C., Buta, R. J., \& de Vaucouleurs, G. 1994, \aj, 108, 2128

\bibitem[Curran et al.(2008)]{curran}
Curran, S. J., Koribalski, B. S., \& Bains, I. 2008, \mnras, 289, 63

\bibitem[Dame et al.(2001)]{dame}
Dame, T. M., Hartmann, D., \& Thaddeus, P. 2001, \apj, 547, 792

\bibitem[De~Robertis \& Shaw(1988)]{derobertis}
De Robertis, M. M. \& Shaw, R. A. 1988, \apj, 329, 629

\bibitem[Deul \& den~Hartog(1990)]{deul}
Deul, E. R. \& den Hartog, R. H. 1990, \aap, 229, 362

\bibitem[Dickman(1978)]{dickman}
Dickman, R. L. 1978, \apjs, 37, 407

\bibitem[Draine(1978)]{draine}
Draine, B. T. 1978, \apjs, 36, 595

\bibitem[Dressler(1980)]{dressler}
Dressler, A. 1980, \apj, 236, 531

\bibitem[Elmegreen(1993)]{elmegreen93}
Elmegreen, B. G. 1993, \apj, 411, 170

\bibitem[Elmegreen \& Elmegreen(1983)]{elmegreen2}
Elmegreen, B. G., \& Elmegreen, D. M. 1983, \mnras, 203, 31

\bibitem[Engelbracht et al.(2007)]{engelbracht}
Engelbracht, C. W., Blaylock, M., Su, K. Y., et al., 2007, \pasp, 119, 994

\bibitem[Ferguson et al.(2000)]{ferguson}
Ferguson, H. C., Dickinson, M. \& Williams, R. 2000, \araa, 38, 667

\bibitem[Few \& Madore(1986)]{fewmadore}
Few, J. M. \& Madore, B. F. 1986, \mnras, 222, 673

\bibitem[Few, Madore \& Arp(1982)]{fma} 
Few, J. M., Madore, B. F., \& Arp, H. J.  1982, \mnras, 199, 633  (FMA)

\bibitem[Fosbury \& Hawarden(1977)]{fh}
Fosbury, R. A. E. \& Hawarden, T. G. 1977, \mnras, 178, 473

\bibitem[Gerin et al.(1991)]{gerin91}
Gerin, M., Casoli, F., \& Combes F. 1991, \aap, 251, 32

\bibitem[Gerin et al.(1988)]{gerin88}
Gerin, M., Combes, F. \& Nakai, N. 1988, \aap, 203, 44

\bibitem[Gerola \& Seiden(1978)]{gerola}
Gerola, H. \& Seiden, P. E. 1978, \apj, 223, 129 

\bibitem[Gordon et al.(2005)]{gordon}
Gordon, K. D., Rieke, G., Engelbracht, C. W., et al. 2005, \pasp, 117, 503

\bibitem[Gottesman \& Mahon(1990)]{mahon}
Gottesman, S.~T., \& Mahon, M.~E.\ 1990, IAU Colloquium 124, Paired and Interacting Galaxies, 209

\bibitem[Graham(1974)]{graham74} 
Graham, J. A. 1974, Observatory, 94, 290

\bibitem[Graham(1982)]{graham82}
Graham, J. A. 1982, \pasp, 94, 244

\bibitem[Heckman et al.(1986)]{heckman}
Heckman, T. M., Smith, E. P., Baum, S. A., et al. 1986, \apj, 311, 526

\bibitem[Hibbard et al.(2001)]{hibbard}
Hibbard, J. E., van der Hulst, J., Barnes, J. \& Rich, R. 2001, \aj, 122, 2969

\bibitem[Hidaka \& Sofue(2002)]{hidaka}
Hidaka, M. \& Sofue, Y. 2002, \pasj, 54, 223

\bibitem[Higdon(1995)]{wof1} 
Higdon, J. L. 1995, \apj, 455, 524

\bibitem[Higdon(1996)]{wof2} 
Higdon, J. L.  1996, \apj, 467, 241

\bibitem[Higdon et al.(1997)]{higdonrandlord}
Higdon, J. L., Rand, R. J., \& Lord, S. 1997, \apj, 489, L133

\bibitem[Higdon \& Wallin(1997)]{wof3} 
Higdon, J. L., \& Wallin, J. F.  1997, \apj, 474, 686 (HW97)

\bibitem[Higdon et al.(2001)]{higdon01}
Higdon, J.~L., Wallin, J.~F., Rand, R.~J., \& Sage, L.\ 2001, Gas and Galaxy Evolution, 240, 860

\bibitem[Hopkins et al.(2003)]{hopkins}
Hopkins, A., Miller, C. J., Nichol, R. C., et al. 2003, \apj, 599, 971

\bibitem[Horellou et al.(1995)]{hor95} 
Horellou, C., Casoli, F., Combes, F., \& Dupraz, C.  1995, \aap, 298, 743

\bibitem[Horellou et al.(1998)]{hor98}
Horellou, C., Charmandaris, V., Combes, F., et al. 1998, \aap, 340, L51

\bibitem[Hsieh et al.(2008)]{hsieh}
Hsieh, Pei-Ying, Matsushita, S., Lim, J., Kohno, K., \& Sawada-Satoh, S. 2008, \apj, 683, 70

\bibitem[Israel(1997)]{israel}	
Israel, F. P. 1997, \aap, 328, 471

\bibitem[Jeske(1986)]{jeske}
Jeske, N. 1986, Ph D  thesis, Univ. California, Berkeley

\bibitem[Jogee et al.(2009)]{jogee}
Jogee, S., Miller, S. H., Penner, K., et al. 2009, \apj, 697, 1971

\bibitem[Kaufman et al.(1999)]{kaufman99}
Kaufman, M. J., Wolfire, M. G., Hollenbach, D. J. \& Luhman, M. L. 1999, \apj, 527, 795

\bibitem[Kennicutt(1989)]{kennicutt89}
Kennicutt, R. C. 1989, \apj, 344, 685

\bibitem[Kennicutt(1998a)]{kennicutt98a}
Kennicutt, R. C. 1998a, \araa, 36, 189

\bibitem[Kennicutt(1998b)]{kennicutt98b}
Kennicutt, R. C. 1998b, \apj, 498, 541

\bibitem[Kennicutt et al.(2003)]{kennicutt03}
Kennicutt, R. C., Armus, L., Bendo, G., et al. 2003, \pasp, 115, 928

\bibitem[Kennicutt et al.(2007)]{kennicutt07}
Kennicutt, R. C., Calzetti, D., Walter, F., et al.  2007, \apj, 671, 333 (K07)

\bibitem[Kennicutt et al.(2009)]{kennicutt09}
Kennicutt, R. C., Hao, C., Calzetti, D., et al. 2009, \apj, 703, 1672

\bibitem[Kerr \& Lynden-Bell(1986)]{kerr}
Kerr, F. J. \& Lynden-Bell, D. 1986, \mnras, 221, 1023

\bibitem[Kim et al.(2010)]{kimostriker}
Kim, C. G., Kim W. T., \& Ostriker, E. C. 2010, \apj, 720, 1454

\bibitem[Komatsu et al.(2011)]{wmap7}
Komatsu, E., Smith, K. M., Dunkley, J., et al. 2011, \apjs, 192, 18

\bibitem[Korchagin et al.(2001)]{korchagin01}
Korchagin, V., Mayya, Y. \& Vorobyov, E. 2001, \apj, 554, 281

\bibitem[Korchagin et al.(1998)]{korchagin98}
Korchagin, V., Mayya, Y., Vorobyov, E., \& Kembhavi, A. 1998, \apj, 495, 757

\bibitem[Korchagin et al.(1999)]{korchagin99}
Korchagin, V., Vorobyov, E., \& Mayya, Y. 1999, \apj, 522, 767

\bibitem[Kuijken \& Gilmore(1989)]{kuijken}
Kuijken, K. \& Gilmore, G. 1989, \mnras, 239, 605 

\bibitem[Leitherer et al.(1999)]{starburst99}
Leitherer, C., Schaerer, D., Goldader, J., et al. 1999, \apjs, 123, 3

\bibitem[Leroy et al.(2009)]{leroy09}
Leroy, A., Bolatto, A., Bot, C., et al. 2009, \apj, 702, 352

\bibitem[Leroy et al.(2005)]{leroy05}
Leroy, A., Bolatto, A. D., Simon, J. D., \& Blitz, L. 2005, \apj, 625, 763

\bibitem[Leroy et al.(2008)]{leroy08}
Leroy, A., Walter, F., Brinks, E., et al. 2008, \aj, 136, 2782

\bibitem[Lynds \& Toomre(1976)]{lynds}
Lynds, R. \& Toomre, A. 1976, \apj, 209, 382

\bibitem[Maddalena \& Morris(1987)]{maddalena}
Maddalena, R. J. \& Morris, M. 1987, \apj, 323, 179

\bibitem[Madden et al.(1997)]{madden}
Madden, S. C., Poglitsch, A., Gies, N., Stacey, G., \& Townes, C. 1997, \apj, 483, 200

\bibitem[Makovoz \& Marleau(2005)]{makovoz}
Makovoz, D., \& Marleau, F. R. 2005, \pasp, 117, 1113

\bibitem[Maloney \& Black(1988)]{maloney}
Maloney, P. \& Black, J. 1988, \apj, 325, 389

\bibitem[Mapelli et al.(2008)]{mapelli}
Mapelli, M., Moore, B., Ripamonti, E., et al. 2008, \mnras, 383, 1223

\bibitem[Marcum et al.(1993)]{marcum}
Marcum, P. M., Appleton, P., \& Higdon, J. L. 1993, \apj, 399, 57

\bibitem[Marston \& Appleton(1995)]{marston}
Marston, A. P. \& Appleton, P.  1995, \aj, 109, 1002

\bibitem[Martin et al.(2005)]{martin}
Martin, D. C., Fanson, J., Schiminovich, D., et al. 2005, \apj, 619, L1

\bibitem[McKee(1990)]{mckee}
McKee, C. F. 1990, in ASP Conf. Ser. 12, The Evolution of the Insterstellar
Medium (San Francisco, CA: ASP), 3

\bibitem[Mebold et al.(1977)]{mebold}
Mebold, U., Goss, W., \& Fosbury, R. A. E.  1977, \mnras, 180, 11P

\bibitem[Moore et al.(1996)]{moore}
Moore, B., Katz, N., Lake, G., Dressler, A. \& Oemler, A. 1996, Nature, 379, 613

\bibitem[Morrissey et al.(2007)]{morrissey07}
Morrissey, P., Conrow, T., Barlow, T., et al. 2007, \apjs, 173, 682

\bibitem[Morrissey et al.(2005)]{morrissey05}
Morrissey, P., Schiminiovich, D., Barlow, T., et al. 2005, \apj, 619, 7

\bibitem[Moshir et al.(1990)]{moshir}
Moshir, M., Kopan, G., Conrow, T., et al. 1990, \baas, 22, 1325

\bibitem[Mullan et al.(2011)]{mullan}
Mullan, B., Konstantopoulos, I., Kepley, A., et al. 2011, \apj, 731, 93

\bibitem[Murray et al.(2010)]{murray}
Murray, N., Quataert, E., \& Thompson, T. A. 2010, \apj, 709, 191

\bibitem[Nagao et al.(2006)]{nagao}	
Nagao, T., Maiolino, R., \& Marconi, A. 2006, \aap, 459, 85

\bibitem[Oort(1954)]{oort}
Oort, J. H. 1954, Bull. Astr. Institute Netherlands, 12, 177

\bibitem[Ramos Almeida et al.(2011)]{ramos}
Ramos Almeida, C. R., Tadhunter, C. N., Inskip, K. J., et al. 2011, \mnras, 410, 1550

\bibitem[Rand(1993)]{rand93}
Rand, R. J. 1993, \apj, 404, 595

\bibitem[Rand(1995)]{rand95}
Rand, R. J. 1995, \aj, 109, 2444

\bibitem[Richard et al.(2010)]{richard}
Richard, S., Brook, C., Martel, H., et al. 2010, \mnras, 402, 1489

\bibitem[Richards et al.(1999)]{richards}
Richards, E., Fomalont, E., Kellermann, K., et al. 1999, \apj, 526, L73

\bibitem[Rownd \& Young(1999)]{rownd}
Rownd, B. K. \& Young, J. S. 1999, \apj, 118, 670

\bibitem[Safronov(1960)]{safronov}
Safronov, V. S.  1960, Ann. d'Astrophys., 23, 979

\bibitem[Saitoh et al.(2009)]{saitoh}
Saitoh, R., Kaisaka, H., Kokubo, E., et al. 2009, \pasj, 61, 481

\bibitem[Sanders \& Mirabel(1996)]{sanders96}
Sanders, D. B., \& Mirabel, I. F. 1996, \araa, 34, 749

\bibitem[Sanders et al.(1984)]{sanders84}
Sanders, D. B., Solomon, P. M., \& Scoville, N. Z. 1984, \apj, 276, 182

\bibitem[Schmidt(1959)]{schmidt}
Schmidt, M. 1959, \apj, 129, 243

\bibitem[Smail et al.(1997)]{smail}
Smail, I., Ivison, R. J., \& Blain, A. W. 1997, \apj, 490, L5

\bibitem[Solomon et al.(1987)]{solomon}
Solomon, P. M., Rivolo, A. R., Barrett, J. \& Yahil, A. 1987, \apj, 319, 730

\bibitem[Solomon \& Sage(1988)]{sage}
Solomon, P. M. \& Sage, L. 1988, \apj, 334, 613

\bibitem[Stacey et al.(1991)]{stacey}
Stacey, G. J., Geis, N., Genzel, R., et al. 1991, \apj, 373, 423

\bibitem[Strong et al.(1988)]{strong}
Strong, A. W., Bloemen, J., Dame, T., et al., 1988, \aap, 207, 1

\bibitem[Struck \& Higdon(1993)]{struckhigdon}
Struck, C. J. \& Higdon, J. L. 1993, \apj, 411, 108

\bibitem[Tacconi et al.(2008)]{tacconi}
Tacconi, L. J., Genzel, R., Smail, I., et al. 2008, \apj, 680, 246

\bibitem[Toomre(1964)]{toomre64}
Toomre, A. 1964, \apj, 139, 1217

\bibitem[Toomre(1974)]{toomre74}
Toomre, A. 1974, in IAU Symp. 58, The Formation and Dynamics of 
Galaxies (Cambridge: Cambridge Univ. Press), 347

\bibitem[Vazquez \& Leitherer(2005)]{vazquez}
Vazquez, G. A., \& Leitherer, C. 2005, \apj, 621, 695

\bibitem[Vogel et al.(1988)]{vogel}
Vogel, S. N., Kulkarni, S. R., \& Scoville, N. Z. 1988, \nat, 334, 402

\bibitem[Wallin \& Struck(1994)]{wallin}
Wallin, J. F. \& Struck, C. 1994, \apj, 433, 631

\bibitem[Wang \& Silk(1994)]{wang}
Wang, B. \& Silk, J. 1994, \apj, 427, 759

\bibitem[Weedman(1983)]{dickhead}
Weedman, D. 1983, \apj, 266, 479

\bibitem[Weil \& Hernquist(1993)]{weil} 
Weil, M. \& Hernquist, L. 1993, \mnras, 261, 804

\bibitem[Werner et al.(2004)]{werner}
Werner, M. W., Roellig, T., Low, F., et al. 2004, \apjs, 154, 1

\bibitem[White \& Rees(1978)]{white}
White, S. D. \& Rees, M. J. 1978, \mnras, 183, 341

\bibitem[Wielen(1977)]{wielen}
Wielen, R.  1977, \aap, 60, 263

\bibitem[Williams et al.(1996)]{williams}
Williams, R. E., Blacker, B., Dickinson, M., et al. 1996, \aj, 112, 1335

\bibitem[Wilson(1995)]{wilson}
Wilson, C. D.  1995, \apj, 488, L97

\bibitem[Wootten \& Thompson(2009)]{wootten}
Wootten, A. \& Thompson, R. 2009, Proc. IEEE, 99, 1463

\bibitem[York et al.(2004)]{york}
York, D., Evensen, N. M., Lopez Martinez, M., and De Basabe Delgado, J. 2004,
Am. J. Phys., 72, 367

\bibitem[Young et al.(1996)]{young}
Young, J. S., Allen, L., Kenney, J., Lesser, A., \& Rownd, B.  1996, \aj, 112, 1903

\end{thebibliography}
\end{document}